\documentclass{article}

\usepackage{arxiv}

\usepackage[utf8]{inputenc} 
\usepackage[T1]{fontenc}    
\usepackage{hyperref}       
\usepackage{url}            
\usepackage{booktabs}       
\usepackage{amsfonts}       
\usepackage{nicefrac}       
\usepackage{microtype}      
\usepackage{lipsum}		
\usepackage{graphicx}
\usepackage{natbib}
\usepackage{doi}
\usepackage{graphicx}
\usepackage{caption}
\usepackage{subcaption}
\usepackage{dcolumn}
\usepackage{bm}
\usepackage{xcolor}
\usepackage{amsmath}
\usepackage{comment} 
\usepackage{amssymb}
\usepackage[T1]{fontenc}
\usepackage{algpseudocode}
\usepackage{authblk}
\usepackage[section]{placeins}

\usepackage[newcommands]{ragged2e}
\usepackage{graphicx,caption}

\usepackage{comment}

\usepackage{graphicx}
\usepackage{dcolumn}
\usepackage{bm}

\usepackage[utf8]{inputenc}
\usepackage[T1]{fontenc}
\usepackage{mathptmx}
\usepackage{etoolbox}

\usepackage{hyperref}
\usepackage{algorithm}
\usepackage{algorithmicx}
\usepackage{algpseudocode}
\usepackage{subcaption}
\usepackage{xcolor}
\usepackage{amsmath}
\usepackage{placeins}
\usepackage{booktabs}
\usepackage{bm}

\newcommand{\tal}{\theta^{k}}

\newcommand{\sbia}{\bm S_i^{k}}

\newcommand{\p}[2]{\cfrac{\partial#1}{\partial#2}}

\makeatletter
\def\@email#1#2{%
 \endgroup
 \patchcmd{\titleblock@produce}
  {\frontmatter@RRAPformat}
  {\frontmatter@RRAPformat{\produce@RRAP{*#1\href{mailto:#2}{#2}}}\frontmatter@RRAPformat}
  {}{}
}%
\makeatother
\begin{document}


\title{Physics Informed Machine Learning with Smoothed Particle Hydrodynamics:\\ Hierarchy of Reduced Lagrangian Models of Turbulence}

\author[1,2,3]{Michael Woodward}
\author[3]{Yifeng Tian}
\author[1,2,3]{Criston Hyett}
\author[3]{Chris Fryer}
\author[1,2]{Mikhail Stepanov}
\author[3]{Daniel Livescu}
\author[1,2]{Michael Chertkov}

\affil[1]{Graduate Interdisciplinary Program in Applied Mathematics, UArizona, Tucson, AZ 85721}
\affil[2]{Department of Mathematics, UArizona, Tucson, AZ 85721}
\affil[3]{Computer, Computational and Statistical Sciences Division, LANL, Los Alamos, NM 87544}

\renewcommand{\shorttitle}{PIML-SPH: Reduced Lagrangian Models}

\maketitle


\begin{abstract}
Building efficient, accurate and generalizable reduced order models of developed turbulence remains a major challenge. This manuscript approaches this problem by developing a hierarchy of parameterized reduced Lagrangian models for turbulent flows, and investigates the effects of enforcing physical structure through Smoothed Particle Hydrodynamics (SPH) versus relying on neural networks (NN)s as universal function approximators. Starting from Neural Network (NN) parameterizations of a Lagrangian acceleration operator, this hierarchy of models gradually incorporates a weakly compressible and parameterized SPH framework, which enforces physical symmetries, such as Galilean, rotational and translational invariances. Within this hierarchy, two new parameterized smoothing kernels are developed in order to increase the flexibility of the learn-able SPH simulators. For each model we experiment with different loss functions which are minimized using gradient based optimization, where efficient computations of gradients are obtained by using Automatic Differentiation (AD) and Sensitivity Analysis (SA). Each model within the hierarchy is trained on two data sets associated with weekly compressible Homogeneous Isotropic Turbulence (HIT): (1) a validation set using weakly compressible SPH; and (2) a high fidelity set from Direct Numerical Simulations (DNS). Numerical evidence shows that encoding more SPH structure improves generalizability to different turbulent Mach numbers and time shifts, and that including the novel parameterized smoothing kernels improves the accuracy of SPH at the resolved scales. 
\end{abstract}

\section{Introduction}
Understanding and predicting turbulent flows is crucial for many engineering and scientific fields, and remains a great unresolved challenge of classical physics \cite{frisch1995}. Turbulent flows are characterized by strong coupling across a broad range of scales and obtaining accurate solutions over all relevant scales currently requires computationally intensive numerical methods, and is often prohibitive in applications \cite{pope_2011}. This motivates the development of reducing the computational cost by only simulating large scale structures instead of the full set of relevant scales \cite{sagaut2006large}. Many data-driven techniques have emerged (or matured) over the last decades, such as the Proper Orthogonal Decomposition  \cite{lumley_book_2012}, Dynamics Mode Decomposition \cite{schmid_2010, dmd_book}, Mori-Zwanzig \cite{lin2021datadriven, tian_2021, mwoodward_mz_bl}, and also many other Machine Learning for Turbulence techniques \cite{CHEN2021, mohan2021learning, lin22_nn_mz, 2021DFD-Yifeng, 2021DFD-Criston,Portwood21}. The main challenge of developing reduced models for turbulence, is in generalizing to flows not seen in training. This motivates our focus in this work on developing \emph{Reduced Lagrangian based Simulators} which encode physical conservation laws independent of the resolution using Smoothed Particle Hydrodynamics (SPH).

SPH \cite{monaghan1977,monaghan1992,monaghan12} has been widely applied to weakly- and strongly compressible turbulence in astrophysics and many engineering applications \cite{shadloo_2015_industry}. It is one of a small set of approaches based on a Lagrangian construction: the fluid quantities follow the flow using particles as opposed to the Eulerian approach which computes flow quantities at fixed locations on a computational mesh. This mesh-free, Lagrangian approximation of Navier-Stokes (NS) is appealing because it naturally unmasks correlations at the resolved scale from sweeping by larger scale eddies \cite{1964Kraichnan,1965Kraichnan} making SPH useful for understanding advection dominated flows, mixing and dispersion, and turbulent transport. One of the main advantages of SPH is that the conservation of mass, energy, and momentum can be enforced within a discrete formulation, ensuring conservation independent of the resolution. Furthermore, the mesh-free nature of SPH is advantageous for highly compressible flows as the Lagrangian particles naturally resolve the variable density regions. Recently \cite{chola_les_sph_2018, tian_22_LLES}, SPH has been connected to Large Eddy Simulations (LES) as a method to coarse grain the Navier-Stokes equations. Furthermore, developing approximation-optimal SPH models (and simulators) for turbulent flows is an ongoing area of research \cite{lind2020_review}, to which this paper contributes by developing two new parameterized smoothing kernels and a physics informed machine learning framework for estimating the parameters of a weakly-compressible SPH formulation fit to Direct Numerical Simulation (DNS) data. 

Numerical simulators were recently blended with modern machine learning tools \cite{ling_kurzawski_templeton_2016,king2018deep,schenck2018_spnet,mohan2018deep,Duraisamy2019,Fonda2019,mohanNN,maulikROM,Ummenhofer2020conv,mohan_livescu,2021Yifeng-PRF} out of which a promising field, coined Physics Informed Machine Learning (PIML), is emerging (and being re-discovered \cite{Lagaris_nn_ode98}). As set, some eight years ago at the first Los Alamos National Laboratory workshop with this name \cite{PIML}, PIML was meant to pivot the mixed community of machine learning researchers on the one hand and scientists and engineers on the other, to discover physical phenomena/models from data. Early work incorporating scientific domain knowledge (e.g. from physics in the form of differential equations) and computational scheme within machine learning algorithms dates back to the 1990s \cite{Lagaris_nn_ode98}. However, in the context of modern machine learning and specifically deep learning, interest in this area has been revived \cite{raissi2, node, rackauckas2020universal, ladicky2015_Rforests}. In part, this is due to the increased computational power afforded by parallelism across both Central, Vector or Tensor Processing Units, along with notable achievements across disciplines (such as scientific applications \cite{sci_ml_applications}, data-compression algorithms \cite{WANG2016}, computer vision \cite{serre_deep_learning_mv}, and natural language processing \cite{nlp_book}). The main computational utilities and strategies of the Physics Informed Machine Learning (PIML) includes embedding physical structure into learn-able models, applying Neural Networks (NNs) as function approximators \cite{hornik_nn_approx89}, differential programming using automatic differentiation \cite{rackauckas2020universal, ad_bucker06}, and optimization tools to minimize a loss function. 

In this manuscript we focus specifically on building PIML scheme(s) which take advantage of the Lagrangian formulation. Other works have pursued similar directions; one of the earliest contributions to this field was made in \cite{ladicky2015_Rforests} where SPH related models were used alongside regression forests (a classical machine learning technique). In \cite{suare2018_evolution} evolutionary algorithms were applied to optimize parameters in SPH flows. Differentiable programming techniques were utilized in \cite{schenck2018_spnet} for Lagrangian robotic control of flows and in \cite{Ummenhofer2020conv} where a continuous Convolutional NN approach was developed. NNs were used in \cite{2021Yifeng-PRF} to train a closure model for Lagrangian dynamics of the velocity gradient tensor in homogeneous isotropic turbulence on the Direct Numerical Simulation (DNS) data. It was shown in \cite{rabault2017performing,lee2017piv,cai2019particle,stulov2021neural} that skillful embedding of NN helps to improve Particle Image Velocimetry techniques to map a Lagrangian representation of the flow into its Eulerian counterpart.  Most recent approach of our team \cite{tian_22_LLES}, which is most related to this work, consisted in developing a fully differentiable, NN-based and Lagrangian Large Eddy Simulator trained on the mix of Eulerian and Lagrangian high-fidelity DNS data.  It was shown in \cite{tian_22_LLES} that the simulator is capable to fit and generalize with respect to Mach numbers, delayed times and advanced turbulence statistics.

We continue the thread of \cite{tian_22_LLES} and develop a hierarchy of learn-able, NN-enforced, Lagrangian simulators to examine reduced order physics at coarse grained scales within the inertial range of homogeneous isotropic turbulent flows. However, in this work we focus on constraining the models using the SPH framework. 
Although including SPH structure will enforce physical constraints, a priori it is not known if this will improve its ability to generalize to different turbulent flows at these scales, since enforcing constraints decreases the expressiveness as compared to the NN based models. Thus, to explore these effects, a hierarchy of parameterized Lagrangian and SPH based fluid models are developed that gradually includes more of the SPH framework, ranging from a Neural Ordinary Differential Equation (ODE) based model \cite{node} to a weakly compressible parameterized SPH formulation. Each model is trained on two sets of the ground truth data: (a) a synthetic weakly compressible SPH simulator, and (b) Eulerian and Lagrangian data from a high-fidelity DNS (similar to the one used in \cite{tian_22_LLES}). Given our focus in this work is on resolving only the large scale structures within the inertial range, the ground truth data is properly coarse-grained, where each model is trained on three different resolutions ($N = 12^3, 16^3, 20^3$). 
An efficient gradient descent is developed using modern optimizers (e.g Adam by \cite{kingma2017adam}), and mixing automatic differentiation (both forward and reverse mode) with the local sensitivity analyses (such as forward and adjoint based methods).

A formulation of this hierarchy along with the learning framework is given in Section \ref{sec:mixed_mode}, where a brief background of the weakly compressible SPH framework used in this work is provided in Section \ref{sec:SPH}. In Section \ref{sec:dns_results}, we first validate the methodology and learning algorithm on the synthetic ground truth SPH data, in which we show the ability to recover parameters within the SPH model as well as to learn the equation of state using NNs (embedded in the SPH framework). Furthermore, in Section \ref{sec:dns_results}, each model is trained on high-fidelity weakly compressible (i.e. low Mach number) DNS data using field and statistical based loss functions. Then a detailed analysis of each model is carried out, from which we show that adding SPH informed structure not only increases the interpretability of the models, but improves generalizability over varying Mach numbers and time scales with respect to both statistical and field based quantitative comparisons. We observe that NNs (considered as universal function approximators \cite{hornik_nn_approx89}) embedded within this structure, such as those used in approximating the Equation of State, also improves this  generalizability over the standard weakly compressible SPH using the cubic smoothing kernel. Moreover, we show that the new proposed smoothing kernels (introduced in Section \ref{sec:mixed_mode}), which when included in the fully informed SPH based model, performs best at generalizing to different DNS flows. 

\section{Smoothed Particle Hydrodynamics}

\label{sec:SPH}

One of the most prominent particle-based Lagrangian methods for obtaining approximate numerical solutions of the equations of fluid dynamics is Smoothed Particle Hydrodynamics (SPH) \cite{monaghan2005}.  Originally introduced independently by \cite{lucy1977} and \cite{monaghan1977} for astrophysical flows, however, over the following decades, SPH has found a much wider range of applications including computer graphics, free-surface flows, fluid-structure interaction,  bio-engineering, compressible flows, galaxies’ formation and collapse, high velocity impacts, geological flows, magnetohydrodynamics, and turbulence  \cite{lind2020_review, shadloo_2015_industry}. Below, we give a brief formulation of a weakly compressible SPH framework and in Section \ref{sec:hiearchy} we use this SPH structure as the basis of our Physics informed machine learning models.

Essentially, SPH is a discrete approximation to a continuous flow field by using a series of discrete particles as interpolation points (using an integral interpolation with smoothing kernel $W$). Using the SPH formalism the Partial Differential Equations (PDEs) of fluid dynamics can be approximated by a system of ODEs for each particle (indexed by $i$), $\forall i \in \{1, 2,... N\} :$ 
\begin{gather} 
    \cfrac{d\bm{r}_i}{dt} = \bm{v}_i,\quad  \\
    \label{wc_sph_av}
    \cfrac{d\bm{v}_i}{dt} = -\sum _{j \neq i}^N m_j \left( \cfrac{P_j}{\rho_j^2} + \cfrac{P_i}{\rho_i^2} + \Pi_{ij} \right) \nabla _i  W_{ij}  + \bm f_{ext}.
\end{gather}
Where $W_{ij} = W(||\bm r_i - \bm r_j||, h)$, $P_i$, and $\rho_i$ represents the smoothing kernel, pressure and density respectively at particle $i$. $\Pi_{ij}$ is an artificial viscosity term \cite{monaghan12} used to approximate the viscous terms. Density and pressure (for the weakly compressible formulation) are computed by $\rho_i = \sum_j m_j W(|\bm r_i - \bm r_j|, h)$, and $P(\rho) = \frac{c ^2 \rho_0}{\gamma} \left[ \left(\frac{\rho}{\rho_0} \right)^{\gamma} - 1 \right]$.

Briefly, Eq.~(\ref{wc_sph_av}) is an approximation of Euler's equations for weakly compressible flows, with an added artificial viscosity term $\Pi_{ij}$.  In this manuscript, a deterministic external forcing $\bm f^i_{ext} = (\theta_{inj}/ke(t)) \bm v_i$ consistent with DNS (as seen in \cite{dlivescu_fext}) provides the energy injection mechanism, where $ke$ is the kinetic energy and $\theta_{inj}$ is an energy injection rate parameter. We also utilize an artificial viscosity $\Pi_{ij}$ described in Eq.~(\ref{eq:artificial_viscosity}), which approximates in aggregate the contributions from the bulk and shear viscosity ($\alpha$), a Nueman-Richtmyer viscosity for handling shocks ($\beta$) \cite{monaghan12, MORRIS1997}, as well as the effective, eddy viscosity effect of turbulent advection from the under-resolved scales, i.e. scales smaller than the mean-particle distance (see Appendix \ref{sec:sph_append} in the appendix for a more detailed discussion).


Fig.~(\ref{fig:sph_flow}) shows consecutive snapshots of an exemplary multi-particle SPH flow in three dimensional space, where coloration is added for visualization purposes. We use a standard set of parameters for weakly-compressible flows (see \cite{cossins2010smoothed}) $\alpha = 1.0$ (bulk-shear viscosity), $\beta = 2 \alpha$ (Nueman-Richtmyer viscosity) , $c = 10$, $\gamma = 7.0$ with energy injection rate  $\theta = 0.5$, and deterministic external forcing, (see Appendix \ref{sec:sph_append} for more details on the parameters and equations such as the equation of state and artificial viscosity). In order to validate the learning algorithm, an inverse problem is solved on "synthetic" SPH data; given a sequence of snapshots of SPH particle flows, estimate the parameters of the SPH model that best fits the SPH flow data over a predefined time scale. The results of this inverse problem can be found in Section \ref{sec:dns_results} (and Appendix \ref{sec:results}). 

\begin{figure}[ht]
        \centering
        \begin{subfigure}[b]{0.2\textwidth}
        \centering \includegraphics[height=0.8\textwidth]{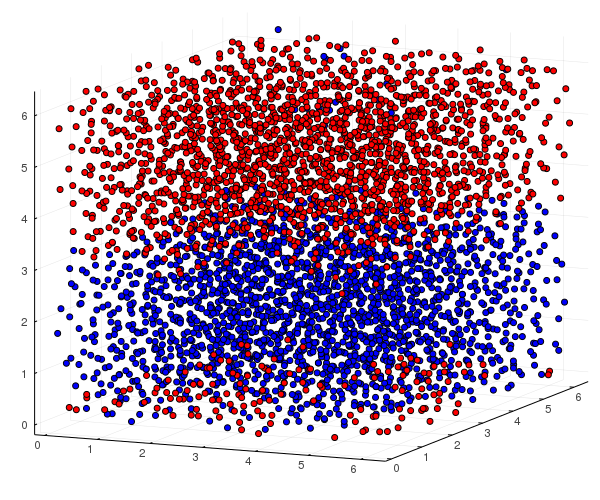}
        \caption{$t_0$}
        \end{subfigure}
        \begin{subfigure}[b]{0.2\textwidth}
        \centering \includegraphics[height=0.8\textwidth]{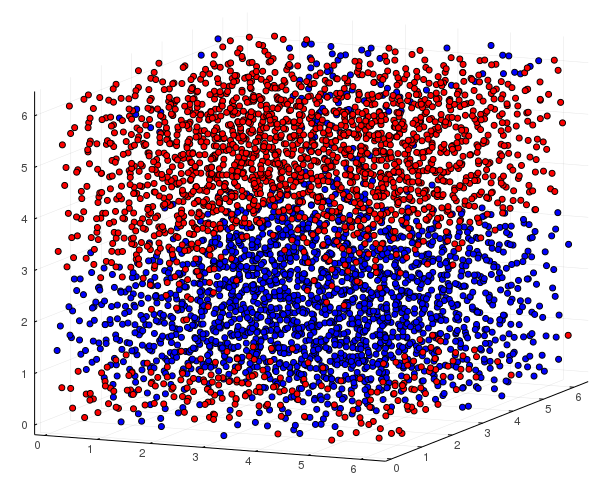}
        \caption{$t_1$}
        \end{subfigure}
        \begin{subfigure}[b]{0.2\textwidth}
        \centering \includegraphics[height=0.8\textwidth]{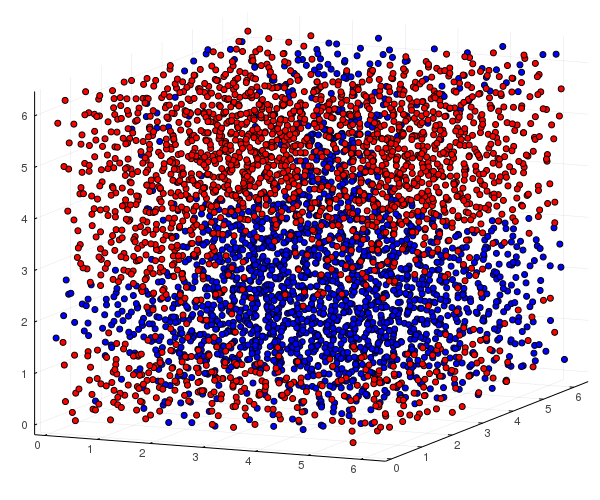}
        \caption{$t_2$}
        \end{subfigure}
        \begin{subfigure}[b]{0.2\textwidth}
        \centering \includegraphics[height=0.8\textwidth]{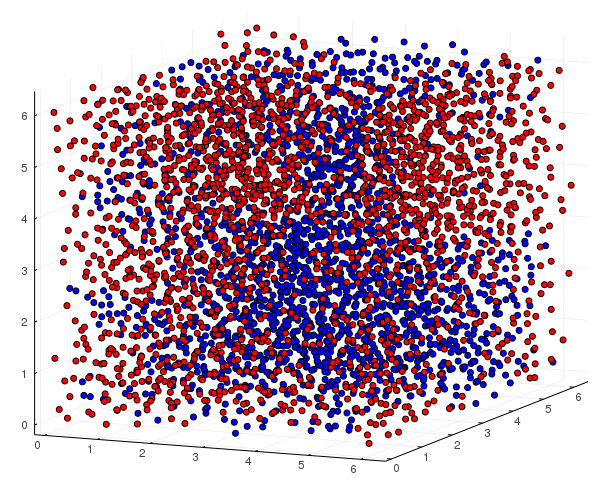}
        \caption{$t_3$}
        \end{subfigure}


        \caption{SPH particles advancing in time driven by external forcing $f_{ext}$ used as training data to validate the learning algorithm, where coloration is added for visualization purposes and $t_{i+1} - t_i \approx 50 \Delta t$.}
        \label{fig:sph_flow}
\end{figure}


\section{Hierarchy of Reduced Lagrangian Models}
\label{sec:hiearchy}

 We develop a hierarchy of parameterized Lagrangian models at coarse grained scales that gradually includes the SPH framework.  The motivation for this is twofold: (1) systematically analyze the effect of including more physical structure, and (2) a priori, we do not know which of the following Lagrangian models (i.e. what level of SPH framework vs. NN parameterizations) will best fit the DNS ground truth data, as well as generalize to different flow regimes not seen in training. The NNs used in this work are Multilayer Perceptrons (MLPs) with hyperbolic tangent activation functions which serve as universal function approximators \cite{hornik_nn_approx89} and are embedded within the ODE structure evolving particles in the Lagrangian frame. It was found through hyper-parameter tuning that 2 hidden layers were sufficient for each model using a NN. Defining $\bm X := \{\bm X_i := [\bm r_i, \bm v_i]^T, \quad \forall i \in \{1,..,N\}\}$, each of the parameterized Lagrangian models take on the general form
\begin{gather}\label{eq:sph_standard}
   \forall i:\quad d \bm X_i/dt = \bm{\mathcal{F}}_i({\bm X}(t, \bm \theta), \bm {\theta}) := 
    (\bm v_i,\ 
    \bm F_i(\bm X, \bm{\theta}))^T.
\end{gather} 
 where the acceleration operator $\bm F$ is uniquely parameterized in the following. 

    \noindent $\bullet$ \underline{\bf NODE:} In this least informed (and most flexible) Neural ODE \cite{node} based Lagrangian model, the entire acceleration operator is approximated by a NN, with the exception of $\bm f_{ext}$. Note that no pairwise interaction between particles is assumed, which increases the flexibility of this model over SPH based parameterizations along with the generic NN structure used to approximate the acceleration of particles. This model is most related to the work done by Chen et al. \cite{node} where we make an additional modification by considering the interaction of particles to be within a local cloud (using a cell linked list algorithm \cite{n_list_dominguez}). We assume that velocities, ${\bm v}_i(t)$, and coordinates, ${\bm r}_i(t)$, of $N$ particles evolve in time according to 
    \begin{eqnarray}\label{eq:bare-Neural-ODE}
        \cfrac{d\bm v_i}{dt} &=& \mathbf{NN}_{ {\bm \theta}}\left(\eta_r(\bm r_{ij}), \eta_v(\bm v_{ij}) \big|\forall j:\ ||\bm r_{ij}||\leq 2h\right) \\
        \nonumber &+& \bm f_{ext}({\theta_{inj}}), 
    \end{eqnarray}
    where $\bm r_{ij} = \bm r_i - \bm r_j$, $\bm v_{ij} = \bm v_i - \bm v_j$,  $\eta_r(\bm r_{ij})$, and $\eta_v(\bm v_{ij})$ are min-max normalizations,  $\mathbf{NN}_{{\bm \theta}} : \mathbb{R}^{2dm} \rightarrow \mathbb{R}^l \rightarrow \mathbb{R}^l \rightarrow \mathbb{R}^d$  ($d=2,3$ is the space dimension, $m$ is the fixed number of particles that are closest to the $i$-th particle in each cloud, and $l$ is the height, or number of nodes, of the hidden layer). Although this $\mathbf{NN}$ is approximating a function which is interpretable (acceleration), the individual parameters of the $\mathbf{NN}$ are not. With some tuning, it was found that $m \in \{20, 21, ... 30\}$ and $l \in \{5, 6, ..., 12 \}$ generally produce the best fit. $\bm \theta$ and $\theta_{inj}$ are the trainable parameters.

    \noindent $\bullet$ \underline{\bf NN summand:}  In the direction of including more of the SPH based physical structure, pairwise interaction is assumed represented via a sum over the $i$-th particle neighborhood (again using a cell linked list algorithm), where the summand term is approximated by a NN. Here, Eq.~(\ref{wc_sph_av}) is modeled by the Lagrangian based ODE
    \begin{gather}\label{eq:NN_sum}
        \cfrac{d\bm v_i}{dt} = \sum_j^N \mathbf{NN}_{{\bm \theta}}\left(\eta_r(\bm r_{ij}), \eta_v(\bm v_{ij}) \right)  + \bm f_{ext}({\theta_{inj}}),
    \end{gather}
    where $\mathbf{NN}_{{\bm \theta}} : \mathbb{R}^{2d} \rightarrow \mathbb{R}^l \rightarrow \mathbb{R}^l \rightarrow \mathbb{R}^d $, $\eta_r(\bm r_{ij})$, and $\eta_v(\bm v_{ij})$ are min-max normalizations. With some tuning, $l \in \{5, 6, ... \}$ generally produce the best fit. $\bm \theta$ and $\theta_{inj}$ are the trainable parameters.

    \noindent $\bullet$ \underline{\bf Rotationally Invariant NN}:  In this formulation, built on the top of the NN summand, we use a neural network of the form ${NN}_{{\bm \theta}} : \mathbb{R}^{4} \rightarrow \mathbb{R}^l \rightarrow \mathbb{R}^l \rightarrow \mathbb{R} $ to approximate the pair-wise part of the acceleration term in Eq.~(\ref{wc_sph_av}), where the rotational invariance is hard coded by construction (about a rotationally invariant basis expansion using the difference vector $\bm r_{ij}$)
    \begin{eqnarray}\label{eq:rot_inv}
        \cfrac{d\bm v_i}{dt} &=& \sum_j^N {NN}_{{\bm \theta}}\left(\cfrac{P_i}{\rho_i^2}, \cfrac{P_j}{\rho_j^2}, \bm r_{ij} \cdot \bm v_{ij}, ||\bm r_{ij}||_2 \right)\bm r_{ij} \\
        \nonumber &+& \bm f_{ext}({\theta_{inj}}).
    \end{eqnarray}
    With some tuning, $l \in \{5, 6, ... \}$ generally produce the best fit. $\bm \theta$ and $\theta_{inj}$ are the trainable parameters.

    \noindent $\bullet$ \underline{\bf $\nabla  P$- NN:}  In this model, we embed a neural network $NN$ within the SPH framework to approximate the gradient of pressure contribution (i.e $\nabla  P$- term in SPH Eq.~(\ref{wc_sph_av})) and explicitly include the artificial viscosity term $\Pi$:
    \begin{eqnarray}\label{eq:grad_p}
        \cfrac{d\bm v_i}{dt} &=& -\sum_j^N m_j \Biggl({NN}_{\bm \theta}\left(\bm r_{ij} \right) + \Pi_{ij}\Biggr) \nabla  W_{ij} \\
        \nonumber &+& \bm f_{ext}({\theta_{inj}}),
    \end{eqnarray}
    where ${NN}_{{\bm \theta}} : \mathbb{R}^d \rightarrow \mathbb{R}^l \rightarrow \mathbb{R}^l \rightarrow \mathbb{R}$. $l \in \{5, 6, ... \}$ generally produce the best fit. $\bm \theta$, $\alpha$, $\beta$ (from $\Pi(\alpha, \beta)$) and $\theta_{inj}$ are the trainable parameters.

    \noindent $\bullet$ \underline{\bf EoS NN:} Approximating the Equation of State (EoS) with a NN in the weakly compressible SPH formulation: 
    \begin{eqnarray}\label{eq:eos}
        \cfrac{d\bm v_i}{dt} &=& -\sum_j m_j \Biggl(\cfrac{Pnn_{{\bm \theta}}(\rho_i)}{\rho_i^2} + \cfrac{Pnn_{{\bm \theta}}(\rho_j)}{\rho_j^2} + \Pi_{ij} \Biggr) \nabla  W_{ij} \\
        \nonumber &+& \bm f_{ext}({\theta_{inj}}).
    \end{eqnarray}
    where $Pnn_{{\bm \theta}}(\rho): \mathbb{R} \rightarrow \mathbb{R}^l \rightarrow \mathbb{R}^l \rightarrow \mathbb{R}$. $l \in \{8, 9, ..., 12\}$ generally produce the best fit. $\bm \theta$, $\alpha$, $\beta$ (from $\Pi(\alpha, \beta)$) and $\theta_{inj}$ are the trainable parameters.

    \noindent $\bullet$ \underline{\bf SPH-Informed: Fixed smoothing Kernel}  In this formulation,  the entire weakly compressible SPH structure is used, and the physically interpretable parameters 
    $\alpha, \beta, \gamma, c, p_0, \theta_{inj}$ from $P(\rho)(c, \gamma, p_0)$ and $\Pi(\alpha, \beta)$ are learned
    \begin{eqnarray}\label{eq:phys}
        \cfrac{d\bm v_i}{dt} &=& -\sum_j m_j \Biggl(\cfrac{P_{i}}{\rho_i^2} + \cfrac{P_{j}}{\rho_j^2} + \Pi_{ij}\Biggr) \nabla  W_{ij} \\
        \nonumber &+& \bm f_{ext}({\theta_{inj}}).
    \end{eqnarray}

    \noindent $\bullet$ \underline{\bf SPH-informed: Including Parameterized $W$}  In this formulation,  the entire weakly compressible SPH structure is used along with a novel parameterized smoothing kernel (described below), and the physically interpretable parameters
    $\alpha, \beta, \gamma, c, p_0, a, b, \theta_{inj}$ from $P(\rho)(c, \gamma, p_0)$, $\Pi(\alpha, \beta)$, and $W(a,b)$ are learned ($a, b$ are parameters defined below for a parameterized smoothing kernel).
    \begin{eqnarray}\label{eq:phys_wab}
        \cfrac{d\bm v_i}{dt} &=& -\sum_j m_j \Biggl(\cfrac{P_{i}}{\rho_i^2} + \cfrac{P_{j}}{\rho_j^2} + \Pi_{ij}\Biggr) \nabla  W_{ij}({a}, {b}) \\
        \nonumber &+& \bm f_{ext}({\theta_{inj}}). 
    \end{eqnarray}

Let us emphasize that, as more of the SPH based structure is added into the learning algorithm, the learned models become more interpretable; i.e. the learned parameters are associated with the actual physical quantities.

The choice of smoothing kernels is important, and effects the consistency and accuracy of results \cite{monaghan2005}, where bell-shaped, symmetric, monotonic kernels are the most popular \cite{FULK1996}, however there is disagreement on the best smoothing kernels to use \cite{Liu2010SmoothedPH}. In this work, we introduce two new smoothing kernels to increase the flexibility of the SPH model as well as to allow the optimization framework to "discover" the best shaped kernel for our application (see Appendix \ref{sec:sph_append} for more details). Our parameterized smoothing kernel of the first type is 
\begin{equation} \label{eq:wab}
    W_1(r; h, {a}, {b}) = \begin{cases}
    \sigma_1(a,b) (1 - (r/(2h))^{{a}})^{{b}}, & 0 \leq r < 2h \\
    0,  & \text{otherwise}
    \end{cases}.
\end{equation}
Here the parameters $a, b$ control the shape of the kernel, $h$ sets the spatial scale of the kernel, and constant, $\sigma$, dependent on the parameters $a,b$ and the problem dimensionality, is chosen to guarantee normalization, $\int d{\bm r} W_1(r;h,{a}, {b})=1$, thus
$$\sigma_1(a,b) = \cfrac{3\Gamma(b + 3/a + 1)}{32 \pi h^3 \Gamma(b+1)\Gamma(3/a + 1)}. $$
Our parameterized smoothing kernel of the second type, which is introduced to make the second derivative at the origin, $r=0$, smooth and thus to examine the effects of the kernel smoothness on the quality of the trained models, is as follows:
\begin{eqnarray}\label{eq:w2ab} \nonumber 
    W_2(r; h, {a}, {b}) & = & 
    \cfrac{1}{h^D \sigma_2(a,b)} \left(\sqrt{a^2+1} - \sqrt{a^2+(r/(2h))^2}\right) \\  
    \nonumber &\times& (1 - (r/(2h))^2)^2 \\  &\times &  \begin{cases} (1+b(r/(2h))^2), \hspace{1mm} 0 \leq r < 2h \\
    0,  \hspace{22mm} \text{otherwise}
    \end{cases};
\end{eqnarray}

\begin{eqnarray}
    \nonumber \sigma_2(a,b) = 4\pi \{[\sqrt{1 + a^2}(32(87 + 22b) + 21a^2[-48(5 + b) \\ \nonumber +
a^2(-380 + 96b + 5a^2[-30 + (46 + 21a^2)b])])]/80640 \\ \nonumber +
[a^4[-32 + 16a^2(-2 + b) + 7a^6b + 10a^4(-1 + 2b)] \\ \nonumber 
(\log{(a^2)} - 2\log{[1 + \sqrt{1 + a^2}}])]/512\},
\end{eqnarray}
where as with the smoothing kernel of the first type $\sigma_2$ is introduced to enforce normalization of the kernel.

\section{Mixed mode gradient based optimization: Efficient parameter estimation}
\label{sec:mixed_mode}

We develop a mixed mode method, mixing local Sensitivity Analysis (SA) with forward and backwards Automatic Differentiation (AD) for efficiently computing the gradients of the loss functions (discussed in Section \ref{sec:methods_append}). 

SA is a classical technique found in many applications, such as gradient-based optimization, optimal control, parameter identification, model diagnostics \cite{donello2020computing, Zhang2014FATODEAL}. There are other ways that gradients can be propagated through numerical simulators, such as using differentiable programming \cite{rackauckas2020universal}, which allows for a simple and flexible implementation of gradient based learning algorithms, however this can require higher memory costs due to storing large computational graphs when computing gradients using reverse mode \cite{node}.  The two main local SA methods are the direct, or forward, method and the adjoint method (see Section \ref{sec:fsa_asa} for more details). Like forward mode AD, the forward SA method is more efficient when the number of parameters is much less than the dimension of the system and the adjoint method is more efficient when the number of parameters is much larger than the dimension of the system \cite{Zhang2014FATODEAL}. Therefore, in regards to this work, when the number of particles $N$ is large, the forward SA method will be more efficient for the models described above (since the NNs are relatively small compared to the dimension of the system when $N$ is large). When differentiating functions within the local SA frameworks we mix forward mode and reverse mode AD \cite{ma2021comparison, ad_bucker06} (depending on the input and output dimension of each function to be differentiated, which can include NNs) for improved efficiency over the fully differentiable programming technique.

We consider loss functions of the form, $L(\bm{X}, \bm \theta) = \int_0^{t_f} \Psi(\bm X, \bm \theta, t) dt$, where ${\bm X}$ and ${\bm \theta}$ are, respectively, the matrix of states $\bm{X} $ (defined above) and the vector of parameters. $\Psi$ is some measure of performance at time $t$. Since our overall goal involves learning Lagrangian and SPH based models for turbulence applications, it is the underlying statistical features and large scale field structures we want our models to learn and generalize with. Thus, two different loss functions are considered; (1) a field based loss $L_f$ which tries to minimize the difference between the large scale structures found in the Eulerian velocity fields, and (2) a statistical based loss $L_{kl}$ which tries to capture the small scale statistical characteristics of turbulent flows using well known single particle statistics \cite{yeung_borgas_2004}. In the experiments below, first only the $L_f$ is used, then a combination of the $L_f$ and $L_{kl}$ are used with gradient descent in order to first guide the model parameters in order to reproduce large scale structures with $L_f$, then later refine the model parameters with respect to the small scale features inherent in the velocity increment statistics by minimizing $L_{kl}$.

The field based loss $L_f$ is introduced by setting $\Psi(\bm X, \bm \theta, t) = \| \bm V^f(t) - \hat{ \bm V}^f(t)\|^2/N_f $, where $ \bm V_i^f = \sum_{j=1}^{N_f} (m_j/\rho_j)\bm v_j W_{ij}(||\bm r^f_i - \bm r_j||, h)$. This uses the same SPH smoothing approximation to interpolate the particle velocity onto a predefined mesh $\bm r^f$ (with $N_f$ grid points). The statistical based loss function $L_{kl}$, using a Kullback–Leibler (KL) divergence, is also introduced by setting $\Psi =  \int_{-\infty}^{\infty} P_{gt}(t, \bm z_{gt}, {\bm x})\log \left(P_{gt}(t, \bm z_{gt}, {\bm x}) / P_{pr}(t, \bm z_{pr}(\bm \theta), {\bm x}) \right) d{\bm x}$, where $\bm z_{gt} = \bm z_{gt}(t)$, and $\bm z_{pr}(\bm \theta) = \bm z_{pr}(\bm \theta, t)$ represent single particle statistical objects over time of the ground truth and predicted data, respectively. For example, we use the velocity increment, $\bm z^i(t) = (\delta u_i, \delta v_i, \delta w_i)$, where $\delta u_i(t) = u_i(t) - u_i(0)$ and $\bm z$ ranges over all particles. Here $P(t, \bm z(t), x)$ is a continuous probability distribution (in $x$) constructed from data $\bm z(t)$ using Kernel Density Estimation (KDE), to obtain smooth and differentiable distributions from data \cite{kde_yen_intro}), that is  $P(\tau, \bm z, \bm x) = (N h_{kde})^{-1}\sum_{i=1}^{N}K\left((z_i - {\bm x})/h_{kde}\right)$ (where $h_{kde}$ is a smoothing parameter selected in this work as $h_{kde}=0.9$ based on Silverman's rule \cite{kde_yen_intro}).

The gradient $ \partial _{\bm \theta}L = \int\limits_0^{t_f} \partial_{\bm X} \Psi(\bm X, \bm \theta, t) d_{\bm \theta} \bm X(\bm \theta, t) + \partial_{\bm \theta}\Psi(\bm X, \bm \theta, t) dt,$ is computed with the forward SA equation by simultaneously integrating the states (Eq. \ref{eq:sph_standard}) along with the sensitivities $\bm S_i^k := d \bm X_i/d\bm {\theta}^k$ according to 
\begin{gather}
\label{eq:fsa_2_1} 
\forall i:\quad \cfrac{d \sbia}{dt} = \cfrac{\partial \bm{\mathcal{F}}_i(\bm X(t), \bm{\theta})}{\partial \bm X_i}\sbia + \cfrac{\partial\bm{\mathcal{F}}_i(\bm X(t), \bm{\theta})}{\partial \theta^{k}}.
\end{gather}
Where a mixed mode AD is used to compute the derivatives of $\partial\bm{\mathcal{F}}_i/{\partial \theta}$ and within $\partial \bm{\mathcal{F}}_i/{\partial \bm X_i}$ (see Appendix \ref{sec:methods_append} for more details).


\section{Results: Training and Evaluating Models}
\label{sec:dns_results}

\begin{figure}[ht]
\centering
\begin{subfigure}[b]{0.48\textwidth}
\centering
\includegraphics[width=0.97\textwidth]{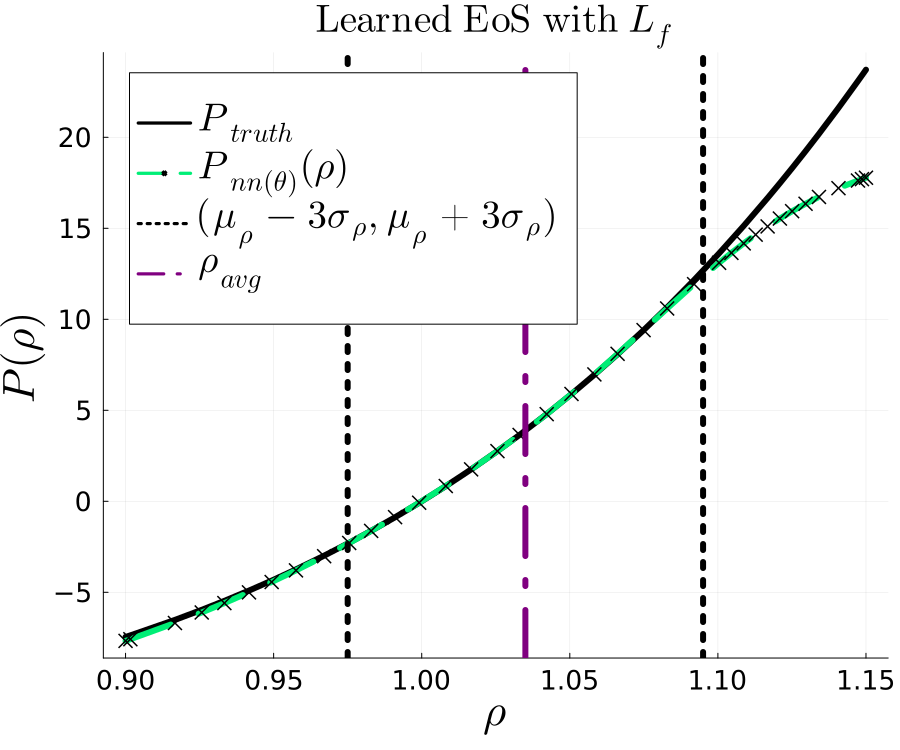}
\end{subfigure}
\begin{subfigure}[b]{0.48\textwidth}
\centering
\includegraphics[width=0.97\textwidth]{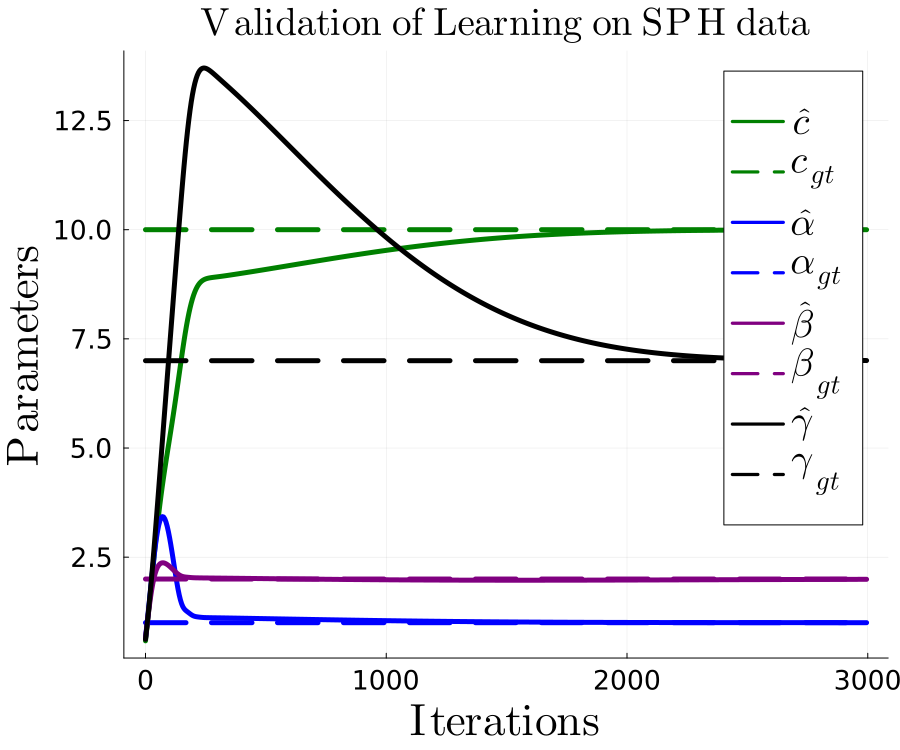}
\end{subfigure}
\caption{Using a NN embedded within the SPH framework to approximate Eq. \ref{eq:eos} using $L_f$. We see that $P(\rho)$ is well approximated, where $P_{truth}$ is the ground truth solution and $P_{nn}$ is the NN approximation. We note that the underlying ground truth data has 99.73\% of density values within the black dotted vertical lines (within 3 standard deviations), and so the NN does well at approximating the ground truth within the data seen in training, however it fails to capture the global shape of the function. Furthermore, we see that the parameters that were used to generate the synthetic SPH training data are learned where the initial guesses are uniformly distributed about (0,1).}
\label{fig:inverse_eos_main}
\end{figure}

First, the methodology is validated on "synthetic" SPH data, by training each model in the hierarchy on the SPH data and testing their ability to interpolate and generalize. For example, in Fig. \ref{fig:inverse_eos_main} we see the ability for NNs embedded within the SPH framework to learn the equation of state. Fig. \ref{fig:inverse_eos_main} provides further validation of the mixed mode learning algorithm for performing parameter estimation by learning the parameterized SPH "physics-informed" from SPH data (see Appendix \ref{sec:results} such as Fig. \ref{fig:inverse_fig_phys_theta} for more details).  Next, we analyze the hierarchy of models trained on weakly compressible (low Mach number) Eulerian based DNS data where the models are evolved on a coarse grained scale using $N = 12^3, 16^3, 20^3$ particles. 

\subsection{DNS data}
Our ``ground truth'' Lagrangian data are generated by tracking simultaneously multiple particles advected by velocity which we extract from the Eulerian DNS (solving the Navier-Stokes equations) of weakly compressible,  thus low Mach number, stationary Homogeneous Isotropic Turbulence. Our numerical implementation of the Eulerian DNS is on a $256^3$ mesh over the three dimensional box, $\Omega=[0,2\pi]^3$. We use sixth-order compact finite differences for spatial discretization and the 4th-order Runge-Kutta scheme for time advancement, also imposing triply-periodic boundary conditions over the box. The velocity field is initialized with 3D Gaussian spectral density enforcing zero mean condition for all components.

A large-scale linear quasi-solenoidal forcing term is applied to the simulation at wavenumber $\vert k\vert<2$ to prevent turbulence from decaying \cite{dlivescu_fext}. The forcing method allows the specification of the Kolmogorov scale at the onset and ensures that it remains close to the specified value. The simulations presented here have $\eta /\Delta x =0.8$, where $\Delta x$ is the grid spacing. Compared to a standard (well-resolved) spectral simulation with $\eta k_{max}=1.5$, where $k_{max}$ is the maximum resolved wavenumber, which has $\eta/\Delta x=1.5/\pi$, the contraction factor is $\approx 0.6$ \cite{dlivescu_fext,baltzer2020} and the maximum differentiation error at the grid (Nyquist) scale is less than $3.5\%$. Compared to a spectral method with $\eta k_{max}=1$, which has $\eta/\Delta x=1/\pi$, the contraction factor is $\approx 0.4$ \cite{dlivescu_fext,baltzer2020} and the maximum differentiation error at the Nyquist scale is less than $0.2\%$.  The initial temperature field is set to be uniform and the initial pressure field is calculated by solving the Poisson equation. More details about the numerical method and setup can be found in Refs. \cite{dlivescu_fext,ryu2014}. The simulation is conducted until the turbulence becomes statistically stationary, which is verified based on the evolution of the kinetic energy and dissipation  \cite{dlivescu_fext,ryu2014}.

Once a statistically-steady state of HIT is achieved, we apply a Gaussian filter to smooth the spatio-temporal Eulerian data for velocity at the resolved scale,  $d$, and then inject the filtered flow with $16^3$ non-inertial Lagrangian fluid particles. We use a Gaussian filter, which is commonly used in LES, with a filtering length scale of the order or larger than the scale $d$ that can be resolved for the particles. In dimensionless units, where the energy containing scale, $L$,  which is also the size of the box, is $L=2\pi$, the smallest scale $d$ we can resolve with this number of particles is  $\pi/8$, i.e. $16$ times smaller than the size of the domain.

The particles are placed in the computational domain,  $[0,2\pi]^3$, where the initial condition is set as an SPH equilibrium solution (particles are evolved according to SPH with no external forcing until particles reach an equilibrium position \cite{cfryer_sph_init}), and then we follow trajectories of the passively advected particles for time, $\tau$, which is of the order of (or longer) than the turbulence turnover time of an eddy of size comparable to the resolved scale, $d$, i.e. $\tau=O(d^{2/3}/\varepsilon^{1/3})$, where $\varepsilon$ is the estimate of the energy flux transferred downscale within the inertial range of turbulence. Note that $d$ is bounded from above by the size of the box, i.e. $L=2\pi$ in the dimensionless units of our DNS setting, and from below by the Kolmogorov (viscous) scale, $\eta=O(\nu^{3/4}/\varepsilon^{1/4})$, where $\nu$ is the (kinematic) viscosity coefficient.

In this work, we consider three turbulence cases for training and testing the model with comparable Reynolds numbers, $Re_\lambda \approx 80$ and turbulent Mach numbers, $M_t= 0.04$, $0.08$, and $0.16$, as shown in Table \ref{table:use-cases} \cite{tian_22_LLES}. The Taylor Reynolds number is calculated from the turbulence Reynolds number, $Re_t$, using the isotropic turbulence formula $Re_{\lambda}=\sqrt{20/3Re_t}$~\cite{Tennekes1978AFirst}, where $Re_t=k_t^2/(\nu \varepsilon)$, with $k_t$ the turbulent kinetic energy based on the filtered velocity.
The turbulent Mach number in DNS is defined as $M_t=(2 k_t)^{1/2}/c_s$. In the limit of low Mach number, for single component flows, density can be expanded as $\rho \approx \rho_0 + \rho_1$, where $\rho_1 \sim M_t^2 \rho_0$ \cite{LivescuARFM}, so that $M_t^2$ is proportional to the fluid density deviation from the uniform distribution. Note that any particle-based modeling of turbulence requires introducing, discussing, and analyzing compressibility simply because any distribution of
particles translates into fluid density which is always spatially non-uniform, even if slightly. Therefore, even if we model fully incompressible turbulence, we should still introduce an effective turbulent Mach number when discussing a particle based approximation, which for the weakly compressible limit can be approximated as $M_t^2 \sim |\rho -\rho_0 |/\rho_0$. 

For the $M_t = 0.08$ case and over each resolution set ($N = 12^3, 16^3, 20^3$), training takes place on the order of the Kolmogorov timescale $t_{\eta}$. When measuring the performance of each model, generalization errors are computed over different turbulent Mach numbers as well as over different time scales (for $M_t = 0.08$) up to the eddy turn over time. Furthermore, when DNS data is used in training, the equation of state used in Eqs. \ref{eq:phys_wab} and \ref{eq:phys} is $P(\rho; c, \gamma, p_0) = c^2 \rho^{\gamma} + p_0$, to be consistent with the DNS formulation used, where the background pressure term $p_0$ is added as a correction for SPH (the pressure gradient term in the SPH framework is not invariant to changes in this background pressure term, see Appendix \ref{sec:sph_append}).

\begin{table}
\centering
\caption{DNS Cases for  training and validation of models. \label{table:use-cases}}
\begin{tabular}{lrrr}
Case number & 1 & 2 & 3 \\
\hline
Turbulent Mach number $M_t$ & 0.08 & 0.16 & 0.04 \\
Taylor Reynolds number $Re_\lambda$ & 80 & 80& 80 \\
Kolmogorov timescale $t_\eta$ & 2.3 & 1.2 & 4.7 \\
Usage & training & validation & validation \\
\hline
\end{tabular}

\end{table}

\subsection{Training and Evaluating Models}
\label{sec:training}
\begin{figure}[htp]
\centering
\begin{subfigure}[b]{0.48\textwidth}
\centering
\includegraphics[width=0.98\textwidth]{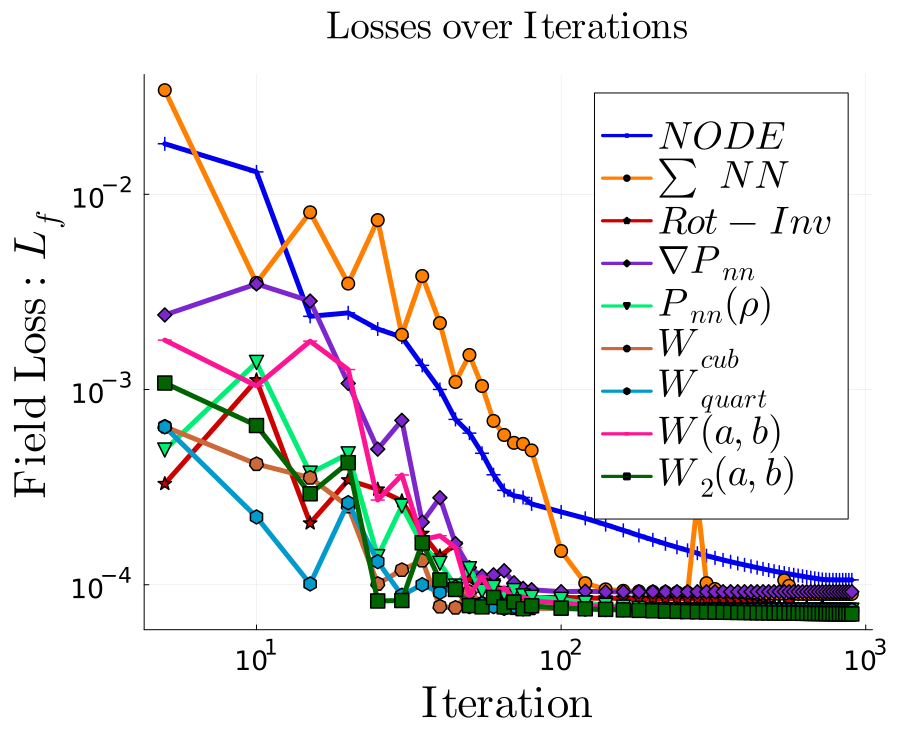}
\end{subfigure}
\caption{Losses converging over iterations when trained on DNS up to the Kolmogorov time scale with $M_t = 0.08$. The physics informed SPH based models achieve the lowest losses. Each model is seen to interpolate onto the DNS field data as seen in the $u$ - component of velocity snapshots in \autoref{fig:gen_t_qual_dns} within in the training window.}
\label{fig:loss_conv}
\end{figure}

\begin{figure*}[htp]
\centering
\begin{subfigure}[b]{0.95\textwidth}
\centering
\includegraphics[width=1\textwidth]{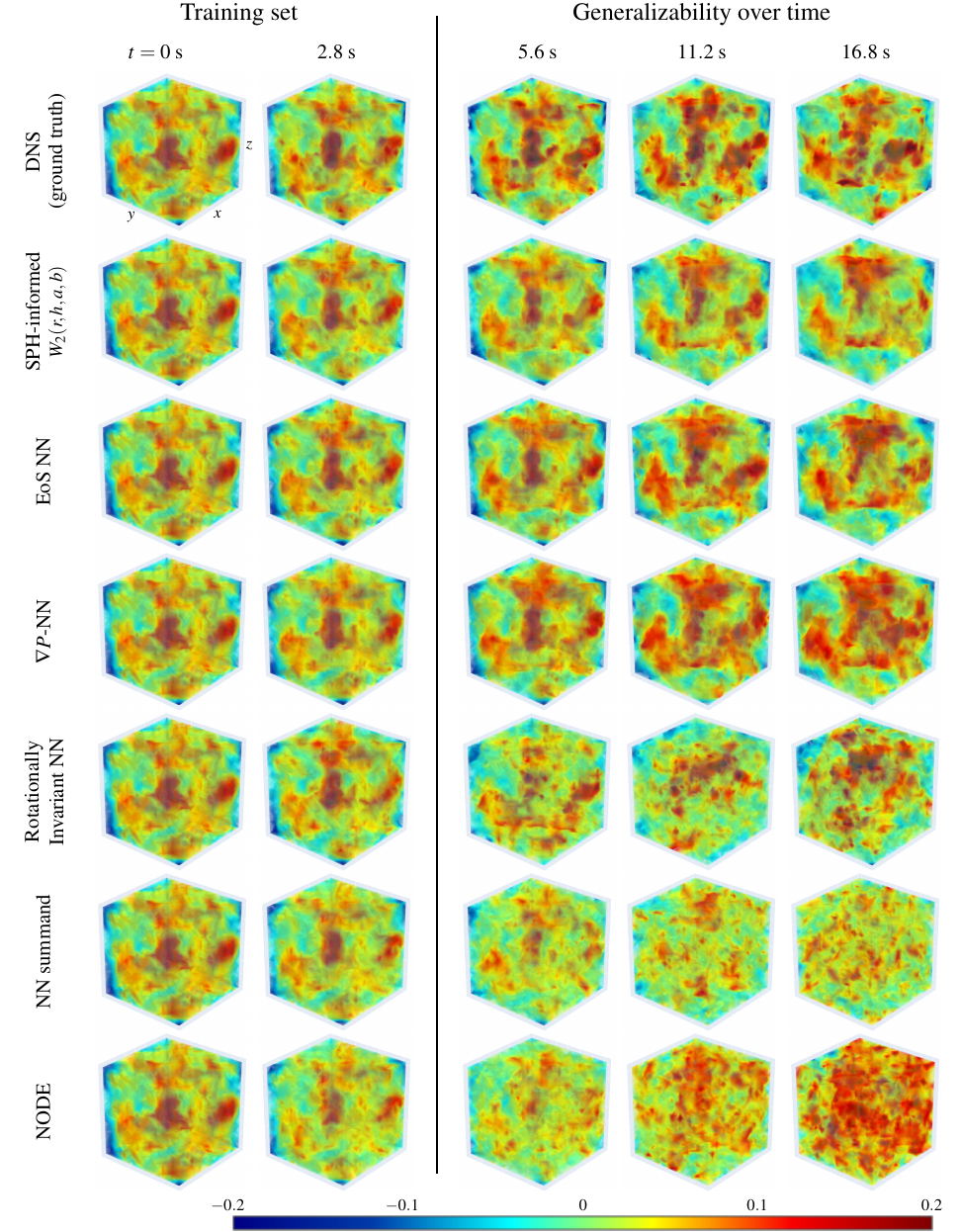}
\end{subfigure}
\caption{Volume plots of snapshots over time comparing Eulerian $u$ velocity component of coarse grained DNS data (at the same resolution of the models and at $M_t = 0.08$) to the predictions made with the trained models. This qualitatively shows that as more SPH structure is included, the better is the ability of the model to generalize all the way to the longest (physically relevant) time scale, $t_{eddy}$ (which is the turnover time scale of the largest eddy of the flow) even when it is trained on the shortest relevant time scale $t_{\eta}$. We see the large scale structures present in the $u$ velocity are best captured with the SPH informed models, and predictions degrade as more reliance is put on using a Neural Network to parameterize the acceleration operator \cite{Supplemental_material}.}
\label{fig:gen_t_qual_dns}
\end{figure*}

Parameter estimation (i.e training) is performed on each model on the order of the Kolmogorov time using the mixed mode gradient based optimization with the field based loss function (and statistical based loss Section \ref{sec:results_append} using velocity increment) until convergence (see Fig. \ref{fig:loss_conv}).  Once all the models are trained, they are used to make forward predictions over larger time scales as seen in Fig. \ref{fig:gen_t_qual_dns} (on the order of the eddy turn over time) and over different turbulent Mach numbers as seen in Table \ref{table:use-cases}. The future state predictions are evaluated with respect to loss functions as a performance measure, and detailed statistical comparisons are given. 

\begin{figure}[htp]
\centering
\begin{subfigure}[b]{0.45\textwidth}
\centering
\includegraphics[width=0.99\textwidth]{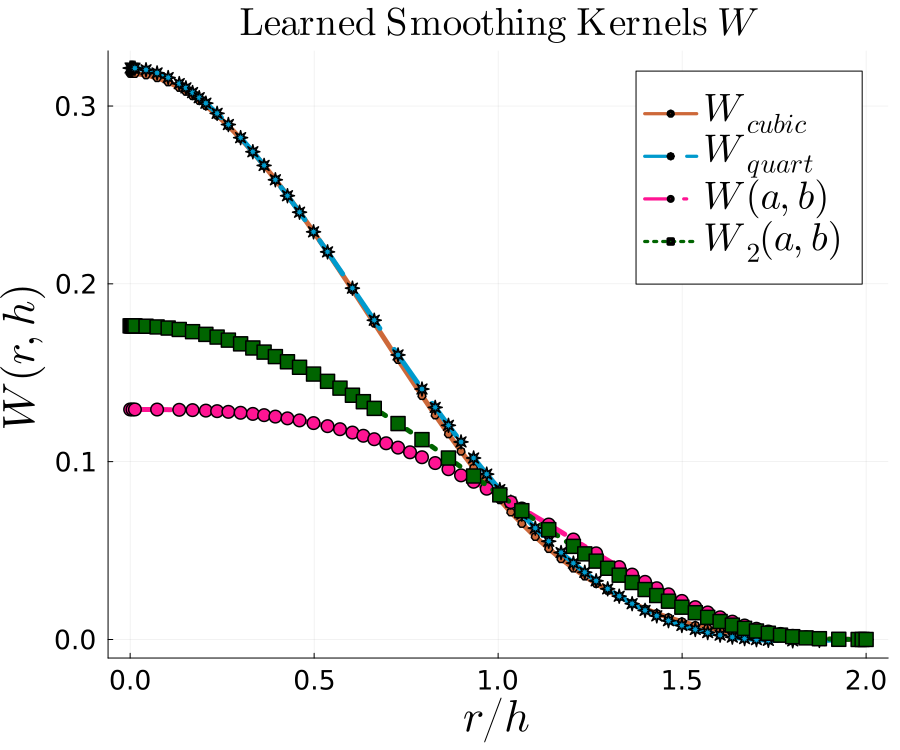}
\end{subfigure}
\begin{subfigure}[b]{0.45\textwidth}
\centering
\includegraphics[width=0.99\textwidth]{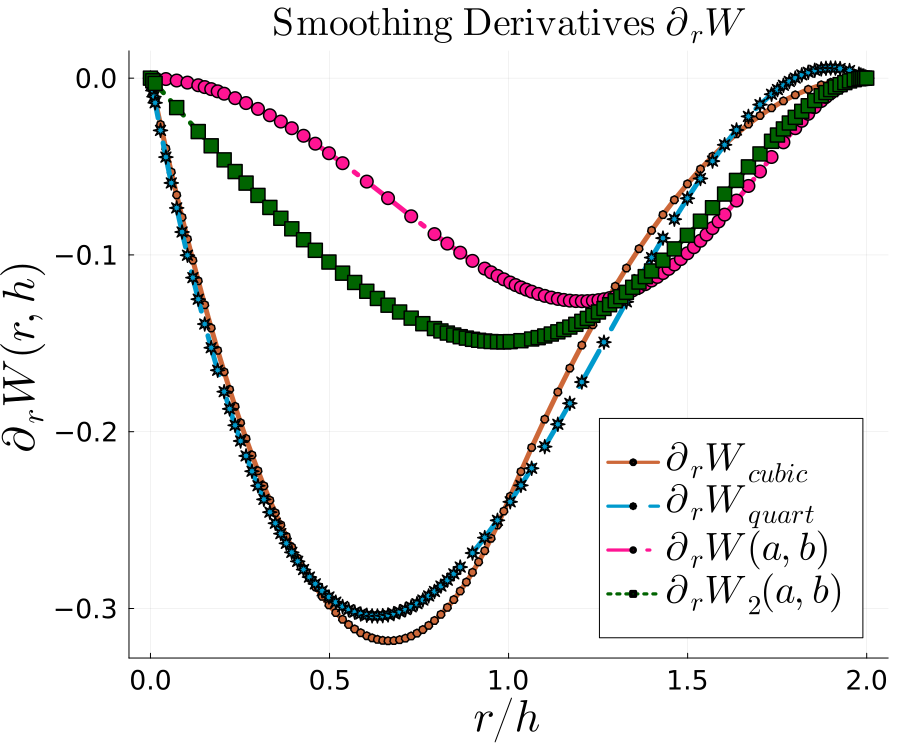}
\end{subfigure}
\caption{Comparing learned smoothing kernels $W_1(r;h,a,b)$ and $W_2(r;h,a,b)$ with standard cubic Eq.~(\ref{cubic}) and quartic Eq.~(\ref{quartic}) kernels. Notice the shape of the learned parameterized smoothing kernels includes a relatively larger contribution from particles farther away and shallower gradients for nearby particles. See Table \ref{table:learned_sph_params} for the learned parameters.}
\label{fig:comp_ws}
\end{figure}

\begin{figure}[htp]
\centering
\begin{subfigure}[b]{0.45\textwidth}
 \centering
\includegraphics[width=1\textwidth]{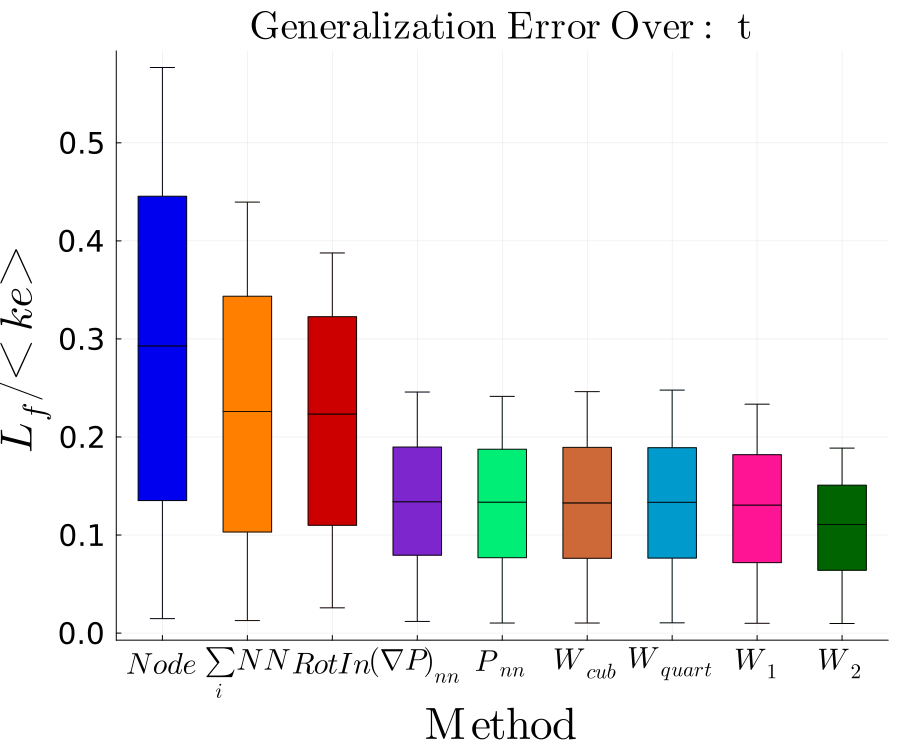}
\caption{Normalized $L_f$ over time}
\end{subfigure}
\begin{subfigure}[b]{0.45\textwidth}
\centering
\includegraphics[width=1\textwidth]{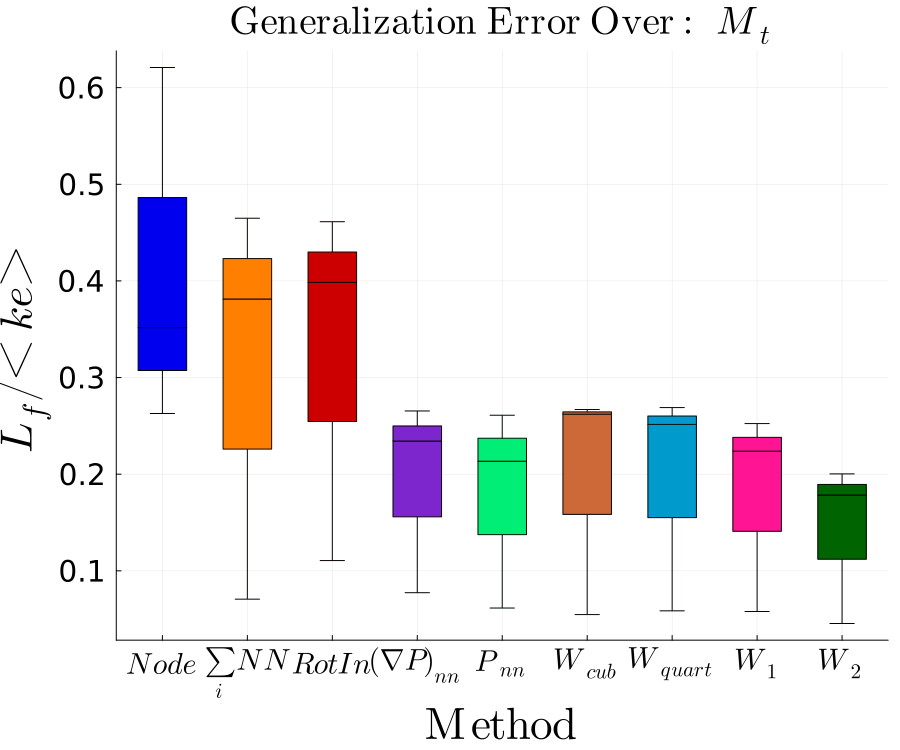}
\caption{Normalized $L_f$ over turbulent Mach numbers $M_t$}
\end{subfigure}
\caption{Measuring the generalization error using the field based loss $L_f$ normalized with respect to the total kinetic energy $\left<ke\right>$ from DNS. (a) The generalization error over $t$ is computed over 20 different time scales ranging from the Kolmogorov time to the eddy turn over time. (b) Generalization error over $M_t$ is computed using 3 different turbulent Mach numbers $M_t$, 0.04, 0.08, and 0.16, integrated up to the eddy turn over time scale. We see that the SPH-informed model with the parameterized smoothing kernel $W_2(a,b)$ performs best at generalizing with time and turbulent Mach numbers. Furthermore, the NNs embedded within the SPH structure, namely $(\nabla P)_{nn}$ and $P_{nn}$, showing improvements over the standard SPH model.}
\label{fig:lf_gen_over_time_mt}
\end{figure}

\begin{figure}[htp]
\centering
\begin{subfigure}[b]{0.45\textwidth}
 \centering
\includegraphics[width=1\textwidth]{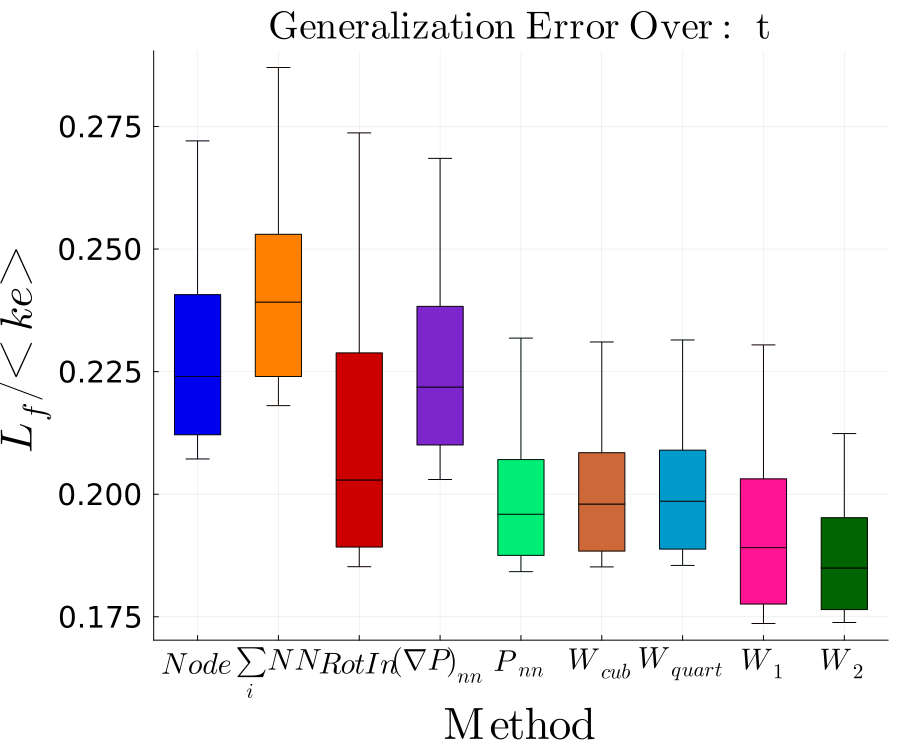}
\caption{Coarser resolution models over time}
\end{subfigure}
\begin{subfigure}[b]{0.45\textwidth}
\centering
\includegraphics[width=1\textwidth]{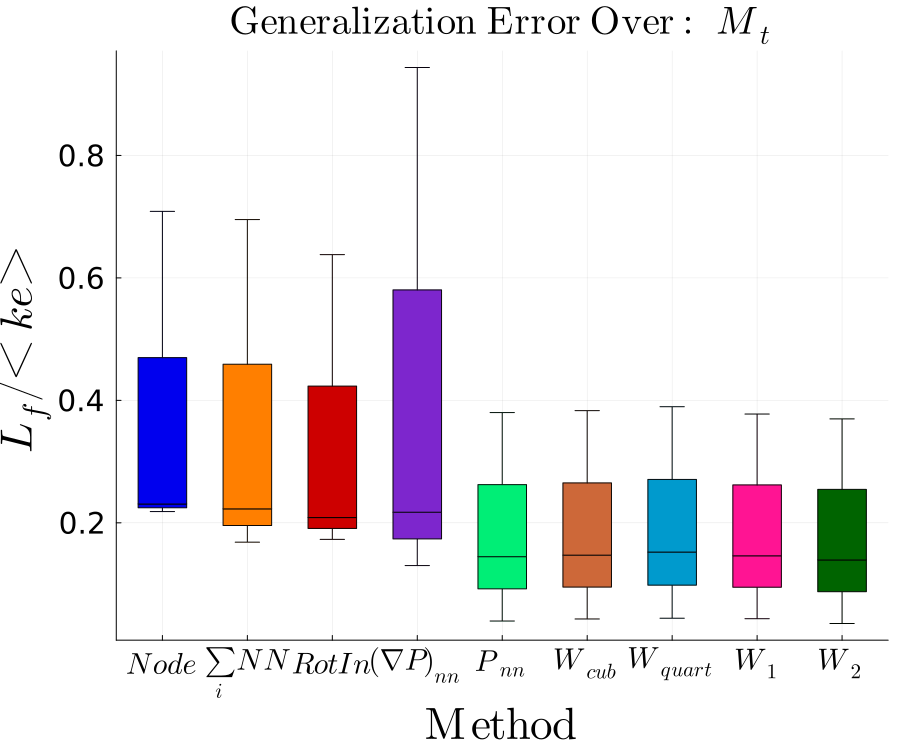}
\caption{Coarser resolution models over turbulent Mach numbers $M_t$}
\end{subfigure}
\caption{Predictive performances of coarser resolution models with $N = 12^3$ particles. Measuring the generalization error using the field based loss $L_f$ normalized with respect to the total kinetic energy $\left<ke\right>$ from DNS. (a) The generalization error over $t$ is computed over 20 different time scales ranging from the Kolmogorov to the eddy turn over time. (b) Generalization error over $M_t$ is computed using 3 different turbulent Mach numbers $M_t$, 0.04, 0.08, and 0.16, integrated up to the eddy turn over time scale. We see that the SPH-informed model with the parameterized smoothing kernel $W_2(a,b)$ performs best at generalizing with respect to time and turbulent Mach numbers.}
\label{fig:lf_gen_over_time_mt_low}
\end{figure}

\begin{figure}[htp]
\centering
\begin{subfigure}[b]{0.45\textwidth}
 \centering
\includegraphics[width=1\textwidth]{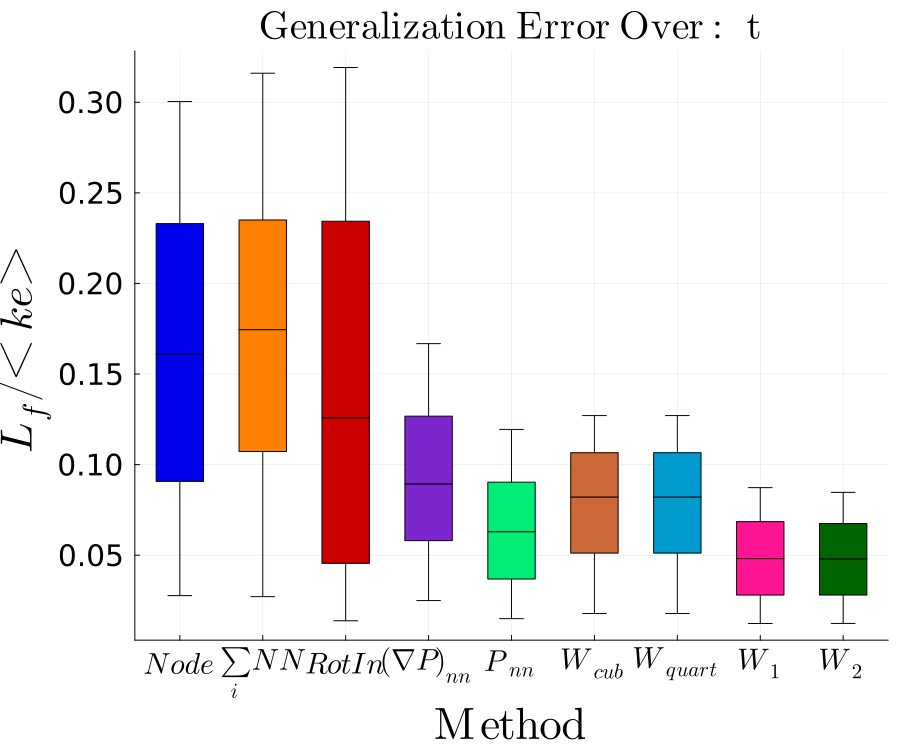}
\caption{Finer resolution models over time}
\end{subfigure}
\begin{subfigure}[b]{0.45\textwidth}
\centering
\includegraphics[width=1\textwidth]{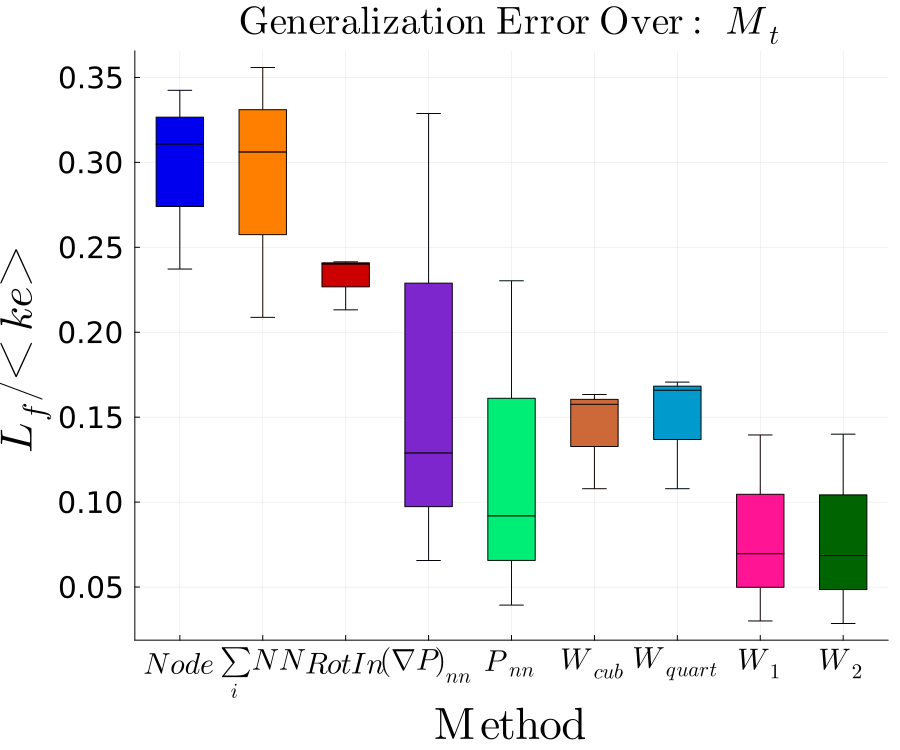}
\caption{Finer resolution models over turbulent Mach numbers $M_t$}
\end{subfigure}
\caption{Predictive performances of finer resolution models with $N = 20^3$ particles. Measuring the generalization error using the field based loss $L_f$ normalized with respect to the total kinetic energy $\left<ke\right>$ from DNS. (a) The generalization error over $t$ is computed over 20 different time scales ranging from the Kolmogorov time to the eddy turn over time. (b) Generalization error over $M_t$ is computed using 3 different turbulent Mach numbers $M_t$, 0.04, 0.08, and 0.16, integrated up to the eddy turn over time scale. We see that the SPH-informed model with the parameterized smoothing kernel $W_2(a,b)$ performs best at generalizing with respect to time and turbulent Mach numbers. We see that the improvements by using the SPH-informed model with parameterized smoothing kernel $W_2(a,b)$ become greater at the finer scale resolutions. We also note that the DNS data are not fully barotropic, like in the weakly compressible SPH framework. Thus, for the training $M_t$, $P_{nn}$ is more flexible and learns a better fit for equation of state than what is used in the standard SPH model. However, this model may suffer from over-fitting as seen in (b). }
\label{fig:lf_gen_over_time_mt_high}
\end{figure}

\begin{figure}[htp]
\centering
\begin{subfigure}[b]{0.45\textwidth}
\centering
\includegraphics[width=0.95\textwidth]{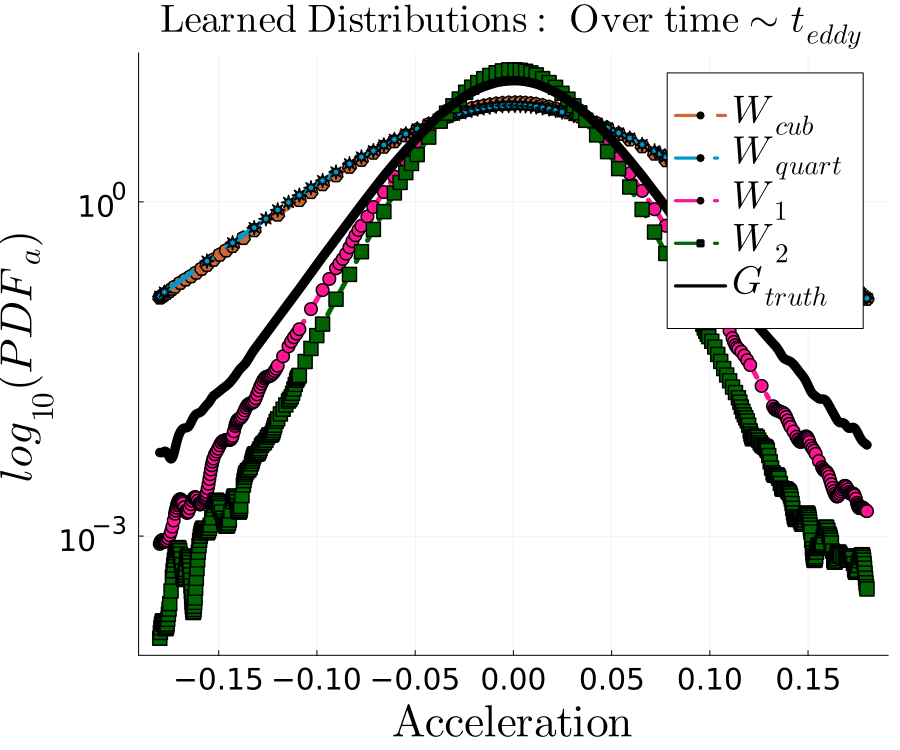}
\end{subfigure}
\begin{subfigure}[b]{0.45\textwidth}
\centering
\includegraphics[width=0.95\textwidth]{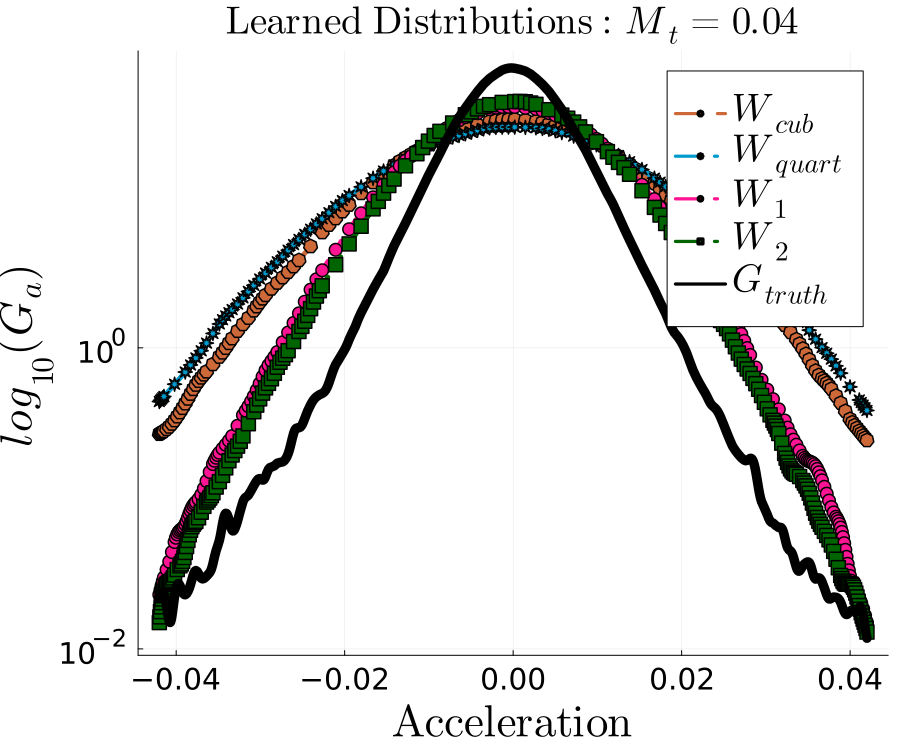}
\end{subfigure}
\begin{subfigure}[b]{0.45\textwidth}
\centering
\includegraphics[width=0.95\textwidth]{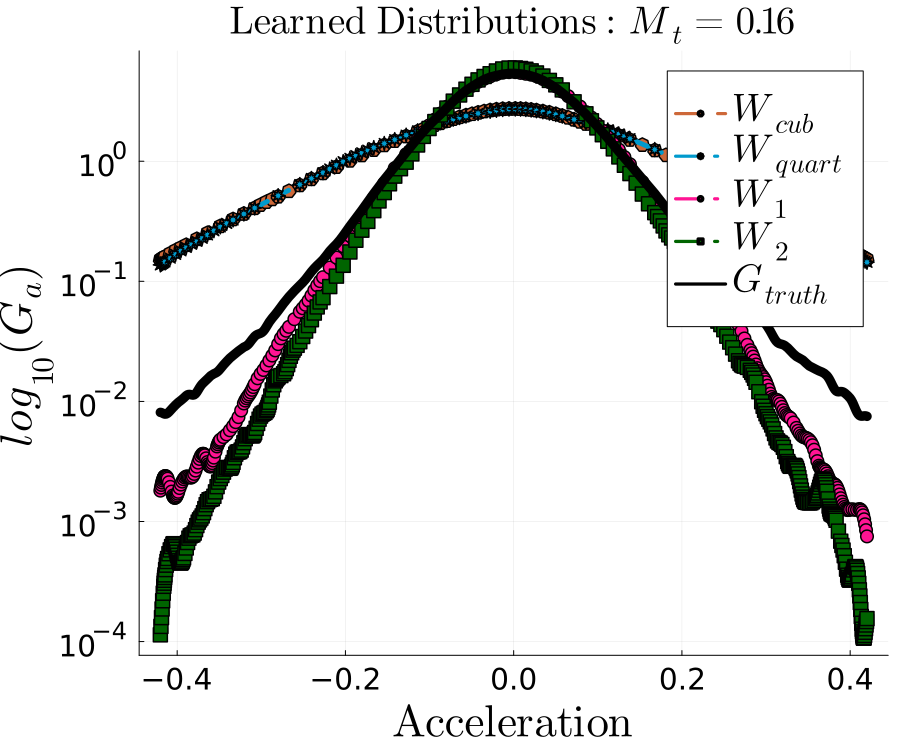}
\end{subfigure}
\caption{Comparing acceleration statistics over time, and turbulent Mach numbers of the trained SPH based models showing the learned parameterized smoothing kernels are a closer match to DNS as compared to the cubic and quartic smoothing kernels over longer time scales as well as at different Mach numbers $M_t = 0.04$ and $M_t=0.16$. However, each model misses the intermittent behavior seen in the long tails from DNS, and performs poorly on the $M_t = 0.04$ indicating limitations of this approach.}
\label{fig:comp_w_ga_mt_t}
\end{figure}

\begin{figure}[htp]
\centering
\begin{subfigure}[b]{0.45\textwidth}
\centering
\includegraphics[width=0.99\textwidth]{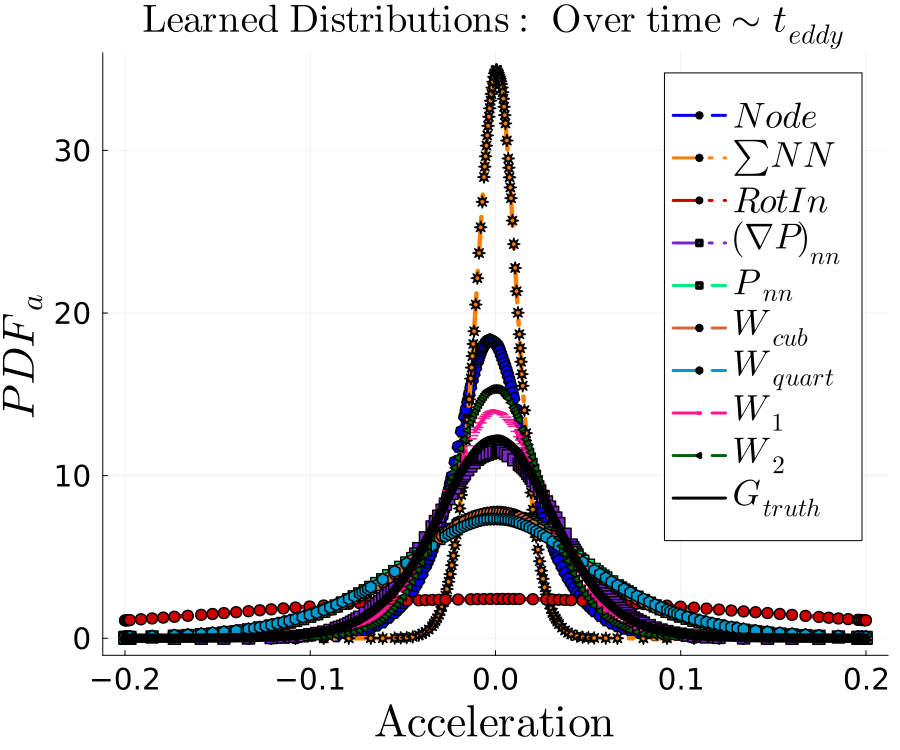}
\end{subfigure}
\begin{subfigure}[b]{0.45\textwidth}
\centering
\includegraphics[width=0.99\textwidth]{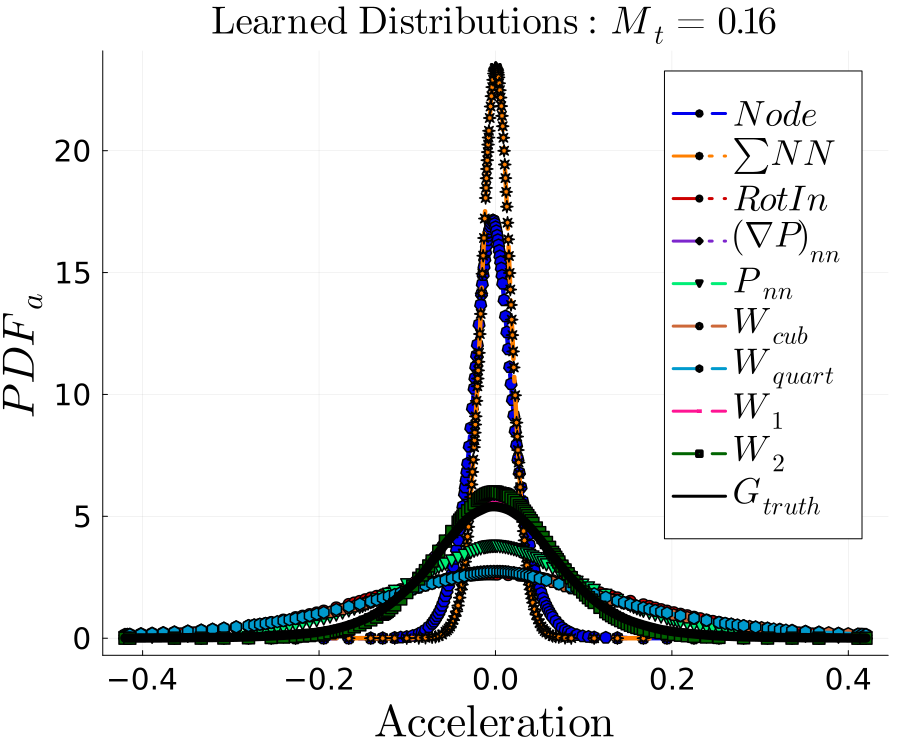}
\end{subfigure}
\begin{subfigure}[b]{0.45\textwidth}
\centering
\includegraphics[width=0.99\textwidth]{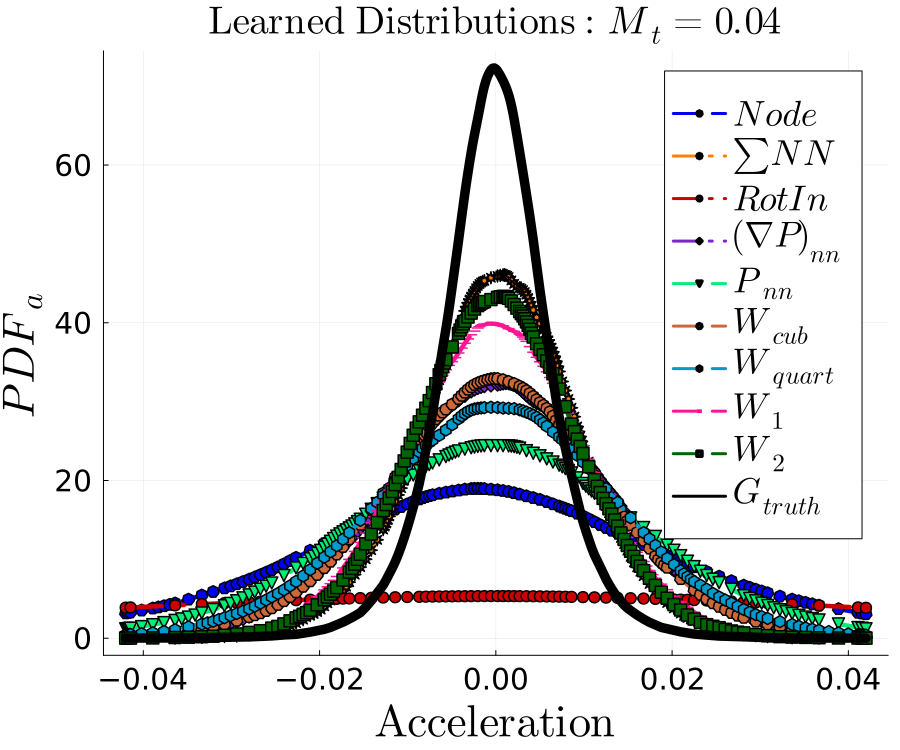}
\end{subfigure}
\caption{Comparing acceleration statistics over time, and turbulent Mach numbers of all trained models showing the learned parameterized smoothing kernels are a closer match to DNS as compared to the cubic and quartic smoothing kernels on over longer time scales as well as at different Mach numbers $M_t = 0.04$ and $M_t=0.16$. Here the smoothing kernels as compared in Fig. \ref{fig:comp_w_ga_mt_t} are reported again to compare all the models}
\label{fig:comp_all_ga_mt_t}
\end{figure}

\begin{figure}[htp]
\centering
\begin{subfigure}[b]{0.48\textwidth}
\centering
\includegraphics[width=1\textwidth]{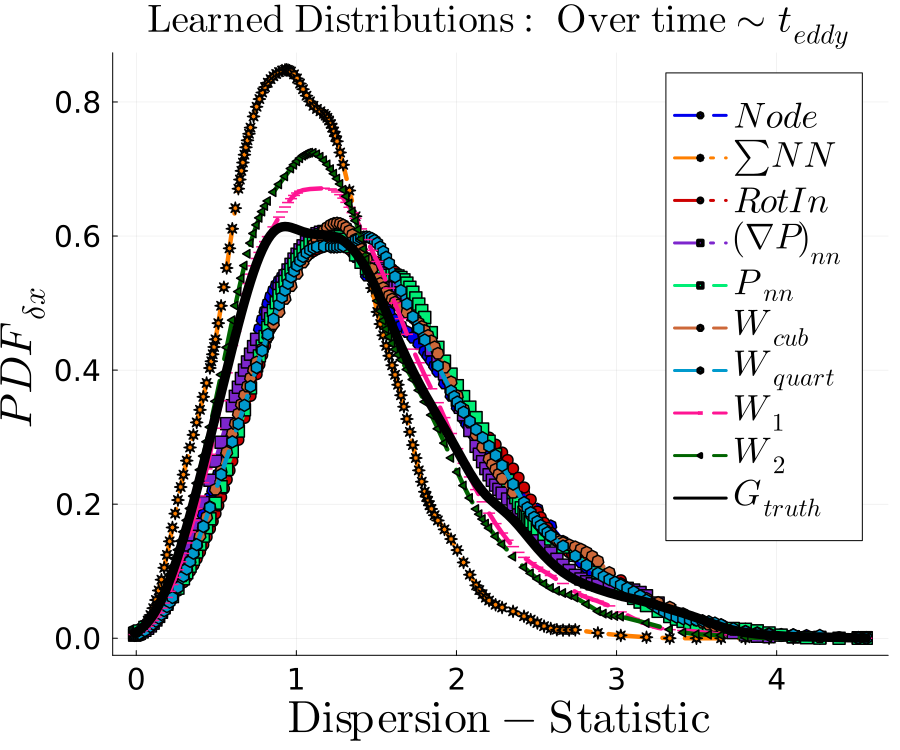}
\end{subfigure}
\begin{subfigure}[b]{0.48\textwidth}
\centering
\includegraphics[width=1\textwidth]{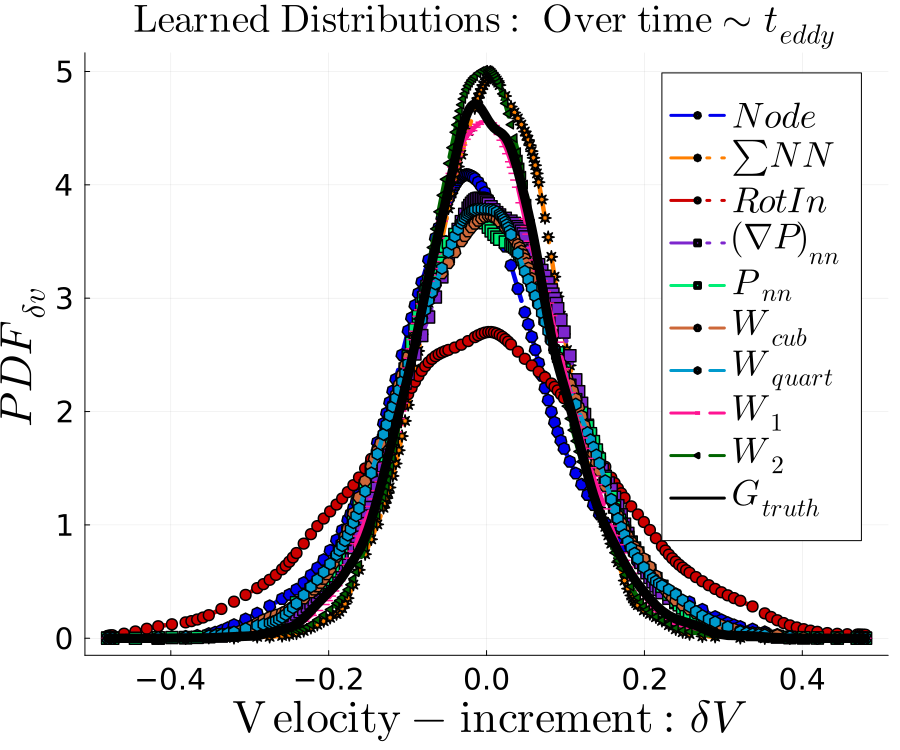}
\end{subfigure}
\caption{A diagnostic check comparing single particle statistics on larger time scales (roughly 20 times longer than seen in training), with the new parameterized smoothing kernels SPH-informed model having the best fit. In this case, each is not seen in training, so represent an external diagnostic, however, in the Appendix \ref{sec:results_append} we see the effects of including the velocity increment statistical based loss. }
\label{fig:sing_part_stats_all}
\end{figure}

\begin{figure}[htp]
\centering
\begin{subfigure}[b]{0.48\textwidth}
\centering
\includegraphics[width=1\textwidth]{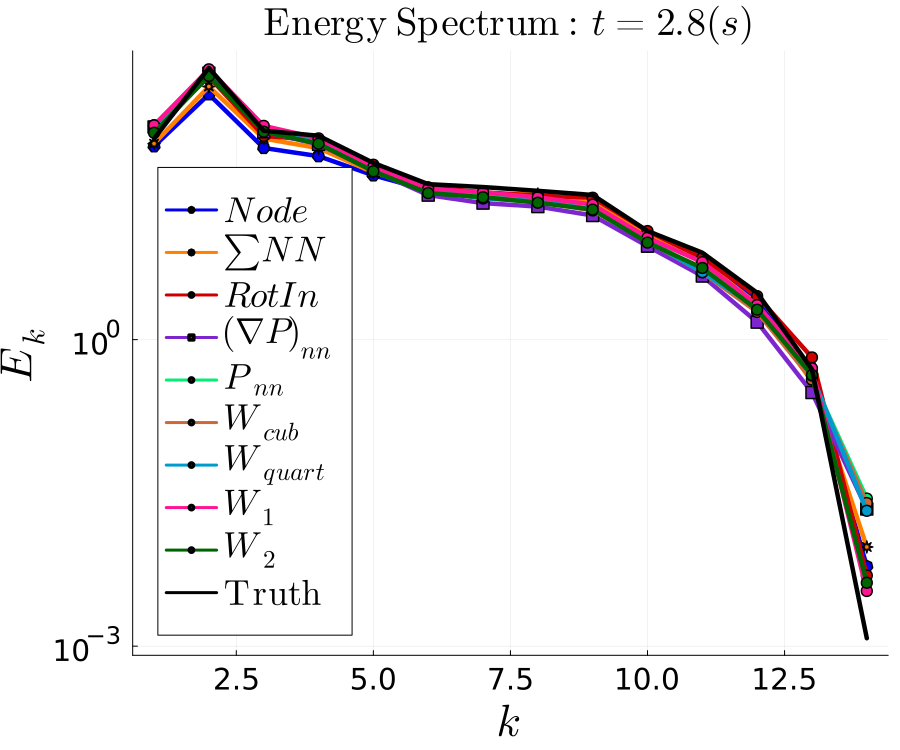}
\end{subfigure}
\begin{subfigure}[b]{0.48\textwidth}
\centering
\includegraphics[width=1\textwidth]{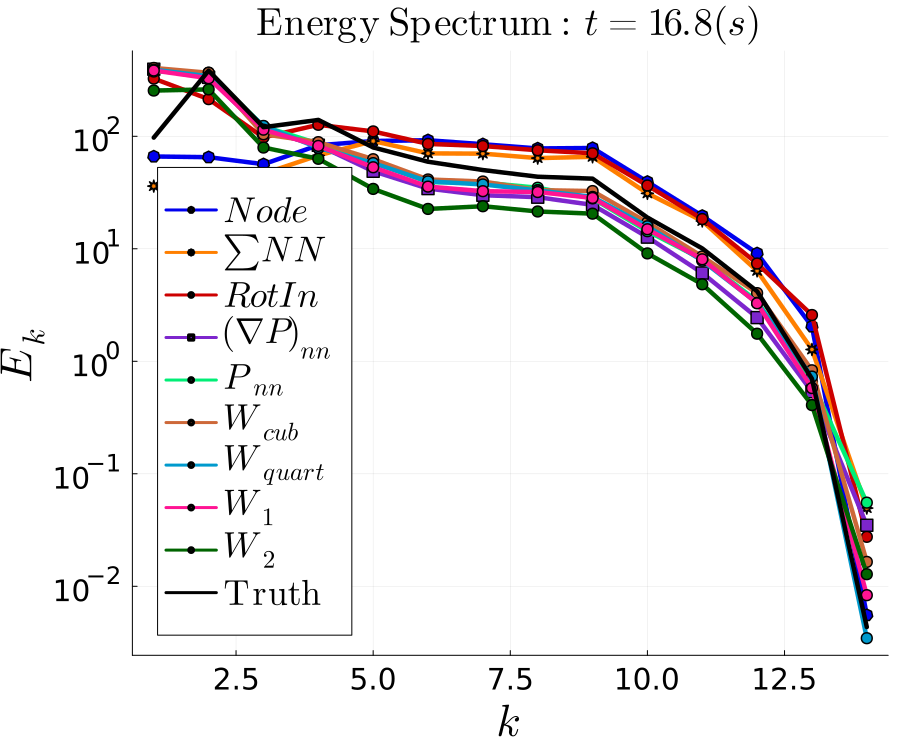}
\end{subfigure}
\caption{Comparing energy spectrum over time; $t = 2.8s \sim t_{\eta}$ is on the Kolmogorov time scale, and $t = 16.8s \sim t_{eddy}$ is at the scale of the eddy turn over time. We see that the SPH based parameterized models captures the energy cascade seen in DNS on both the training time scale and in generalizing to larger time scales. Furthermore, as seen qualitatively seen in Fig. \ref{fig:gen_t_qual_dns}, the less informed models do not capture the cascade and show more energy in the small scale structures and less in the large scales. However, the fully parameterized SPH-informed models show a dissipation rate from the large to small scales to be larger than DNS.}
\label{fig:ek_over_t_1}
\end{figure}

The shapes of the novel learned parameterized smoothing kernels are reported in Fig \ref{fig:comp_ws}. We use the field based loss normalized with respect to the total kinetic energy from DNS as a quantitative measure comparing each model over larger time scales and different turbulent Mach numbers as seen in Figs. \ref{fig:lf_gen_over_time_mt}, \textcolor{blue}{ \ref{fig:lf_gen_over_time_mt_high}, \ref{fig:lf_gen_over_time_mt_low}}. The errors in translational and rotational symmetries are recorded in Table \ref{table:rot_tran}, which shows that as more SPH based structure is included, conservation of linear and angular momenta is enforced.  Furthermore, the single particle statistics, acceleration pdfs, and energy spectrum (as seen in Figs. \ref{fig:comp_w_ga_mt_t}, \ref{fig:comp_all_ga_mt_t}, \ref{fig:sing_part_stats_all}, and \ref{fig:ek_over_t_1} respectively) are used to evaluate the statistical performance of each model as external diagnostics (not used in training). The results in these figures show that the SPH informed model using the novel parameterized smoothing kernel $W_2$ (Eq.~(\ref{eq:phys_wab})) performs best at generalizing, with respect to the statistical and field based performance evaluations, to larger time scales and different turbulent Mach numbers, as well as enforces physical symmetries and improves interpretability over the less informed models.

Discussing in slightly more detail, Fig. \ref{fig:lf_gen_over_time_mt} clearly shows that there is a gradual improvement in generalizability as more physics informed SPH based structure is included in the model. Although the most generalizable model is the fully parameterized SPH informed model using the new parameterized smoothing kernel $W_2(a,b)$, there is a close match between $P_{nn}$ (using a NN to approximate the equation of state within the SPH framework Eq. \ref{eq:eos}) and $(\nabla P)_{nn}$ (using a NN to approximate the SPH based pressure gradient Eq. \ref{eq:grad_p}) and the SPH-$W_2(a,b)$ model. Thus, within the SPH-informed models, using a NN embedded within the SPH framework (such as in approximating the Equation of State and pressure gradient term) can actually improve generalizability over the uninformed models and the standard SPH-informed model when using the classical cubic or quartic smoothing kernels.  However, relying on a NN to approximate the full acceleration operator without including any conservation laws, although still being able to interpolate, does not generalize nearly as well as the SPH-informed models.

In Figs. \ref{fig:comp_w_ga_mt_t} and \ref{fig:comp_all_ga_mt_t}, the acceleration statistics comparing smoothing kernels and each model respectively are reported for one eddy turn over time with $M_t = 0.08$ (as seen in training) and $M_t = 0.16$, and $M_t = 0.04$. In this figure, we see a clear improvement in generalization with respect to the acceleration statistics as more SPH structure is used and improving the accuracy of the SPH framework using the novel parameterized smoothing kernels. A similar trend is observed in the single particle statistics as seen in Fig. \ref{fig:sing_part_stats_all}. However, even the SPH informed models seem to struggle to provide a close match to DNS acceleration when generalizing to $M_t = 0.04$ flow, indicating a limitation to this approach. However, we should note that the only parameter that was changed in making forward state predictions for different turbulent Mach numbers was $\theta_{inj}$, in order to be consistent across models and with the DNS external forcing. Finally, in Fig. \ref{fig:ek_over_t_1}, the energy spectrum shows some of the key statistical differences in the model predictions, that the less informed models, although able to interpolate on the training window like the other models, predict more energy to be distributed in the small scales and less in the large scales as compared to DNS (and as seen in the volume plots of the $u$-component of velocity field Fig. \ref{fig:gen_t_qual_dns}); whereas the SPH informed models provide a closer match to the large to small scale energy cascade.  For further analysis of the smoothing kernels and experiments using a combination of $L_f$ and $L_{kl}$ see Appendix \ref{sec:results_append}.

\begin{table}[!htbp]
  \caption{Rotational and Translational Invariance errors in trained models}
  \label{table:rot_tran}
  \centering
  \begin{tabular}{llll}
    \toprule
    Model  \hspace{26mm} &   Rotational  \hspace{6mm}  &   Translational   \\
    \midrule
    NODE                        &  $3.5 \times 10^{-4}$    &  $4.1 \times 10^{-4}$      \\
    NN summand                  &  $3.4 \times 10^{-5}$    &  $7.9 \times 10^{-33}$   \\
    Rot-Invariant NN   &  $1.5 \times 10^{-32}$   &  $3.3 \times 10^{-32}$  \\
    $\nabla  P$ - NN            &  $2.6 \times 10^{-6}$    &  $4.5 \times 10^{-32}$    \\
    EoS NN                      &  $4.1 \times 10^{-32}$   &  $1.2 \times 10^{-31}$     \\
    SPH-informed: $W_{cubic}$   &   $3.5 \times 10^{-32}$  &  $8.5 \times 10^{-32}$     \\
    SPH-informed: $W_{quartic}$ &   $2.1 \times 10^{-31}$  &  $5.4 \times 10^{-31}$     \\
    SPH-informed: $W(a,b)$      &   $2.4 \times 10^{-32}$  &  $1.1 \times 10^{-31}$     \\
    SPH-informed: $W_2(a,b)$    &   $3.4 \times 10^{-32}$  &  $1.2 \times 10^{-31}$     \\
    \bottomrule
  \end{tabular}
\end{table}

\FloatBarrier

Table \ref{table:run_time} Shows a comparison of run-times and memory used for training each model on an Intel Core i9-10900X CPU, applying one step of the training algorithm (see Section \ref{sec:methods_append}) using 2D models with 1024 particles. In the numerical experiments presented in this manuscript convergence was typically observed between 300-1000 iterations for both 2D and 3D training sets. Thus, in summary of training the 2D models, NODE required up to 1.3-4.3 CPU hours, NN-Sum required up to 3.5-11.5 CPU hours, Rot-Inv up to 2.4-8.1 CPU hours, $\nabla  P$ - NN required up to 2.5-8.1 CPU hours, EoS NN required up to 0.7-2.5 CPU hours, and SPH-informed models required up to 0.5-1.7 CPU hours. Scaling up to 3D flows with 4096 particles required up to 240 CPU hours for the most expensive NN-Sum model to achieve convergence. 

\begin{table}[hbt]
  \caption{Comparing (2D) models: Run-times and Memory, $t \sim t_{\lambda}$: One iteration}
  \label{table:run_time}
  \centering
  \begin{tabular}{llll}
    \toprule
    Model with FSA  \hspace{4mm} &   Run-time (s)     &  Memory  (GiB)  \\
    \midrule
    NODE    &  15.50    &   9.62     \\
    NN summand    &  41.60   &  15.61 \\
    Rotationally Invariant NN    &  29.16  &   9.70  \\
    $\nabla  P$ - NN    &  29.37      &   9.51   \\
    EoS NN    &  8.87      &  5.19    \\
    SPH-informed    &   6.03     &  3.52    \\
    \bottomrule
  \end{tabular}
\end{table}


\section{Conclusions}
\label{sec:conclusion}

Combining SPH based modeling, deep learning, automatic differentiation, and local sensitivity analysis, we have developed a learn-able hierarchy of parameterized "physics-explainable" Lagrangian models, and trained each model on both a validation set using weakly compressible SPH data and a high fidelity DNS data set (at three different resolutions) in order to find which model minimizes generalization error over larger time scales and different turbulent Mach numbers. We proposed two new parameterized smoothing kernels, which, once trained on DNS data, improve the accuracy of the SPH predictions compared to DNS, with the second kernel, $W_2(r;h,a,b)$ (which is smooth at the origin), performing best. 

Starting from a Neural ODE based model, we showed that incrementally adding more physical structure into the Lagrangian models using SPH has several important benefits: 
\begin{itemize}

    \item \underline{Improves Generalizability}: as seen in Section \ref{sec:extrapolate} and Section \ref{sec:dns_results}, where we test the ability of the models to predict flows under different conditions not seen in training. The general trend emerged: as more physics informed SPH based structure was embedded in the model, the lower the generalization errors became (both with respect to the loss function used in training and with the external statistical diagnostics). Furthermore, using NNs embedded within the SPH framework to approximate unknown functions can improve generalizability over the standard formulation, but the fully parameterized SPH including the novel parameterized smoothing kernels outperformed the rest.  
    
    \item \underline{Learn new Smoothing Kernels}:  a key ingredient in the construction of the SPH method relies on the properties of the smoothing kernel. Two novel parameterized and learn-able smoothing kernels  -- non-smooth and smooth at the origin, respectively -- were developed. We showed that introducing the parameterization freedom  in the kernels increases flexibility of the physics informed SPH based models and improves generalizability. 
    
    \item \underline{Enforces Physical Symmetries}: the parameterized framework automatically enforces the Gallilelian invariance and allows to keep conservation of linear and angular momenta (translational and rotational invariances) under control across the scales of coarse-graining, see \autoref{table:rot_tran}.
    
    \item \underline{Improves Interpretability}: as the learned parameters become physically meaningful -- in a critical contrast to parameters of the NN models which are physics-agnostic. For example, and as seen in Section \ref{sec:results_append}, our approach is capable to reconstruct with minimal error (counted against the ground truth) physical parameters, associated with the equation of state, artificial shear and bulk viscosity and external forcing. 
    
    \item \underline{More Efficient to Train}: as seen in \autoref{table:run_time}, where we compare different models within the generalized SPH hierarchy. 
    
    \item \underline{Robust with respect to different levels of coarse-graining}: the reported results hold under different resolutions within the inertial range of scales, namely at $N = 12^3, 16^3, 20^3$ as seen in Fig. \ref{fig:lf_gen_over_time_mt_low} and Fig. \ref{fig:lf_gen_over_time_mt_high}.
    
\end{itemize}

In future works, we plan to go beyond the weakly compressible SPH Lagrangian modeling discussed so far and include compressible effects such as shocks. Specifically, we aim to further investigate SPH as a reduced-order Lagrangian model of highly compressible turbulent flows, and further investigate the ability to improve and optimize the SPH framework using the two new parameterized smoothing kernels  proposed in this work, parameterized artificial viscosities and regularization terms.


The Julia source code of our parameterized Lagrangian simulators, the gradient based learning algorithm, sensitivity calculations for each model in the above hierarchy, and the post processing tools can be found in \url{https://github.com/mwoodward-cpu/LearningSPH}.

\subsubsection*{Acknowledgments}

This work has been authored by employees of Triad National Security, LLC which operates Los Alamos National Laboratory (LANL) under Contract No. 89233218CNA000001 with the U.S. Department of Energy/National Nuclear Security Administration. We acknowledge support from LANL’s Laboratory Directed Research and development (LDRD) program, project number 20190058DR and computational resources from LANL's Institutional Computing (IC) program.

The work of Michael Woodward on the project was supported by an Office of Science Graduate Student Research (SCGSR) Program of the Department of Energy (DOE) fellowship, administered by the Oak Ridge Institute for Science and Education (ORISE) for the DOE. ORISE is managed by the Oak Ridge Associated Universities (ORAU) under contract number DE‐SC0014664.

\bibliographystyle{abbrvnat}
\bibliography{./arxiv_prf.bib}

\appendix

\section{
SPH (in more details)}
\label{sec:sph_append}

\subsection{Basic formulation}
In this section we give a summary of SPH, most of which can be found in \cite{monaghan1992, monaghan12, cossins2010smoothed, cfryer_sph_init}. SPH is a discrete approximation to a continuous flow field by using a series of discrete particles. Starting with the trivial identity 

\begin{equation*}
A(\mathbf{r}) = \int_{V} A(\mathbf{r}') \delta (\mathbf{r} - \mathbf{r}') d\mathbf{r}',
\end{equation*}
where $A$ is any scalar or tensor field. Using the smoothing kernel $W$ (for interpolation onto smooth "blobs" of fluid) and after a Taylor expansion it can be shown that (according to  symmetry of smoothing kernel \cite{cossins2010smoothed})
\begin{equation*}
A(\mathbf{r}) = \int_{V} A(\mathbf{r}') W(|\mathbf{r} - \mathbf{r}'|, h) d\mathbf{r}' + \mathcal{O}(h^2).
\end{equation*}
where $W$ is constrained to behave similar to the delta function, 
$$ \int_{V}W(\mathbf{r},h)d\mathbf{r} = 1,\quad \lim_{h \rightarrow 0} W(\mathbf{r},h) = \delta(\mathbf{r}).  $$

\noindent{The choice of smoothing kernels is important, and effects the consistency and accuracy of results \cite{monaghan2005}, where bell-shaped, symmetric, monotonic kernels are the most popular \cite{FULK1996}, however there is still disagreement on the best smoothing kernels to use (but generally should satisfy the conditions outlined in \cite{Liu2010SmoothedPH}). Commonly used are the B-spline smoothing kernels with a finite support (approximating a Gaussian kernel). The cubic smoothing kernel used in this work has the following form}:

\begin{equation}\label{cubic}
 w(q) =  \sigma \begin{cases} 
      \frac{1}{4}(2-q)^3 - (1-q)^3 & 0 \leq q < 1 \\
      \frac{1}{4}(2-q)^3  & 1\leq q\leq 2 \\
      0 & 2\leq q 
   \end{cases}
\end{equation}
where, $W(|\bm r - \bm r_j|, h) = h^{-d}w(q)$ with $q = |\bm r - \bm r_j|/h$, and $\sigma = \sigma(d) = \left[ 1/\pi \hspace{2mm} \text{if} \hspace{2mm} d = 3, \hspace{1mm} 10/7\pi  \hspace{2mm} \text{if} \hspace{2mm} d = 2  \right]$ is a normalizing constant to satisfy the integral constraint on $W$ (see \cite{cossins2010smoothed, FULK1996, monaghan12} for more details). The finite support allows one to use neighborhood list algorithms discussed below to utilize computational advantages (only requires a local cloud of interacting particles instead of all the particles in the computational domain that would be required with a Gaussian kernel). The quartic smoothing kernel (see \cite{liu_sph_overview}) used in this work is 

\begin{equation}\label{quartic}
 w(q) =  \sigma \begin{cases} 
      (2/3 - 9/8 q^2 + 19/24 q^3 - 5/32 q^4) & 0 \leq q < 2 \\
      0 & 2\leq q 
   \end{cases}
\end{equation}
Where $\sigma = 315/(208 \pi h^3)$ for $d = 3$. 

In Section \ref{sec:mixed_mode}, two new parameterized smoothing kernels are introduced, where the shapes are described by the algebraic equations Eqs. \ref{eq:wab} and \ref{eq:w2ab}. This is done in order to increase the flexibility of the SPH model as well as to allow the optimization framework to "discover" the best shaped kernel for our application. Both kernels are introduced to cover a wide range of possible shapes, not necessarily bell-shaped (see Fig. \ref{fig:range_of_ws} below), although both satisfy the conditions for approximating the delta function. The main difference between the two is that $W_2(a,b)$ has a smoother second derivative and may reduce the likelihood of particle pairing as compared to $W_1(a,b)$, however, we postpone a deeper analysis of this comparison until future work, in which we will analyze the effects to changing kernels with standardized compressible flows (such as the Sod shock and Sedov blast wave \cite{Price_2012}).

\begin{figure}[ht]
\centering
\begin{subfigure}[b]{0.48\textwidth}
\centering
\includegraphics[width=1\textwidth]{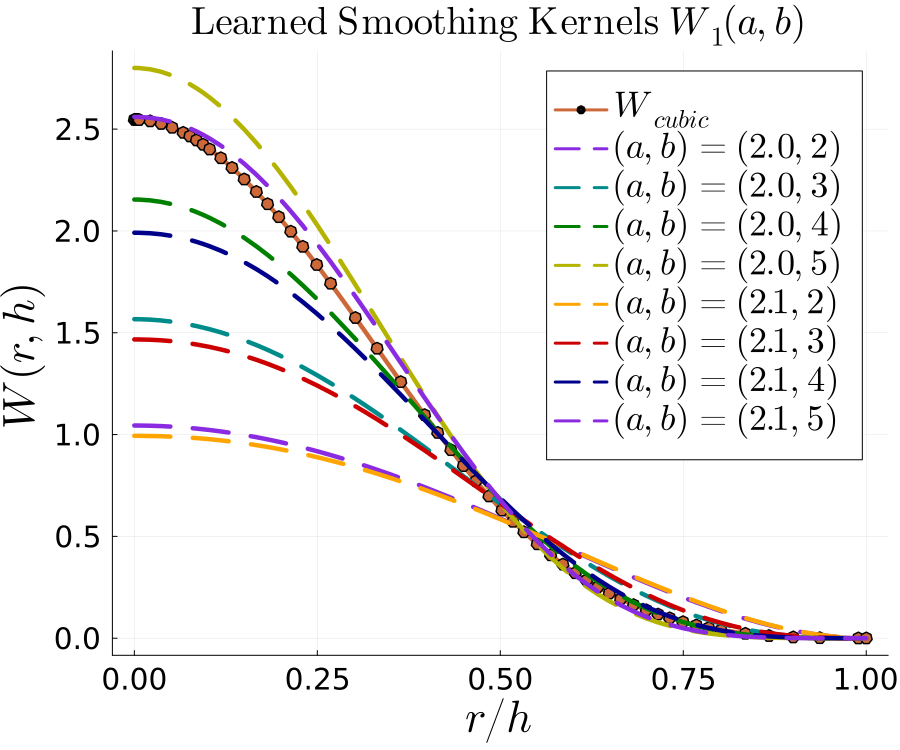}
\end{subfigure}
\begin{subfigure}[b]{0.48\textwidth}
\centering
\includegraphics[width=1\textwidth]{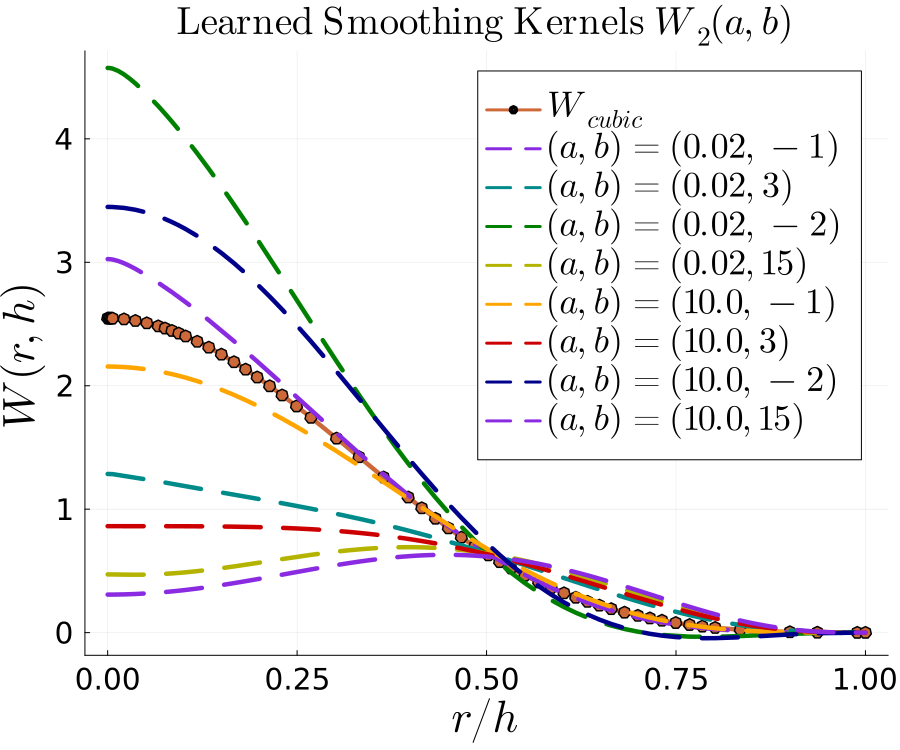}
\end{subfigure}
\caption{Varying shapes for the two new parameterized smoothing kernels as compared to the cubic and Wendland kernel for $d = 3$. The optimization framework is applied in order to discover the best fit shape (see \ref{fig:comp_ws}). }
\label{fig:range_of_ws}
\end{figure}

SPH can be formulated through approximating integral interpolants of any scalar or tensor field $A$ by a series of discrete particles

\begin{eqnarray}\label{eq:sph_discretized}
 \nonumber \left< A(\mathbf{r}) \right> &=& \int_{V} A(\mathbf{r}') W (|\mathbf{r} - \mathbf{r}'|,h) \mathbf{dr}' \\
 &\approx& \sum_{i} m_i \cfrac{A(\mathbf{r}_i)}{\rho(\mathbf{r}_i)} W (|\mathbf{r} - \mathbf{r}_i|,h),
\end{eqnarray}

\noindent{(i.e a convolution of $A$ with $W$) where $\mathbf{dr}'$ denotes a volume element and $W(r, h)$ is the smoothing kernel. Each particle represents a continuous "blob" of fluid and carries the fluid quantities in the Lagrangian frame (such as pressure $P_i$, density $\rho_i$, velocity $\bm v_i$, etc.)}\\

The convenience of this method becomes apparent when the differential operators are approximated (see \cite{liu_sph_overview} for a more detailed derivation). Using the integral interpolation,
\begin{equation*}
     \left< \nabla_{\bm r} A(\mathbf{r}) \right> = \int_{V} \nabla_{\bm r} A(\mathbf{r}') W (||\mathbf{r} - \mathbf{r}'||_2,h) d\mathbf{r}'.
\end{equation*}
Now, using the particle approximation
\begin{equation*}
     \left< \nabla_{\bm r} A(\mathbf{r}) \right> \approx \sum_{i} m_i \cfrac{A(\mathbf{r}_i)}{\rho(\mathbf{r}_i)} \nabla_{\bm r} W (||\mathbf{r} - \mathbf{r}_i||_2,h)),
\end{equation*}

where we see that in this direct approach to approximate the gradient operator we only need to know the gradient of the smoothing kernel (which is usually fixed beforehand). Multiple methods have been proposed and different methods are best suited for different problems. Similar approximations hold for taking the divergence or curl of a vector field \cite{monaghan1977}. The most common "symmetrized" approximation of the gradient operator is derived from the following identity,
\begin{equation*}
    \nabla_{\bm r} A(\bm r_i) = \rho \left( \cfrac{A(\bm r_i)}{\rho^2} \nabla_{\bm r} \rho  - \nabla_{\bm r} \left(\cfrac{A(\bm r_i)}{\rho} \right) \right), 
\end{equation*}
and is approximated with particles as
\begin{equation}\label{eq:common_approx_grad_part}
    \nabla_{\bm r} A_i \approx \rho_i \sum_j^N m_j \left( \cfrac{A_j}{\rho_j^2}  - \cfrac{A_i}{\rho_i^2} \right) \nabla_{{\bm r}_i} W_{ij}. 
\end{equation}

\subsection{Approximation of Flow Equations}
\label{sec:sph_approximation_flow}

The above integral interpolant approximations using series of particles can be used to discretize the equations of motion (as seen in Eq.~(\ref{wc_sph_av}) with more details found in  \cite{monaghan1977, monaghan12}. Each particle carries a mass $m_i$ and velocity $\bm v_i$, and other properties (such as pressure, density etc.). We can use Eq.~(\ref{eq:sph_discretized}) to estimate the density everywhere by 
\begin{gather*}
    \rho(\bm r_i) = \sum_j m_j W(|\bm r_i - \bm r_j|, h),
\end{gather*}
where although the summation is over all particles, because the smoothing kernel has compact support the summation only needs to occur over the smoothing radius (here $2h$ as seen in Eq.~(\ref{cubic}). Another popular way to approximate the density is through using the continuity equation and approximating the divergence of the velocity field in different ways \cite{monaghan1992}. In what follows we use the notation $A_i = A(\bm r_i)$. Using the gradient approximation defined above (Eq.~(\ref{eq:common_approx_grad_part}), the pressure gradient could be estimated by using 
$$ \rho_i \nabla_{\bm r} P_i = \sum_j m_j(P_j - P_i)\nabla_{{\bm r}_i} W_{ij},$$
where $W_{ij} = W(|\bm r_i - \bm r_j|, h)$. However, in this form the momentum equation $d_t \bm v = - \frac{1}{\rho}\nabla_{\bm r} P$ does not conserve linear and angular momentum \cite{monaghan1992}. To improve this, a symmetrization is often done to the pressure gradient term by rewriting $\frac{\nabla_{\bm r} P}{\rho} = \partial_{\bm r} \left( \frac{P}{\rho} \right) + \frac{P}{\rho^2} \nabla_{\bm r} \rho$. This results in a momentum equation for particle $i$ discretized as 
$$\cfrac{d \bm v_i}{dt} = -\sum_j m_i \left(\cfrac{P_j}{\rho_j^2} + \cfrac{P_i}{\rho_i^2} \right) \nabla_{{\bm r}_i} W_{ij}, $$
which produces a symmetric central force between pairs of particles and as a result linear and angular momentum are conserved \cite{monaghan1992}, however is not invariant to changes in background pressure $p_0$. Next, including an artificial viscosity term and external forcing, the full set of ODEs approximating PDEs governing fluid motion (namely Euler's equations with an added artificial viscosity $\Pi$) is
\begin{gather*}
    \cfrac{d\mathbf{r}_i}{dt} = \mathbf{v}_i \hspace{3mm} \forall i \in \{1, 2,... N\} \\
    \label{wc_sph_av-1}
    \cfrac{d\mathbf{v}_i}{dt} = -\sum _{j \neq i}^N m_j \left( \cfrac{P_j}{\rho_j^2} + \cfrac{P_i}{\rho_i^2} + \Pi_{ij} \right) \nabla_{{\bm r}_i}  W_{ij}  + \bm f_{ext}.
\end{gather*}

\noindent{In this work, we start by using the weakly compressible formulation by assuming a barotropic fluid, where equation of state (EoS) is given by}

\begin{equation*}
P(\rho) = \frac{c ^2 \rho_0}{\gamma} \left[ \left(\frac{\rho}{\rho_0} \right)^{\gamma} - 1 \right],
\end{equation*}

\noindent{as in \cite{monaghan12}, where $\rho_0$ is the initial reference density, and $\gamma = 7$ is used. In future work, we plan on including the energy equation to extend these methods for highly compressible applications.}\\

There are many different forms of artificial viscosity that have been proposed \cite{MONAGHAN1997}. In this work, we use the popular formulation of $\Pi_{ij}$ that approximates the contribution from the bulk and shear viscosity along with an approximation of Nueman-Richtmyer viscosity for handling shocks \cite{monaghan12, MORRIS1997}: 
\begin{equation}\label{eq:artificial_viscosity}
   \Pi_{ij} = \begin{cases} 
      \cfrac{-\alpha c_{ij} \mu_{ij}+\beta \mu_{ij}^2}{\rho_{ij}}, & \mathbf{v}_{ij} \cdot \mathbf{r}_{ij} < 0 \\
      0, & \text{otherwise} 
   \end{cases},
   \end{equation}
where $c_{ij} = 0.5 (c_i + c_j)$ and $c_i = \sqrt{d P(\rho_i) / d \rho}$ represents the speed of sound of particle $i$ and 
$$ \mu_{ij} = \cfrac{h \mathbf{v}_{ij} \cdot \mathbf{r}_{ij}}{|\mathbf{r}_{ij}|^2 + \epsilon h^2},\quad
\rho_{ij} = 0.5(\rho_i + \rho_j),$$
This artificial viscosity term was constructed in the standard way following \cite{MONAGHAN1997, MONAGHAN1983}: The linear term involving the speed of sound was based on the viscosity of an ideal gas. This term scales linearly with the velocity divergence, is negative to enforce $\Pi_{ij} > 0$, and should be present only for convergent flows ($\bm v_{ij} \cdot \bm r_{ij} < 0$). The quadratic term including $(\bm v_{ij} \cdot \bm r_{ij})^2$ is used to prevent penetration in high Mach number collisions by producing an artificial pressure roughly proportional to $\rho |\bm v|^2$ and approximates the von Neumann-Richtmyer viscosity (and should also only be present for convergent flows). There are several advantages to this formulation of $\Pi_{ij}$; mainly it is Galilean and rotationally invariant, thus conserves total linear and angular momentum. A more detailed derivation is found in \cite{cossins2010smoothed} (along with other formulations of artificial viscosities). 

In practice the summation Eq.~(\ref{wc_sph_av}) over all particles is carried out through a neighborhood list algorithm (such as the cell linked list algorithm with a computational cost that scales as $\mathcal{O}(N)$ \cite{n_list_dominguez}). We also note that Eq.~(\ref{wc_sph_av}) can also be derived from Euler-Lagrange equations after defining a Lagrangian, see \cite{cossins2010smoothed} (for respective analysis of the inviscid case, when the artificial viscosity term is neglected), then an artificial viscosity $\Pi_{ij}$ term can be incorporated by using the SPH discretizations, see \cite{monaghan2005} for details.

Although the above SPH framework is the most common and contains the same elements as most formulations \cite{monaghan12, monaghan2005, Price_2012, liu_sph_overview, LIND2012, MARRONE2011}, additional terms can be included to address certain applications and situations.  For example, a particle regularization can be included in an attempt to correct issues occurring in certain applications involving particle instabilities and anisotropic distributions which can decrease the accuracy of SPH. In the works of Lind et al. \cite{LIND2012}, a particle shifting is introduced for incompressible SPH solvers to reduce noise in the pressure field by using Fick's law of diffusion to shift particles in a manner that prevents highly anisotropic distributions. In Marrone et al. \cite{MARRONE2011}, the authors introduce a density diffusion term for simulating violent impact flows. However, in this work, we do not incorporate any of these additional particle type regularization terms (on top of the artificial viscosity term which also acts to regularize the particle distributions) for two main reasons: (1) it is not expected that these additional particle regularization terms will have a significant effect for the resolutions considered in this study (as it was found that the artificial viscosity term used was sufficient), and also because (2) the primary focus of this manuscript is to develop a hierarchy of parameterized reduced Lagrangian models for turbulent flows, and to investigate the effects of enforcing physical structure within a Lagrangian framework through SPH versus relying on neural networks (NN)s as universal function approximators.

\subsection{External Forcing}
\label{sec:ext_force}
\hspace{\parindent} In order to approach a stationary homogeneous and isotropic turbulent flow, a deterministic forcing is used (for simplifying the learning algorithms), which is commonly used in CFD literature, (e.g. as in \cite{dlivescu_fext} which is what is used in generating the ground truth DNS data)  for analyzing stationary homogeneous and isotropic turbulence. Then, 
$$\bm f_{ext}^i =  \cfrac{\theta_{inj}}{KE} \bm v_i,\quad KE = \cfrac{0.5}{N}\sum_{k=1}^{N} \rho_k(u_k^2 + v_k^2 + w_k^2)$$
is the kinetic energy computed at each time step, $\theta_{inj}$ represents the rate of energy injected into the flow.


\subsection{Numerical Algorithm for Forward Solving SPH}
\label{sec:numerics}

We use a Velocity Verlet (leap frog) numerical scheme for generating the SPH ground truth data, and for making prediction steps required in our gradient based optimization described in Eq.~(\ref{sec:mixed_mode}. Using the notation,
$$\bm{X} = \{\bm {(r_i, v_i)} | \forall i \in \{1, ..., N\}\}, \hspace{2mm}
\bm{\rho} = \{\bm {\rho}_i | \forall i \in \{1, ..., N\}\},$$ 
$$\cfrac{d \bm r_i}{dt} = \bm v_i,\quad \cfrac{d \bm v_i}{dt} = \bm {{F}}_i(\bm {\rho}, \bm {X}),$$

we proceed according to the following algorithm 
\begin{algorithmic}[1]
\State Compute $\bm{\rho}^k$ using Eq.~(\ref{eq:sph_discretized}),
\State Compute $\bm {{F}}^k_i(\bm {\rho}^k, \bm {X}^k)$ using Eq.~(\ref{wc_sph_av}),
\State $\bm v^{k + \frac{1}{2}}_i = \bm v_i^k +  \cfrac{\Delta t}{2} \bm {{F}}_i^{k}$,
\State $\bm r^{k + 1}_i = \bm r_i^k +  \Delta t \bm {v}_i^{k + \frac{1}{2}}$,
\State Compute $\bm{\rho}^{k+1}$ using Eq.~(\ref{eq:sph_discretized}),
\State Compute $\bm {{F}}^{k+\frac{1}{2}}_i(\bm {\rho}^{k+1}, \bm {X}^{k+\frac{1}{2}})$ using Eq.~(\ref{wc_sph_av}),
\State $\bm v^{k + 1}_i = \bm v_i^{k+\frac{1}{2}} +  \cfrac{\Delta t}{2} \bm {{F}}_i^{k+\frac{1}{2}}$,
\end{algorithmic}
repeated for each time step, $k \in \{0, \Delta t, ..., T\}$, where the time step, $\Delta t$, is chosen according to the Courant-Friedrichs-Lewy (CFL) condition, $\Delta t \leq 0.4 h/c$. This algorithm has the following physical interpretation: it prevents spatial information transfer through the code at a rate greater than the local speed of sound (small in the almost incompressible case considered in this manuscript).

\section{
Methods}
\label{sec:methods_append}

In this section, we provide further details into the methods of this work, including loss functions, sensitivity analysis, and the learning algorithm. 

\subsection{Loss functions}
\label{sec:Loss}

In this section, we consider three different loss functions: trajectory based (Lagrangian), field based Eulerian, and Lagrangian statistics based, described in the following three subsections. Since our overall goal involves learning Lagrangian and SPH based models for turbulence applications, it is the underlying statistical features and large scale field structures we want our models to learn and generalize with. This is discussed further in Section \ref{sec:results} and Appendix \ref{sec:dns_results}, where we compare the statistical and field based generalizability of each model within the hierarchy.

\subsubsection{Trajectory Based Loss Function}
\label{sec:traj_based}

A naive loss function to consider is the Mean Squared Error $(MSE)$ of the difference in the Lagrangian  particles positions and velocities,  as they evolve in time,
$$L_{tr}(\bm \theta) =  MSE(\bm X, \hat{\bm X}(\bm \theta)) = \| \bm X - \hat{ \bm X}(\bm \theta)\|^2/N,$$
where $\bm X$ and $\hat{ \bm X}$  are the particle states -- the ground truth and the predicted, respectively. Minimizing this loss function will result in discovering optimal parameters such that the predicted trajectories gives the best possible match (within the model) for each of the particles. However, the Lagrangian tracer particles are non-inertial, whereas the SPH particles have mass, thus the trajectory based loss function is not consistent with the data.

\subsubsection{Field Based Loss Function}

The field based loss function tries to minimize the difference between the large scale structures found in the Eulerian velocity fields, 
\begin{equation*}
L_f(\bm \theta) = MSE(\bm V^f, \hat{ \bm V}^f) =  \| \bm V^f - \hat{ \bm V}^f\|^2/N_f, 
\end{equation*}
where $ \bm V_i^f = \sum_{j=1}^{N_f} (m_j/\rho_j)\bm v_j W_{ij}(||\bm r^f_i - \bm r_j||, h)$
uses the same SPH smoothing approximation to interpolate the particle velocity onto a predefined mesh $\bm r^f$ (with $N_f$ grid points). Lets recall that SPH is, by itself, an approximation for the velocity field, therefore providing a strong additional motivation for using the field based loss function.

\subsubsection{Lagrangian Statistics Based Loss Function}

In order to approach learning Lagrangian models that capture the statistical nature of turbulent flows, one can use well established statistical tools/objects, such as single particle statistics \cite{yeung_borgas_2004}. In this direction, consider the time integrated Kullback–Leibler divergence (KL) as a loss function 
\begin{equation*} 
    L_{kl}(\bm \theta) =  \int_{t = 0}^{t_f} \int_{-\infty}^{\infty} P_{gt}(t, \bm z_{gt}, {\bm x})\log \left(\cfrac{P_{gt}(t, \bm z_{gt}, {\bm x})}{P_{pr}(t, \bm z_{pr}(\bm \theta), {\bm x})} \right) d{\bm x} dt,
\end{equation*}
where $\bm z_{gt} = \bm z_{gt}(t)$, and $\bm z_{pr}(\bm \theta) = \bm z_{pr}(\bm \theta, t)$ represent single particle statistical objects over time of the ground truth and predicted data, respectively. For example, we can use the velocity increment, $\bm z^i(t) = (\delta u_i, \delta v_i, \delta w_i)$, where $\delta u_i(t) = u_i(t) - u_i(0)$ and $\bm z$ ranges over all particles. Here $P(t, \bm z(t), x)$ is a continuous probability distribution (in $x$) constructed from data $\bm z(t)$ using Kernel Density Estimation (KDE), to obtain smooth and differentiable distributions from data,  as discussed in \cite{kde_yen_intro}), that is  $$P(\tau, \bm z, \bm x) = (N h)^{-1}\sum_{i=1}^{N}K\left((z_i - {\bm x})/h\right),$$
where $K$ is the smoothing kernel (chosen to be the normalized Gaussian in this work). In the experiments below, a combination of the $L_f$ and $L_{kl}$ are used with gradient descent in order to first guide the model parameters in order to reproduce large scale structures with $L_f$, then later refine the model parameters with respect to the small scale features inherent in the velocity increment statistics by minimizing $L_{kl}$.

\subsubsection{Forward and Adjoint based Methods Supplemental}
\label{sec:fsa_asa}
Let us, first, introduce some useful notations for our parameterized SPH informed models;
$\bm X_i = 
( \bm r_i,\ 
\bm v_i)^{T}, 
\quad
{\bm X} = \{\bm X_i | i =1,...N\}
$, where, $\bm r_i = (x_i, y_i, z_i)$, and, $\bm v_i = (u_i, v_i, w_i)$, are the position and velocity of particle $i$ respectively. Also, $\bm {\theta} = [\theta^1, ..., \theta^p]^T $ where $p$ is the number of model parameters. Now, each parameterized Lagrangian model in the hierarchy can be stated in the ODE form
\begin{gather}\label{eq:sph_standard_sup}
   \forall i:\quad d \bm X_i/dt = \bm{\mathcal{F}}_i({\bm X}(t, \bm \theta), \bm {\theta}) = 
    (\bm v_i,\ 
    \bm F_i(\bm X, \bm{\theta}))^T,
\end{gather} 
where $\bm F_i(\bm X, \bm \theta) = \dot{\bm v}_i$ is the parameterized acceleration operator as defined in the above hierarchy. Forward and Adjoint based Sensitivity Analyses (FSA, ASA), analogous to forward and reverse mode AD respectively, can be used to compute the gradient of the loss function (see Section \ref{sec:Loss}), as seen in Eq.(\ref{eq:L_p_1})
\begin{equation}\label{eq:L_p_1}
    \partial _{\bm \theta}L = \int\limits_0^{t_f} \partial_{\bm X} \Psi(\bm X, \bm \theta, t) d_{\bm \theta} \bm X(\bm \theta, t) + \partial_{\bm \theta}\Psi(\bm X, \bm \theta, t) dt,
\end{equation}
 Briefly, FSA computes the sensitivity equations (SE) for $\sbia = d \bm X_i/d \theta^{k}$, by simultaneously integrating a system of ODEs: ($\forall i,\ \forall k=1,\cdots,p:\quad$)
\begin{gather}
\label{eq:fsa_2_1_sup} 
\cfrac{d \sbia}{dt} = \cfrac{\partial \bm{\mathcal{F}}_i(\bm X(t), \bm{\theta})}{\partial \bm X_i}\sbia + \cfrac{\partial\bm{\mathcal{F}}_i(\bm X(t), \bm{\theta})}{\partial \theta^{k}}
\end{gather}
which has a computational cost that scales linearly with the number of parameters (for derivations of FSA and ASA see Appendix \ref{sec:methods_append}).
$\sbia$ is computed by solving the IVP Eq.~(\ref{eq:fsa_2_1}) with the initial condition $\sbia(0) = 0$ (assuming $\bm X(0)$ does not depend on $\bm \theta$). After solving for $\sbia$, the gradient is computed directly from  Eq.~(\ref{eq:L_p_1}) which is used with standard optimization tools (e.g. Adam algorithm of \cite{kingma2017adam}) to update the model parameters iteratively to minimize the loss (see Appendix \ref{sec:methods_append}). As opposed to the FSA method, the ASA approach avoids needing to compute $d_{\bm \theta}\bm X$ by instead numerically solving a system of equations for the adjoint equation (AE) Eq.~(\ref{eq:lambda_t_simp}) backwards in time, according to Section \ref{sec:asa_append}. Once $\bm \lambda$ is computed through integration backwards in time, the gradient of the loss function can be computed according to,
    $\partial_{\bm \theta}L = -\int_0^{t_f} \bm \lambda^T \partial \bm{\mathcal{F}} / \partial \bm \theta dt$.

The computational cost of solving the AE is independent of the number of parameters, however, for high dimensional time dependent systems, the forward-backward workflow of solving the AE imposes a significant storage cost since the AE must be solved backwards in time \cite{donello2020computing, nouri2021skeletal}. Through hyper-parameter tuning we found that the number of parameters $\mathcal{O}(1) \lesssim p \lesssim \mathcal{O}(9 * 10^2)$ for the best fit models within the hierarchy, and so for $N \gtrsim 1000$, the dimension of the system is much larger than $p$ so FSA is more efficient, and therefore FSA forms our main structure to compute the gradient. The gradient is computed over all parameters found from using SA over all particles; alternatively stochastic gradient descent (SGD) could be used, however, for each batch all the particle would need to be simulated forward in time (since each particle is interacting with all neighboring particles) requiring a full forward solve for each batch, defeating any computational advantages of SGD.

\subsubsection{Mixed Mode AD}

In both FSA and ASA described above, the gradient of $\bm{\mathcal{F}}_i$ with respect to the parameters, $\partial\bm{\mathcal{F}}_i(\bm X(\tau), \bm{\theta})/\partial \bm \theta$, and the Jacobian  matrix, $\{\partial \bm{\mathcal{F}}_i(\bm X(\tau), \bm{\theta})/\partial \bm X_j| \forall i,j \}$, need to be computed. In this manuscript, we accomplish this with a mixed mode approach, i.e. mixing forward and reverse mode AD within the FSA framework, where the choice is based on efficiency. This is determined by the input and output dimensions of the function being differentiated. Depending on the above model used, several functions need to be differentiated and a mixture of forward and reverse mode can be implemented within the FSA system for optimizing efficiency. For example, when computing  $\partial \bm{\mathcal{F}}_i(\bm X(\tau), \bm{\theta})/\partial \theta^{\alpha}$, with AD where, $\bm{\mathcal{F}}_i(\bm \theta)
: \mathbb{R}^p \rightarrow \mathbb{R}^{2d}$, if $p\gg 2d$, then reverse mode AD is more efficient than forward mode \cite{ad_bucker06}.

\subsection{Forward Sensitivity Analysis}
\label{sec:fsa_append}
\hspace{\parindent} In general, the loss functions in this work can be defined as 
\begin{equation*}
    L(\bm{X}, \bm \theta) = \int_0^{t_f} \Psi(\bm X, \bm \theta, t) dt.
\end{equation*}
The forward SA (FSA) approach simultaneously integrates the state variables along with their sensitivities (with respect to parameters) forward in time to compute the gradient of $L$; 
\begin{equation}\label{eq:L_p_append}
    d_{\bm \theta}L = \int_0^{t_f} \partial_{\bm X} \Psi(\bm X, \bm \theta, t) d_{\bm \theta} \bm X(\bm \theta, t) + \partial_{\bm \theta}\Psi(\bm X, \bm \theta, t) dt.
\end{equation}
Where, through using the chain rule, we see that the sensitivities of the state variables with respect to the model parameters ($ d_{\bm \theta}\bm X)$ are required to compute the gradient of the loss. Assuming that the initial conditions of the state variables do not depend on the parameters, then $\partial{\bm X}(0)/\partial\theta^{\alpha} = 0$. Now, define the sensitivities as $\sbia := d \bm X_i/d \tal$. Then, from Eq.~(\ref{eq:sph_standard}) we derive
\begin{gather}
\cfrac{d \sbia}{dt} = \cfrac{d \bm{\mathcal{F}}_i(\bm X(t), \bm{\theta})}{d \tal},
\end{gather}
then resulting, after applying the chain rule, in 
\begin{gather}
\label{eq:fsa_2_append} 
\cfrac{d \sbia}{dt} = \p{\bm{\mathcal{F}}_i(\bm X(t), \bm{\theta})}{\bm X_i}\sbia + \p{\bm{\mathcal{F}}_i(\bm X(t), \bm{\theta})}{\theta^{\alpha}}.
\end{gather}
Since the initial condition $\bm X(0)$ does not depend on $\bm \theta$, then $ \sbia(0) = 0$. Now, computing the gradient of the loss function reduces to solving a forward in time Initial Value Problem (IVP) by integrating simultaneously the state variables $\bm X_i$ 
defined in the main text, and sensitivities $\sbia$, defined in Eq.~(\ref{eq:fsa_2_append}).

In order to integrate Eq.~(\ref{eq:fsa_2_append}) the gradient of $\bm{\mathcal{F}}_i$ with respect to the parameters, both $\partial{\bm{\mathcal{F}}_i(\bm X(\tau), \bm{\theta})}/\partial\theta^{\alpha}$ and the Jacobian matrix, $\partial{\bm{\mathcal{F}}_i(\bm X(\tau), \bm{\theta})}/\partial \bm X_i$ need to be computed. In this work, this is done with a mixed mode approach. $\partial {\bm{\mathcal{F}}_i(\bm X(\tau), \bm{\theta})}/\partial \theta^{\alpha}$, with $\bm{\mathcal{F}}_i(\bm \theta)
: \mathbb{R}^k \rightarrow \mathbb{R}^{2d}$, is computed with AD (the choice of forward or reverse mode is determined by the dimension of the input and output space), where $k$ is the number of parameters and $d$ is the dimension. For example, if $k \gg 2d$ (as is the case when $NN$s are used), reverse mode AD is more efficient than forward mode \cite{ma2021comparison}. The Jacobian matrix is computed and obtained through mixing symbolic differentiation packages (or analytically deriving by hand), as well as mixing AD. For example, when there are NNs used for the parameterization of the right hand side, then according to expression for the Jacobian from the main text, AD derivatives will need to be computed on different functions each with potentially different dimensions of input and output space.  The AD packages used in this work were both ForwardDiff.jl \cite{RevelsLubinPapamarkou2016} for forward mode and Zygote.jl \cite{Zygote.jl-2018} for reverse mode. The Jacobian matrix for 2D problems is

\begin{gather*}
    \p{\mathcal{F}_i(\bm X(t), \bm{\theta})}{\bm X_i} = \begin{pmatrix}
      [0]_2  \hspace{3mm} I_2   \\
      \p{\bm F_i(\bm X(t), \bm{\theta})}{\bm X_i}\\
    \end{pmatrix}
\end{gather*}
\begin{gather*}
    = \begin{pmatrix}
        & & & \\
     & [0]_2  & I_2 &\\
        & & & \\
    \p{F^x_i}{x_i^1} & \p{F^x_i}{x_i^2} & \p{F^x_i}{x_i^3} & \p{F^x_i}{x_i^4}\\
     \p{F^y_i}{x_i^1} & \p{F^y_i}{x_i^2} & \p{F^y_i}{x_i^3} & \p{F^y_i}{x_i^4}
    \end{pmatrix}.
\end{gather*}
where, the individual derivatives $\p{F^y_i}{x_i^d}$ are computed with AD (Reverse or Forward mode depending on model chosen from Section \ref{sec:hiearchy}, and a similar formulation is carried out in 3D).

\subsubsection{Adjoint Method}
\label{sec:asa_append}

\hspace{\parindent} This section provides an outline of the Adjoint SA (ASA) method used in this work (for more details, see \cite{bradley_adjoint} \cite{donello2020computing}). However, we note that the main results of the text did not require using the adjoint method because the FSA was found to be more efficient (since a relatively small number of parameters were needed compared to the number of particles as discussed in Section \ref{sec:mixed_mode}), but we include this as a reference to the associated source code that has the option of using ASA in case the number of parameters becomes large enough ($p >> 2*D*N$). Again, the goal is to compute the gradient of the loss function. This is a continuous time dependent formulation, where the goal is to minimize a loss function $L(\bm X(\bm \theta, t), \bm \theta)$ which is integrated over time,
$L(\bm{X}, \bm \theta) = \int_0^{t_f} \Psi(\bm X, \bm \theta, t) dt$,
subject to the physical structure constraints (ODE or PDE),
$H(\bm X, \bm {\dot X}, \bm \theta, t) = 0$,
and the dependence of the initial condition,
$g(\bm X(0), \bm \theta) = 0$, on parameters.
Here, $H$ is the explicit ODE form obtained through the SPH discretization equations Eq.~(\ref{eq:sph_standard}) (discretization of PDE flow equations),
\begin{equation}\label{eq:H}
 H(\bm X, \bm {\dot X}, \bm \theta, t) = \bm {\dot X}(t) - \bm{\mathcal{F}}(\bm X(t), \bm \theta).
\end{equation}

A gradient based optimization algorithm requires that the gradient of the loss function,
$$ d_{\bm \theta}L(\bm X, \bm \theta) = \int_0^{t_f} \partial_{\bm X} \Psi(\bm X, \bm \theta, t) d_{\bm \theta} \bm X(\bm \theta, t) + \partial_{\bm \theta}\Psi(\bm X, \bm \theta, t) dt,$$
be computed. The main difference in the FSA and ASA approach is that in the ASA calculating $d_{\bm \theta}\bm X$ is not required (which avoids integrating the additional $k$ ODEs as in FSA). Instead, the adjoint method develops a second ODE (size of which is independent of $k$) in the adjoint variable $\bm \lambda$ as a function of time (which is then integrated backwards in time).

The following provides a Lagrange multiplier approach to deriving this ODE in $\bm \lambda$. First define
\begin{eqnarray}\label{eq:lagrange_append}
    \nonumber \mathcal{L} &=& \int_0^{t_f}(\Psi(\bm X, \bm \theta, t) + \bm \lambda^T(t) H(\bm X, \bm {\dot X}, \bm \theta, t))dt \\
     &+& \mu^T g(\bm X(0), \bm \theta),
\end{eqnarray}
where $\bm \lambda$ and $\mu$ are the Lagrange multipliers. Now, since $H$ and $g$ are zero everywhere, we may choose the values of $\bm \lambda$ and $\mu$ arbitrarily; also as a consequence of $H$ and $g$ being everywhere zero, we note that $\nabla_{\bm \theta} \mathcal{L} = \nabla_{\bm \theta} L$. 

Now, taking the gradient of Eq.~(\ref{eq:lagrange_append}), and simplifying
\begin{gather*}
 \nabla_{\bm \theta} \mathcal{L} = \int_0^{t_f} (\partial_{\bm X} \Psi d_{\bm \theta} \bm X + \partial_{\bm \theta}\Psi + \bm \lambda^T ( \partial_{\bm X}H d_{\bm \theta} \bm X \\
 + \partial_{\bm{\dot{X}}}H d_{\bm \theta}\bm{\dot{X}} + \partial_{\bm \theta}H)) dt \\
 + \mu^T (\partial_{\bm X(0)} g d_{\bm \theta}\bm X(0) + \partial_{\bm \theta}g).
\end{gather*}
Integrating the equation by parts, and eliminating, $d_{\bm \theta}\bm{\dot{X}}$, we arrive at
\begin{gather*}
    \nabla_{\bm \theta} \mathcal{L}= \int_0^{t_f} \Biggl[ \left( \partial_{\bm X} \Psi + \bm \lambda^T(\partial_{\bm X}H - d_t\partial_{\bm{\dot{X}}}H) - \bm{\dot{\lambda}^T}\partial_{\bm{\dot{X}}}H \right)d_{\bm{\theta}}\bm X \\
     + \partial_{\bm \theta} \Psi + \bm{\lambda}^T\partial_{\bm \theta}H] \Biggr] dt + \bm \lambda^T \partial_{\bm{\dot{X}}}H d_{\bm \theta}\bm X \bigg\rvert_{t_f} \\
     +(-\bm \lambda^T\partial_{\bm{\dot{X}}}H + \mu^Tg)\bigg\rvert_0 d_{\bm \theta}\bm X(0) + \mu^T \partial_{\bm \theta}g .
\end{gather*}
Since the choice of, $\bm \lambda^T$, and, $\mu$, is arbitrary, we set, $\bm \lambda^T(T) = 0$, and, $\mu^T = (\bm \lambda^T \partial_{\bm{\dot{X}}}H)\rvert_0 (g\rvert_{\bm X(0)})^{-1}$, in order to avoid needing to compute, $d_{\bm \theta}\bm X (T)$, and thus canceling the second to the last term in the latest (inline) expression. Now, assuming that the initial values of the state variables, $\bm X(0)$, do not depend on the parameters, then $d_{\bm \theta}\bm X(0) = 0$, and, $g = 0$. Furthermore, we use a loss function $\Psi$ that does not depend on $\bm \theta$ explicitly, so that, $\partial_{\bm \theta}\Psi = 0$. And finally, we can avoid computing $d_{\bm \theta} \bm X$ at all other times $t>0$ by setting
\begin{equation*}
\partial_{\bm X} \Psi + \bm \lambda^T (\partial_{\bm X} H - d_t \partial_{\bm{\dot{X}}}H) - \bm{\dot{\lambda}}^T \partial_{\bm{\dot{X}}}H = 0.
\end{equation*}
The resulting equation for the time derivative of ${\bm\lambda}$ can be re-stated as the following Adjoint Equation (AE)
\begin{equation}\label{eq:lambda_t_simp}
    \bm{\dot{\lambda}}^T = \partial_{\bm X} \Psi - \bm \lambda^T \p{\bm{\mathcal{F}}}{\bm X}, \hspace{8mm}
    \bm \lambda({t_f}) = 0,
\end{equation}
where we also used that according to Eq.~(\ref{eq:H}), $\partial_{\bm X} H = -\partial_{\bm X}\bm{\mathcal{F}}$ and $\partial_{\bm{\dot{X}}}H = I_{2d}$.

Combining all of the above, we see that the simplified equation for the gradient of the loss function is
$$\nabla_{\bm \theta} L = \nabla_{\bm \theta} \mathcal{L} = \int_0^{t_f} \bm \lambda^T \partial_{\bm \theta}H dt.  $$
Which, according to Eq.~(\ref{eq:H}), becomes 
\begin{gather*}
    \nabla_{\bm \theta}L = -\int_0^{t_f} \bm \lambda^T \p{\bm{\mathcal{F}}}{\bm \theta}dt.
\end{gather*}
Therefore, in order to compute the gradient, the IVP expression  Eq.~(\ref{eq:lambda_t_simp}) needs to be integrated backwards in time for $\bm \lambda(t)$, starting from $\bm \lambda({t_f}) = 0$.  Similar to the FSA formulation found above, both $\partial_{\bm X}\bm{\mathcal{F}}$ and $\partial_{\bm \theta}\bm{\mathcal{F}}$, are computed with a mixture of forward and reverse mode AD tools, depending on the dimension of the input and output dimension of the functions to be differentiated.

\subsection{Learning Algorithm}

In what follows, we combine all of the computational tools and techniques introduced so far to outline the mixed mode learning algorithm.

\begin{algorithm}[H]
    \centering
    \caption{Mixed Mode Learning Algorithm}\label{algorithm1}
    \begin{algorithmic}[1]
    \State Given Ground Truth:  $\{\bm X(t_0), \bm X(t_1), ..., \bm X(t_f)\}$ SPH data \;
    \State Select model: $d_t\bm{X}(t) = \bm {\mathcal{F}}(\bm X(t), \bm \theta)$ from hierarchy \\
    \State Select SA method (Forward or Adjoint)\\
    \State Choose Loss methods (combination of Trajectory, Field, or Probabilistic) \\
    \State Choose optimizer: \hspace{2mm} e.g   $Opt = \{RMSprop,  ADAM\}$ \\

        --------------------------------------------------------------------------------------------\\
        
        \vspace{0.5cm}

        \For{$k\gets 1, ...,n$}
        \State 		Prediction step: Verlet integration of model $\hat{\bm X} = Verlet (\bm{\mathcal{F}}, \bm \theta_k, \bm X(t_0), t_0, t_f)$ \\
        	\State	\hspace{3mm} Simultaneously compute $ \p{\bm{\mathcal{F}}}{\hat{\bm X}_i}$, $ \p{\bm{\mathcal{F}}}{\bm \theta}$ with mixed mode AD.\\
        	\State Simultaneously integrate system of ODEs for sensitivities $\bm S_i^{\alpha}$; \\
            \State $$ d_t \bm S_i = \p{\bm{\mathcal{F}}_i(\hat{\bm X}(t), \bm{\theta_k})}{\hat{\bm X}_i}\bm S_i + \p{\bm{\mathcal{F}}_i(\hat{\bm X}(t), \bm{\theta_k})}{\bm \theta_k} $$
            \State Compute $\nabla L$ 
        \EndFor
		
		\vspace{2mm}

		\State update $\bm \theta$ using optimizer;\\
		\State \hspace{4mm} $\bm \theta_{k+1} = Opt(\bm \theta_k)$
    \end{algorithmic}
    \label{alg:mixed}
\end{algorithm}

\section{Evaluating models: additional results}
\label{sec:results_append}
In this section, we include additional results of the trained models, for example using the combination of $L_f$ and $L_{kl}$. 

\subsection{Comparing Smoothing Kernels}
\label{sec:comp_new_W}
\begin{figure}[htp]
\centering
\begin{subfigure}[b]{0.42\textwidth}
\centering
\includegraphics[width=0.99\textwidth]{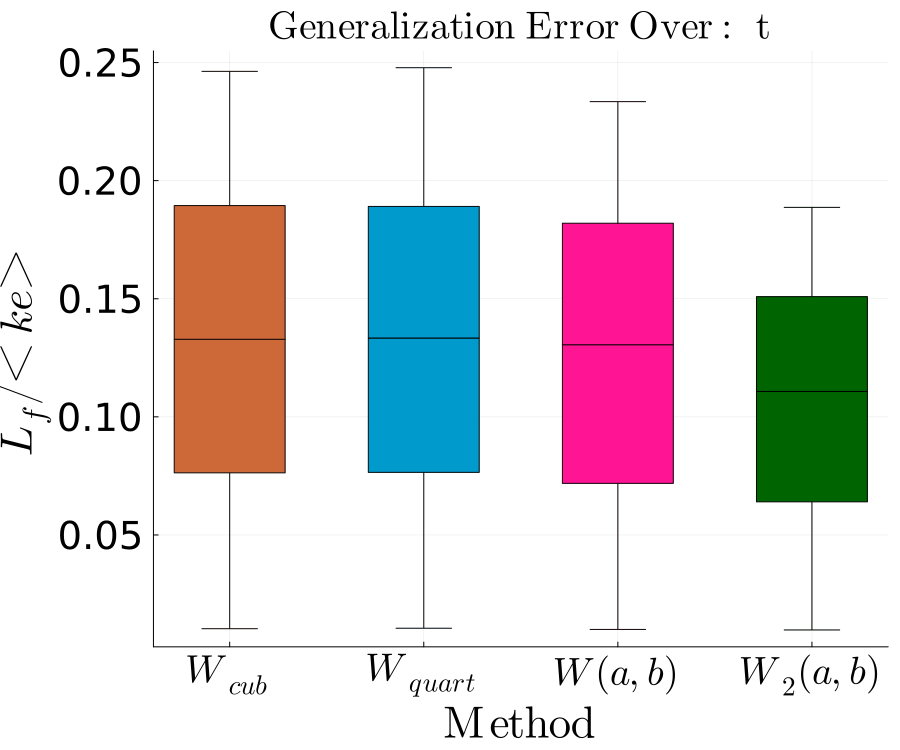}
\caption{Over time}
\end{subfigure}
\begin{subfigure}[b]{0.42\textwidth}
\centering
\includegraphics[width=0.99\textwidth]{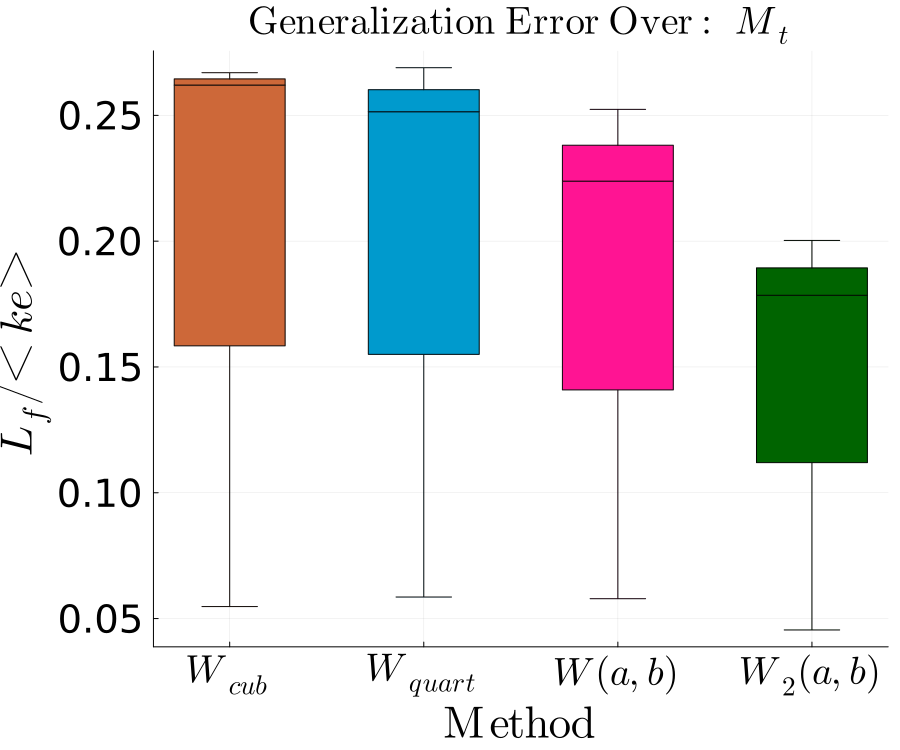}
\caption{Over $M_t$}
\end{subfigure}
\caption{(a) Measuring (using the $L_f$ normalized with the average kinetic energy) the generalization error over (a) 20 longer time scales up to the eddy turn over time and (b) over 3 different $M_t \in \{0.04, 0.08, 0.16\}$ at the eddy turn over time scale of each fully parameterized SPH informed models, comparing two new parameterized smoothing kernels with two classic smoothing kernels. This shows that the new parameterized smoothing kernels outperform in their ability to generalize to other DNS flows as compared to the the cubic and quartic at this spatial resolution using $N=4096$ particles.}
\label{fig:comp_w_loss_gen}
\end{figure}

\begin{figure}[htp]
\centering
\begin{subfigure}[b]{0.45\textwidth}
\centering
\includegraphics[width=0.99\textwidth]{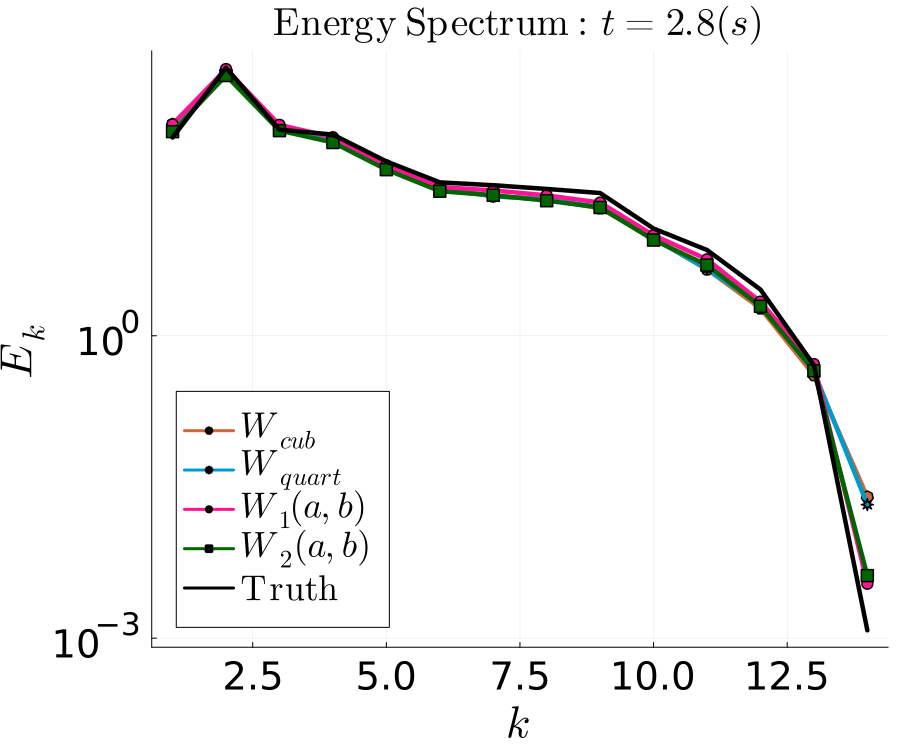}
\end{subfigure}
\begin{subfigure}[b]{0.45\textwidth}
\centering
\includegraphics[width=0.99\textwidth]{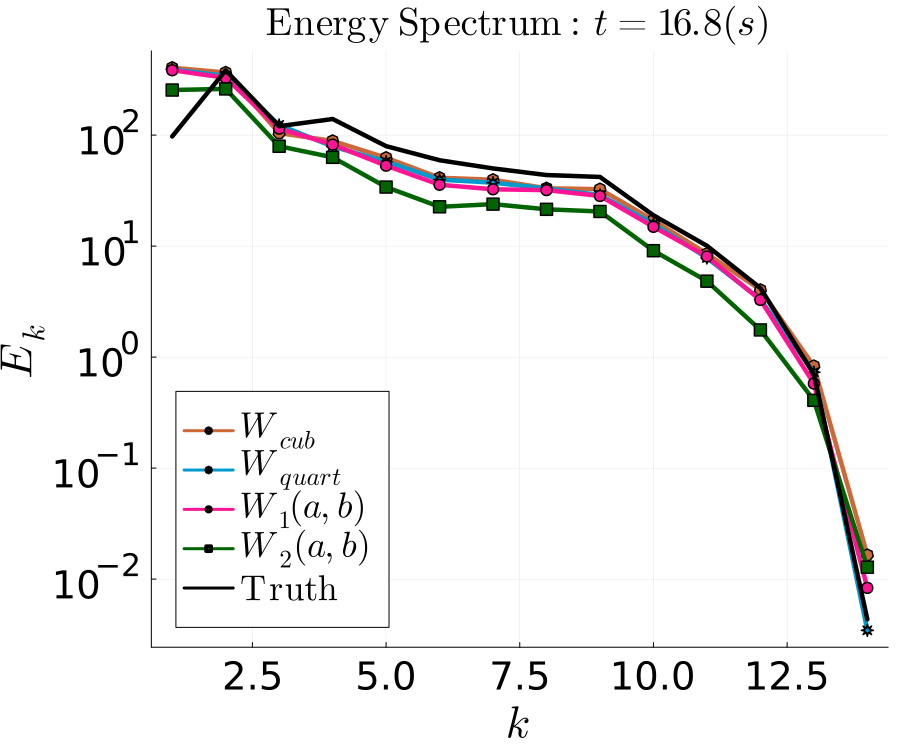}
\end{subfigure}
\caption{Plotting Energy spectrum over different time scales and comparing each smoothing kernel. Each model does well at capturing energy spectrum on the Kolmogorov time scale, however, on the eddy turn over time, the quartic and $W(a,b)$ seem to perform marginally better. }
\label{fig:comp_w_Ek}
\end{figure}

\begin{figure}[htp]
\centering
\begin{subfigure}[b]{0.48\textwidth}
\centering
\includegraphics[width=1\textwidth]{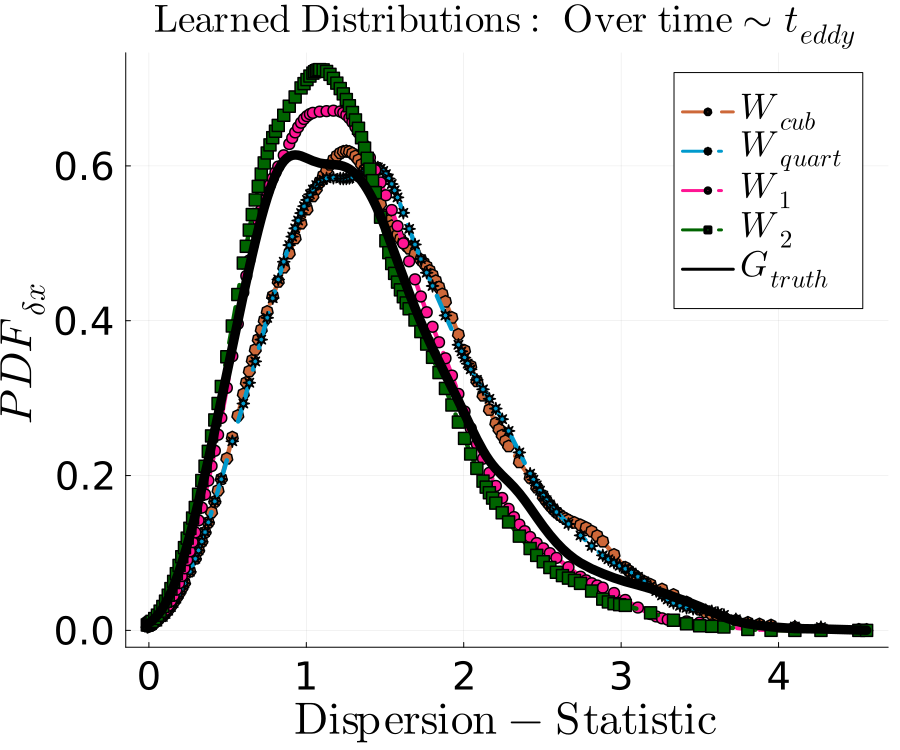}
\end{subfigure}
\begin{subfigure}[b]{0.48\textwidth}
\centering
\includegraphics[width=1\textwidth]{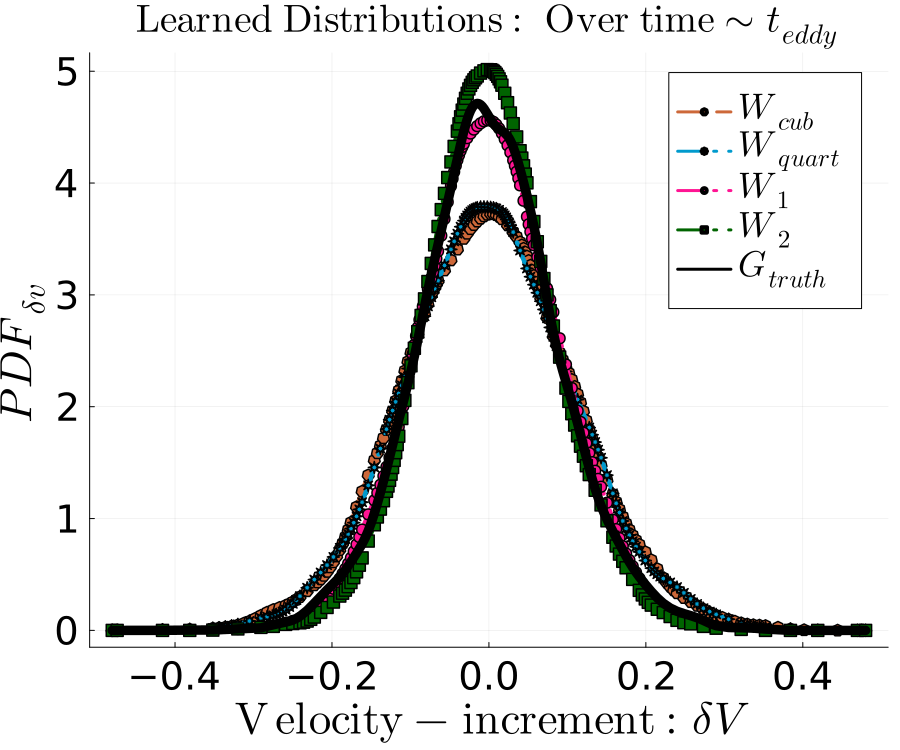}
\end{subfigure}
\caption{A diagnostic check comparing single particle statistics on larger time scales (roughly 20 times longer than seen in training), with the new parameterized smoothing kernels SPH-informed model having the best fit. Each is not seen in training, so represent an external diagnostic. }
\label{fig:sing_part_stats_ws}
\end{figure}

In Fig. \ref{fig:comp_ws} we compare four different smoothing kernels included when learning the fully parameterized SPH model; two standard and fixed smoothing kernels, namely the cubic Eq.~(\ref{cubic}) and quartic Eq.~(\ref{quartic}) splines (which follow the Gaussian bell shape); and two novel parameterized smoothing kernels that are learned from DNS data when training the fully parameterized SPH-informed model Eq.~(\ref{eq:phys_wab}). We provide numerical evidence that shows including these novel parameterized smoothing kernels in training improves the ability of the SPH informed model to solve the interpolation problem and generalize to flows not seen in training as in Fig. \ref{fig:comp_w_loss_gen}. This result seems to suggest that an improvement to the SPH construction can be made, at least in the weakly compressible setting at the resolved scales as seen in this work, by considering smoothing kernels which are not of the classical bell shape, but rather of the more flexible and parameterized form proposed in Eq.~(\ref{eq:wab}) and Eq.~(\ref{eq:w2ab}). Future work could be done in this direction to further analyze and compare these new smoothing kernels with regards to convergence and consistency. We further compare these smoothing kernels from the trained SPH models by making forward predictions and evaluating the acceleration statistics Fig. \ref{fig:comp_w_ga_mt_t} and energy spectrum Fig. \ref{fig:comp_w_Ek} and single particle statistics Fig. \ref{fig:sing_part_stats_ws}; from which we can conclude that the new parameterized smoothing kernels, especially $W_2$, has the best generalizability, i.e improves the accuracy of SPH over the standard cubic and quartic smoothing kernels to match DNS at different time scales and turbulent Mach numbers with respect to field based and statistical measures. The learned parameters are tabulated in Fig. \ref{table:learned_sph_params}.

The learned parameters found from training the SPH based models are found in Table \ref{table:learned_sph_params}.

\begin{table}[htp]
  \caption{Learned SPH parameters}
  \label{table:learned_sph_params}
  \centering
  \begin{tabular}{lllllllll}
    \toprule
    Model  \hspace{4mm} &   $\hat{c}$     &  $\hat{\alpha}$    &  $\hat{\beta}$  &  $\hat{\gamma}$   &  $\hat{p_0}$   &  $\hat{\theta}$  & $\hat{a}$  & $\hat{b}$  \\
    \midrule
    SPH-informed: $W_{cubic}$   &   $0.55$  & $0.19$  & $-0.29$  &  $1.69$  & $-0.3$  & $0.0013$  &  &      \\
    SPH-informed: $W_{quartic}$ &   $0.53$  &  $0.19$  & $-0.25$  & $1.82$  & $-0.28$  & $0.0013$  &  &    \\
    SPH-informed: $W(a,b)$      &   $0.48$  &  $0.21$  & $0.62$  &  $2.05$  & $-0.24$  &  $0.0009$  & $2.82$  & $3.05$     \\
    SPH-informed: $W_2(a,b)$    &   $0.50$  &  $0.19$  & $0.44$  &  $2.05$  &  $-0.26$  &  $0.0006$  &  $1.43$  & $0.53$      \\
    \bottomrule
  \end{tabular}
\end{table}


\subsection{Generalizability: Training with Statistical Loss}
\begin{figure}[htp]
\centering
\begin{subfigure}[b]{0.48\textwidth}
\centering
\includegraphics[width=1\textwidth]{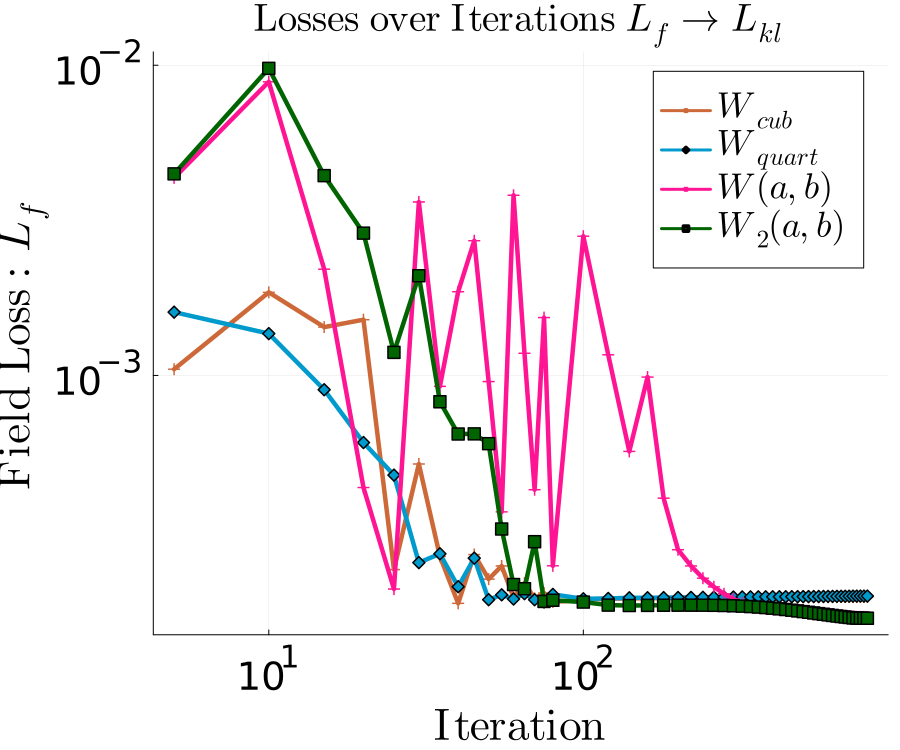}
\end{subfigure}
\caption{The field based loss converging, using $L_f$ to guide large scale structures then switching to $L_{kl}$ to capture small scale statistical characteristics found the velocity increment distribution. Fig. \ref{fig:skl1_comp_W_loss} shows the shape of the smoothing kernels is similar as with only using $L_f$.}
\label{fig:skl1_comp_W_loss_conv}
\end{figure}

\begin{figure}[htp]
\centering
\begin{subfigure}[b]{0.48\textwidth}
\centering
\includegraphics[width=1\textwidth]{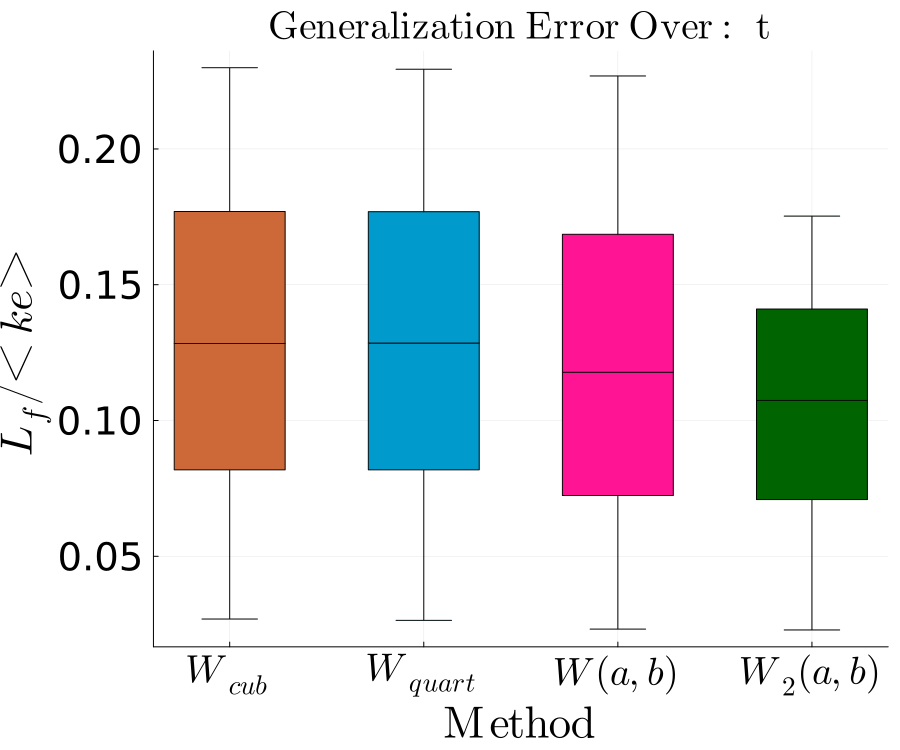}
\end{subfigure}
\begin{subfigure}[b]{0.48\textwidth}
\centering
\includegraphics[width=1\textwidth]{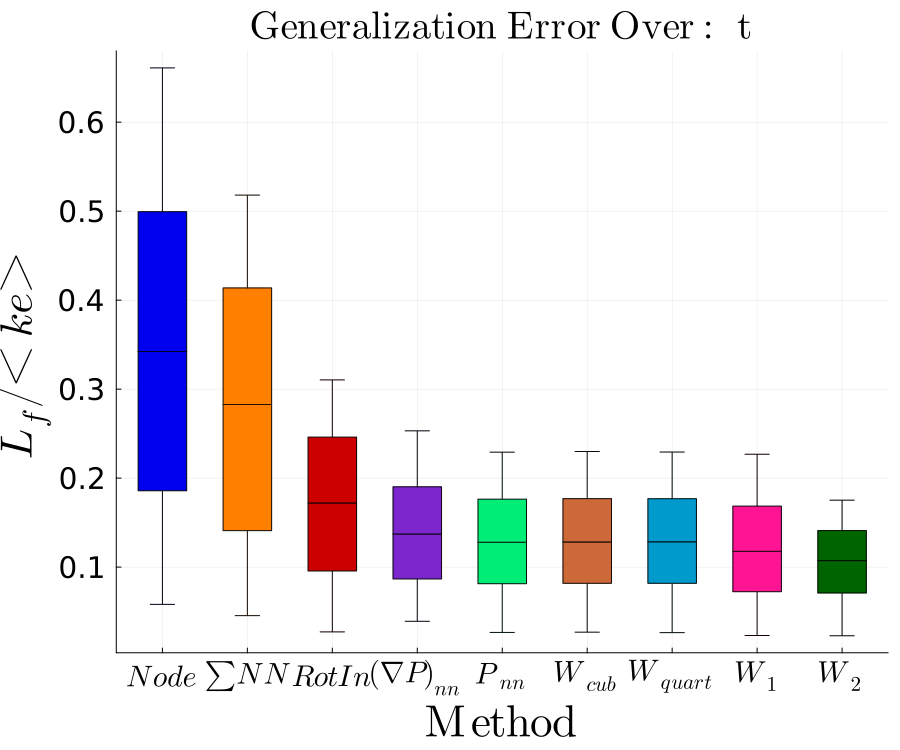}
\end{subfigure}
\caption{Measuring the relative error of $L_f$ as a percentage of total kinetic energy. The generalization error over $M_t$ is computed over 3 different turbulent Mach numbers $M_t \in \{0.04, 0.08, 0.16\}$ and on the order of the eddy turn over time scale. We see that the parameterized smoothing kernel $W_2$ outperforms under this field based measure. }
\label{fig:Lf_skl1_gen}
\end{figure}

\begin{figure}[htp]
\centering
\begin{subfigure}[b]{0.48\textwidth}
\centering
\includegraphics[width=1\textwidth]{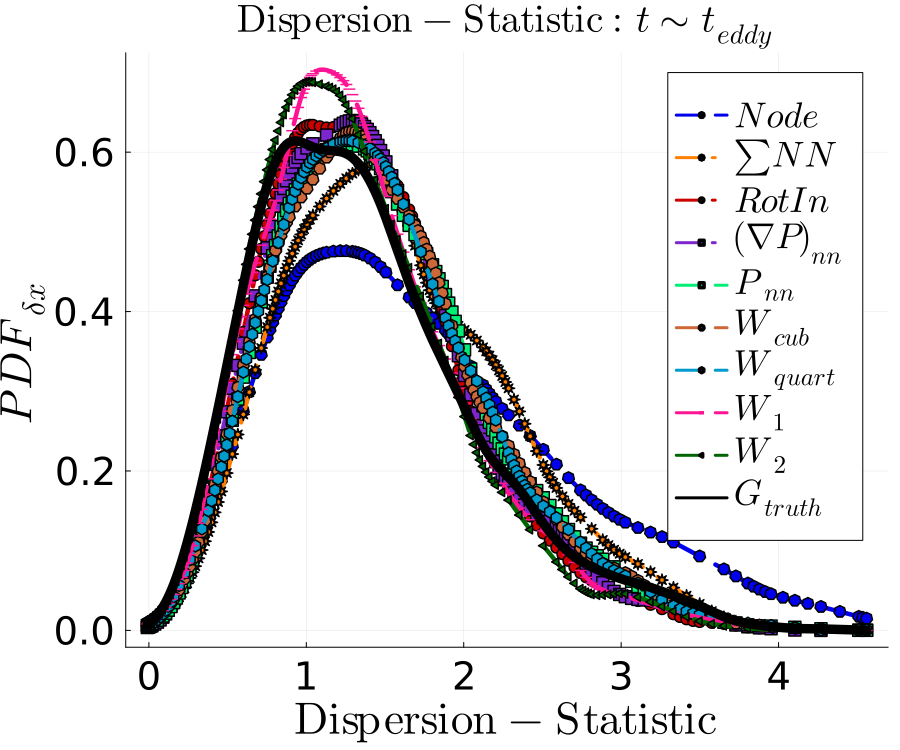}
\end{subfigure}
\begin{subfigure}[b]{0.48\textwidth}
\centering
\includegraphics[width=1\textwidth]{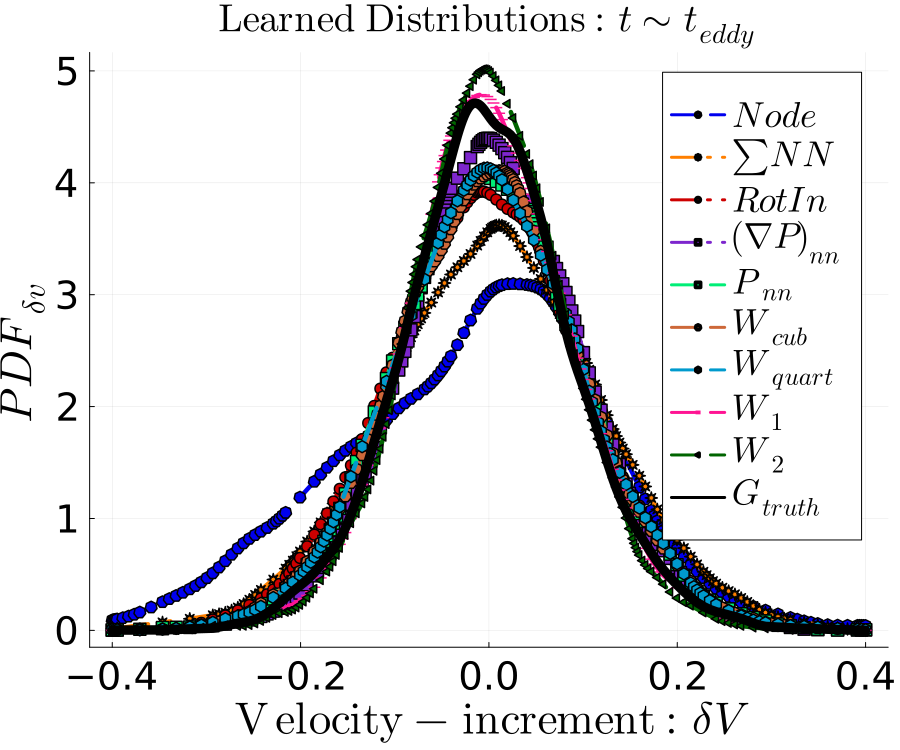}
\end{subfigure}
\caption{Single particle statistics on larger time scales (roughly 20 times longer than seen in training), with the new parameterized smoothing kernels having the best fit. In this case, the velocity increment distribution is used in learning with $L_{kl}$. }
\label{fig:sing_part_t50_comp}
\end{figure}

\begin{figure}[htp]
\centering
\begin{subfigure}[b]{0.48\textwidth}
\centering
\includegraphics[width=1\textwidth]{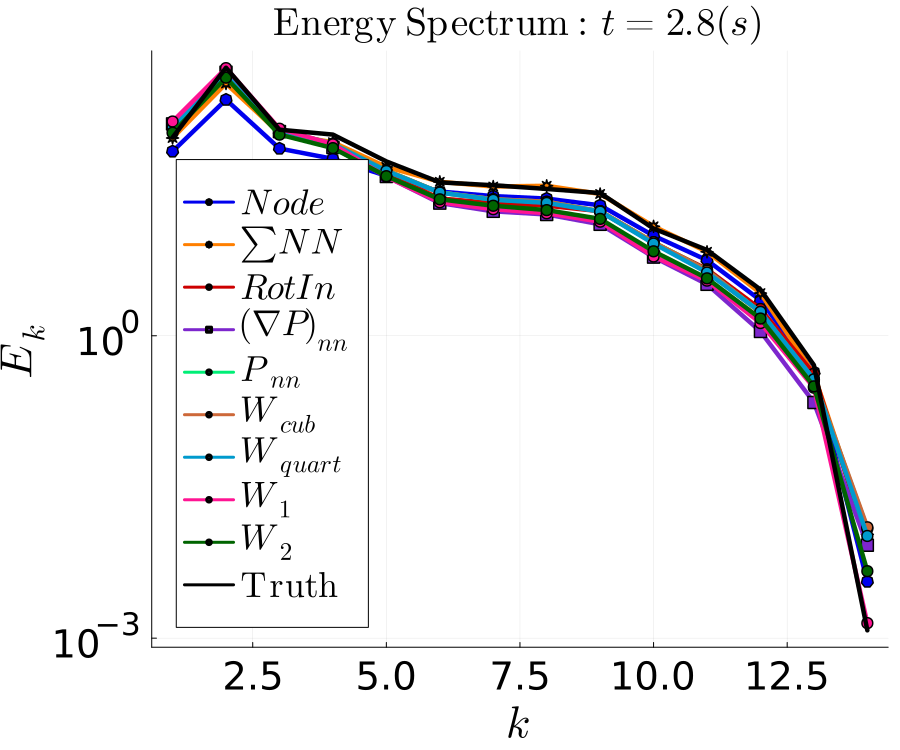}
\end{subfigure}
\begin{subfigure}[b]{0.48\textwidth}
\centering
\includegraphics[width=1\textwidth]{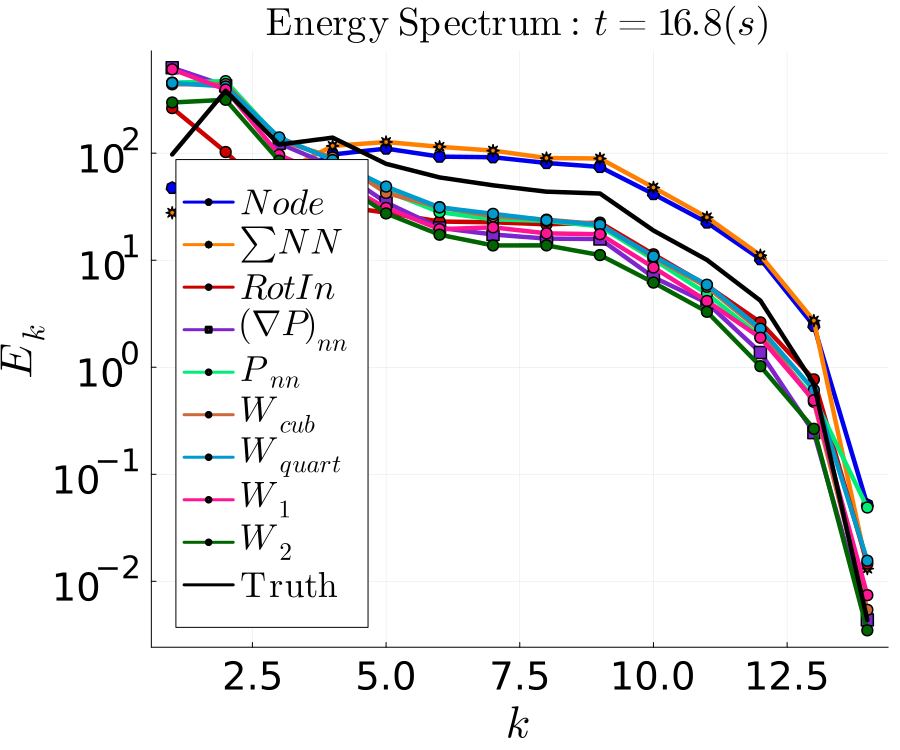}
\end{subfigure}
\caption{Comparing energy spectrum over time; $t = 2.8s \sim t_{\eta}$ is the Kolmogorov time scale, and $t = 16.8s \sim t_{eddy}$ is at the scale of the eddy turn over time. We see that when, using $L_f$ then $L_{kl}$, as more SPH structure is included in the models the better it is at capturing energy contained in the large scales, however the less informed models overestimate the energy contained in small scales and the SPH-informed models underestimate the energy contained in the small scales.}
\label{fig:ek_over_t_lfklt}
\end{figure}

In this section we show numerical evidence of the generalizability of the models when using the $L_{kl}$ defined above. This is not reported in the main text as there are only slight differences seen in the results. However, the largest difference is seen in Fig. \ref{fig:Lf_skl1_gen} when compared to Fig. \ref{fig:lf_gen_over_time_mt}, namely, the generalization error in the rotational invariant model (when measured with $L_f/\left<ke\right>$) has improved by roughly $50\%$. The first part of training is guided by $L_f$, then switched to $L_{kl}$ in order to first guide the parameters to reproduce large scale structures, then converge the loss using the $L_{kl}$ in order to learn the small scale characteristics seen in the velocity increment statistics (see Fig. \ref{fig:skl1_comp_W_loss_conv}). Furthermore, this was done in order to drive the distributions close enough to avoid large sensitivities of the KL divergence with respect to the predicted and target distributions. In what is shown below, we see that by using the statistics based loss, the models perform better at generalizing with respect to the single particle statistics (as seen in Fig. \ref{fig:sing_part_t50_comp} when only $L_f$ was used. However, now the energy spectrum (Fig. \ref{fig:ek_over_t_lfklt}) of each model performs worse at generalizing to longer time scales. This could be due to the fact that the models are overfit to reproduce the velocity increment of the DNS particles and degrade with respect to other measures. This suggests that the field based loss is a better choice to capture the energy spectrum, and the $L_{kl}$ is better to improve generalizability with respect to single particle statistics. The main conclusion between experimenting with both loss formulations is that the SPH based parameterization using $W_2$ performs best with respect to both statistical and field based measures at generalizing over longer time scales and different Mach numbers.

\begin{figure}[htp]
\centering
\begin{subfigure}[b]{0.48\textwidth}
\centering
\includegraphics[width=1\textwidth]{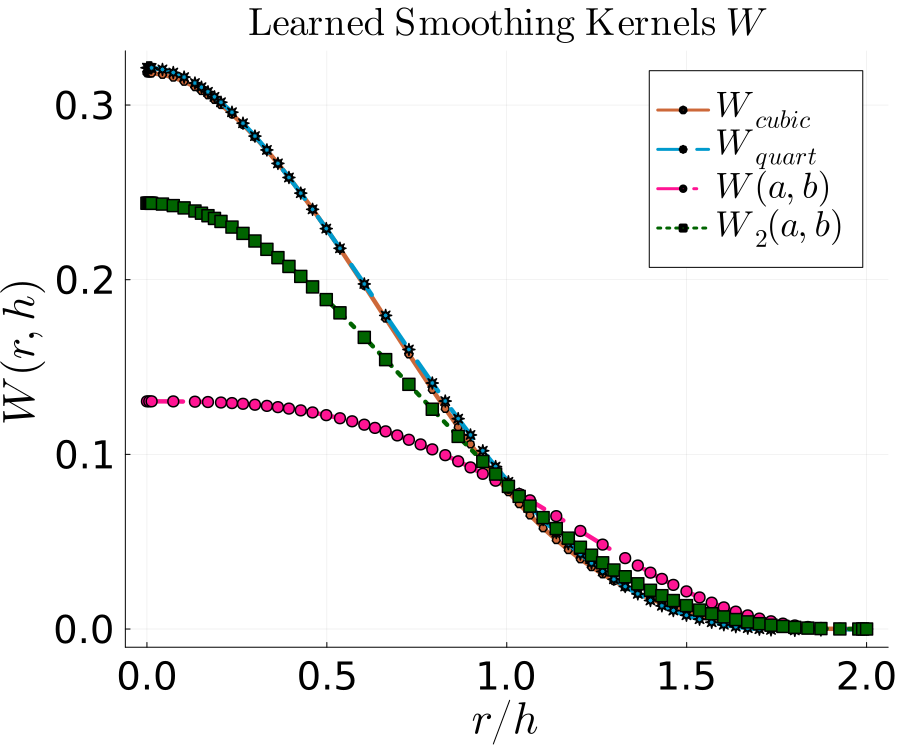}
\end{subfigure}
\begin{subfigure}[b]{0.48\textwidth}
\centering
\includegraphics[width=1\textwidth]{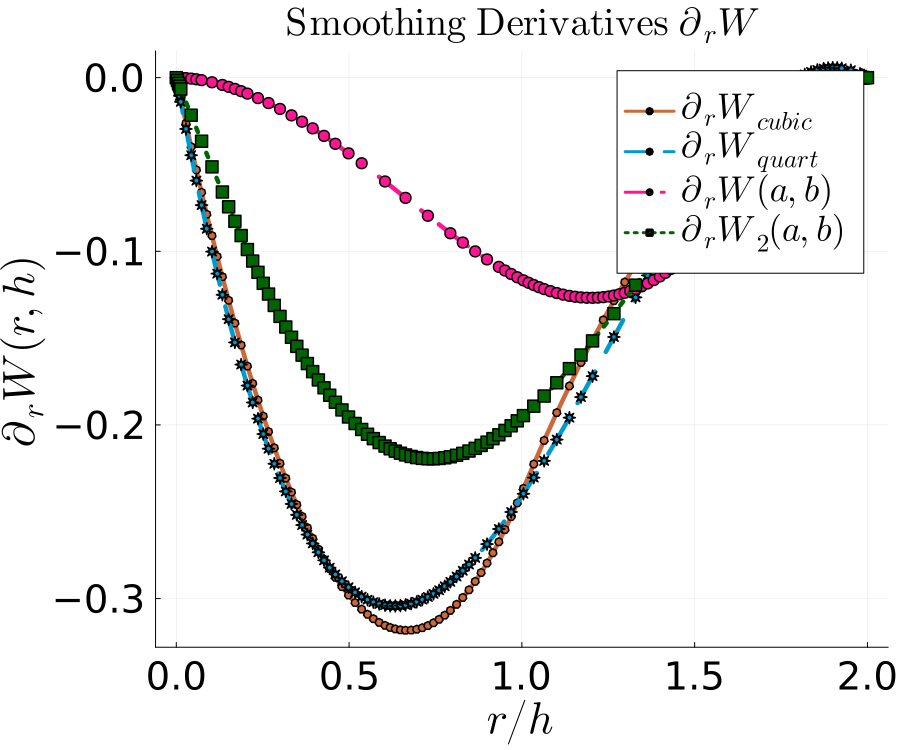}
\end{subfigure}
\caption{Learned parameterized smoothing kernels compared to cubic and quartic smoothing kernels.}
\label{fig:skl1_comp_W_loss}
\end{figure}

\FloatBarrier




\section{
Validating methodology: Training and Evaluating Models on SPH data}
\label{sec:results}

In this Section, we test the learning algorithm by training the hierarchy of models on "synthetic" SPH data, and see if the learned parameters of the fully parameterized SPH model converge to the true parameters, as well as test if the other models can interpolate onto the flow. Furthermore, we demonstrate the effects of incrementally embedding physical structure into the Lagrangian models. We show the ability of the mixed mode FSA + AD method to learn the parameterized Lagrangian and SPH based simulators on SPH flow data (as seen in Section \ref{sec:SPH}) with the field based loss and a mixture of field based and statistical based loss functions. We show that each  model is capable of solving the interpolation problem, which we evaluate using several quantitative and qualitative diagnostics, but as more physical structure is embedded in the model, the better is the generalizability to flows not seen in training and the better it conserves linear and angular momentum. 

Each parameterized Lagrangian model within the hierarchy (see Section \ref{sec:hiearchy}) is trained under equivalent conditions: (a) on the same SPH samples (see Section \ref{sec:SPH}) of fixed temporal duration (which we choose to be equal to the time scale required for a pair of neighboring particles to separate by the distance which is on average a factor of $O(1)$ larger than the pair's initial separation and henceforth denoted $t_{\lambda}$); (b) with the same loss function $L_{kl} + L_f$; (c) with the FSA method; and (d) with a deterministic forcing $\bm f_{ext}$ (Section \ref{sec:ext_force} with constant rate of energy injection). 

\subsection{
Interpolation of Models: SPH data}
\begin{figure}[ht]
\centering
\begin{subfigure}[b]{0.48\textwidth}
\centering
\includegraphics[width=1\textwidth]{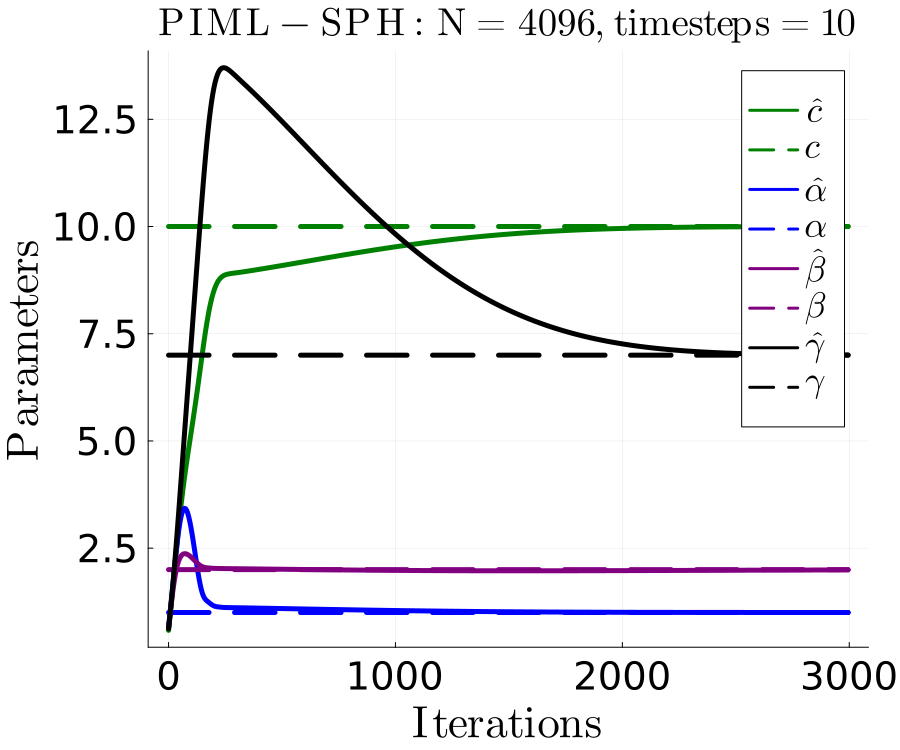}
\caption{Learning physical parameters}
\end{subfigure}
\begin{subfigure}[b]{0.48\textwidth}
\centering
\includegraphics[width=1\textwidth]{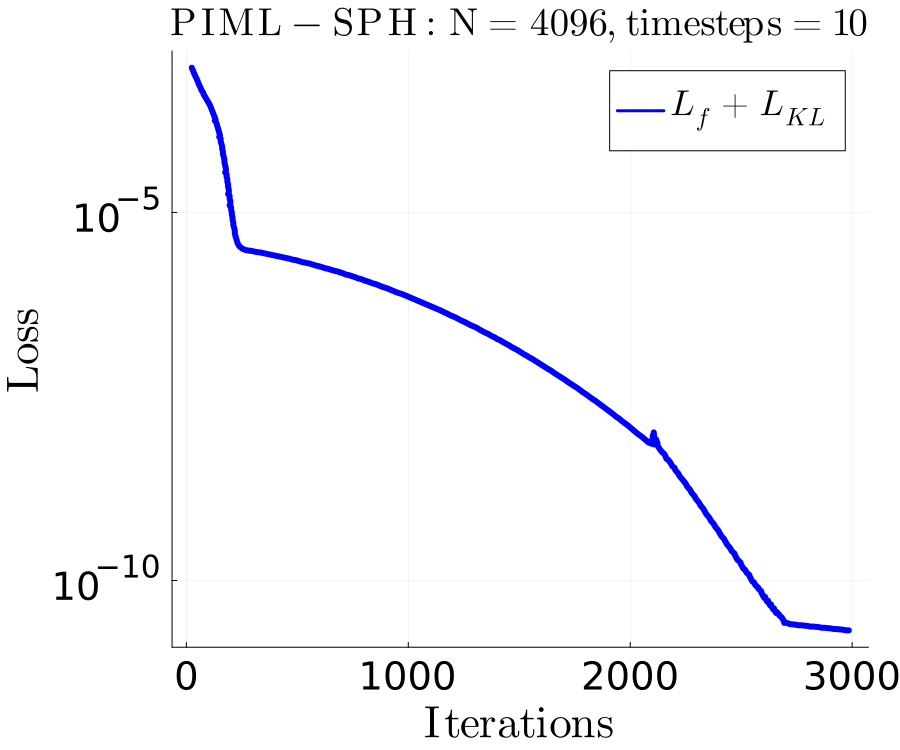}
\caption{Loss function converging}
\end{subfigure}
\caption{Solving an inverse problem for the fully physics informed model on 3D SPH flow with deterministic external forcing on 4096 particles over the SPH based parameters. The solid lines show the SPH model parameters (initially chosen to be uniformly distributed about (0,1)) converging to the dashed lines representing the ground truth parameters. Here the $L_{kl} + L_f$ loss function (see Section \ref{sec:Loss}) is used (where $L_f$ is used in pre-training up to 2100 iterations as seen in the small increase in the loss as the $L_{kl}$ is added in). }
\label{fig:inverse_phys_lf_kl}
\end{figure}

\begin{figure}[ht]
\centering
\begin{subfigure}[b]{0.48\textwidth}
\centering
\includegraphics[width=0.99\textwidth]{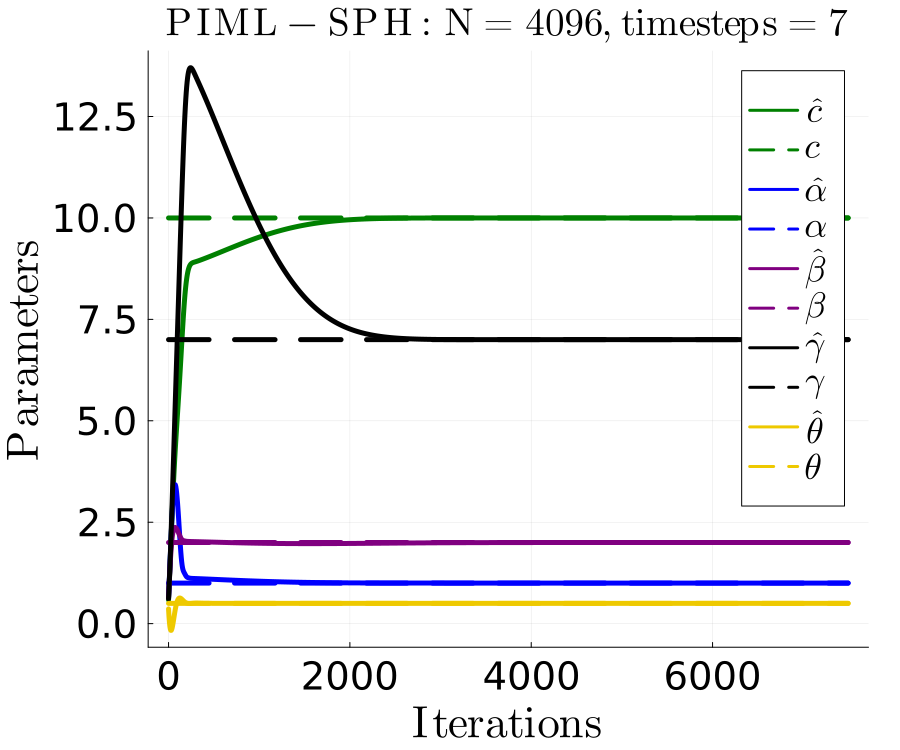}
\caption{Learning physical parameters}
\end{subfigure}
\begin{subfigure}[b]{0.48\textwidth}
\centering
\includegraphics[width=0.99\textwidth]{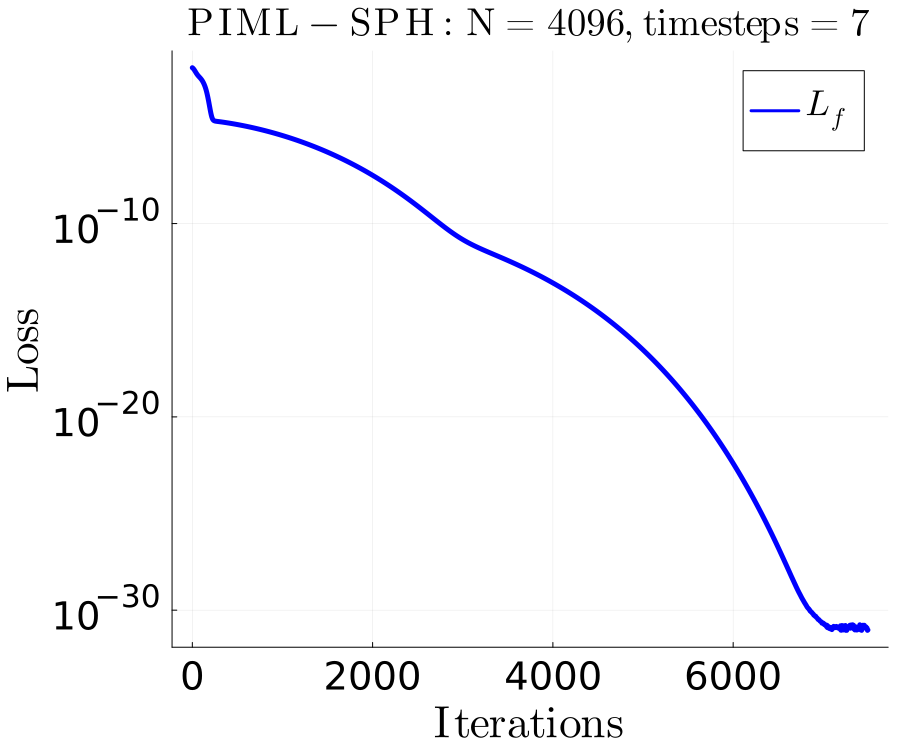}
\caption{Loss function converging}
\end{subfigure}
\caption{Solving the inverse problems for 3D SPH flow using $L_f$ with a linear deterministic external forcing on 4096 particles over physical parameters including the rate of energy injection $\theta_{inj}$. Notice the field based loss (using MSE) converging to machine precision, validating the learning algorithm and its stability. We also note that the smoothness of the convergence of the loss is most likely due to the relatively small number of parameters in this constrained optimization problem leading to a smooth loss landscape with an identifiable global optimum within a reasonable neighborhood of initial parameters. For example, when NNs were embedded withing in the ODE structure, the loss functions were observed to have some fluctuation along a downward trend towards convergence.}
\label{fig:inverse_fig_phys_theta}
\end{figure}

In Fig. \ref{fig:inverse_phys_lf_kl} (and Fig. \ref{fig:inverse_fig_phys_theta}) we see that the physics informed parameterized SPH simulator is learn-able; the estimated physical parameters $\hat{\alpha}, \hat{\beta}, \hat{c}$, $\hat{\gamma}$, and energy injection rate $\hat{\theta_{inj}}$ converging to the true values where the initial guess for each parameter is uniformly distributed about $(0, 1)$. Furthermore, these results show that the Weakly compressible SPH flow is more sensitive to the parameters $\alpha$ and $\beta$, which control the strength of the artificial viscosity term, compared to parameters $c$ and $\gamma$, which control the local slope and shape of the EoS. This is due to the weakly compressible nature of the flow being less sensitive to changes in the speed of sound. Fig. \ref{fig:inverse_eos_main} illustrates the ability of the mixed mode method applied to the EoS-NN model to learn (approximate) physically interpretable functions (namely the barotropic equation of state $P(\rho)$) using NNs embedded within an SPH model. However, we notice that the learned EoS $P_{nn}(\rho)$ begins to deviate from the $P_{truth}(\rho)$ outside the domain of densities that are seen in training. 

\begin{figure}[ht]
\centering
\begin{subfigure}[b]{0.48\textwidth}
\centering
\includegraphics[width=1\textwidth]{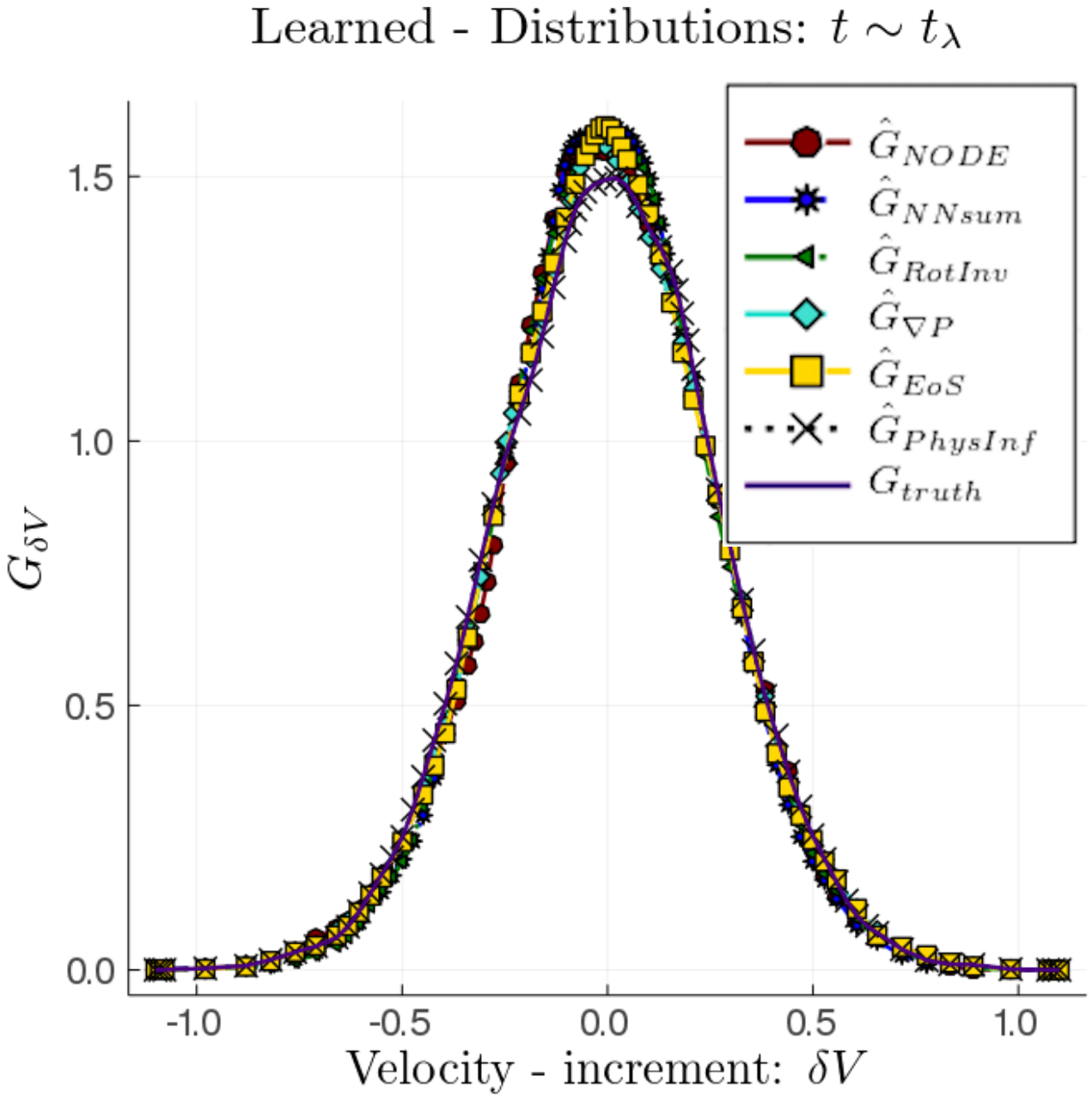}
\end{subfigure}
\begin{subfigure}[b]{0.48\textwidth}
\centering
\includegraphics[width=1\textwidth]{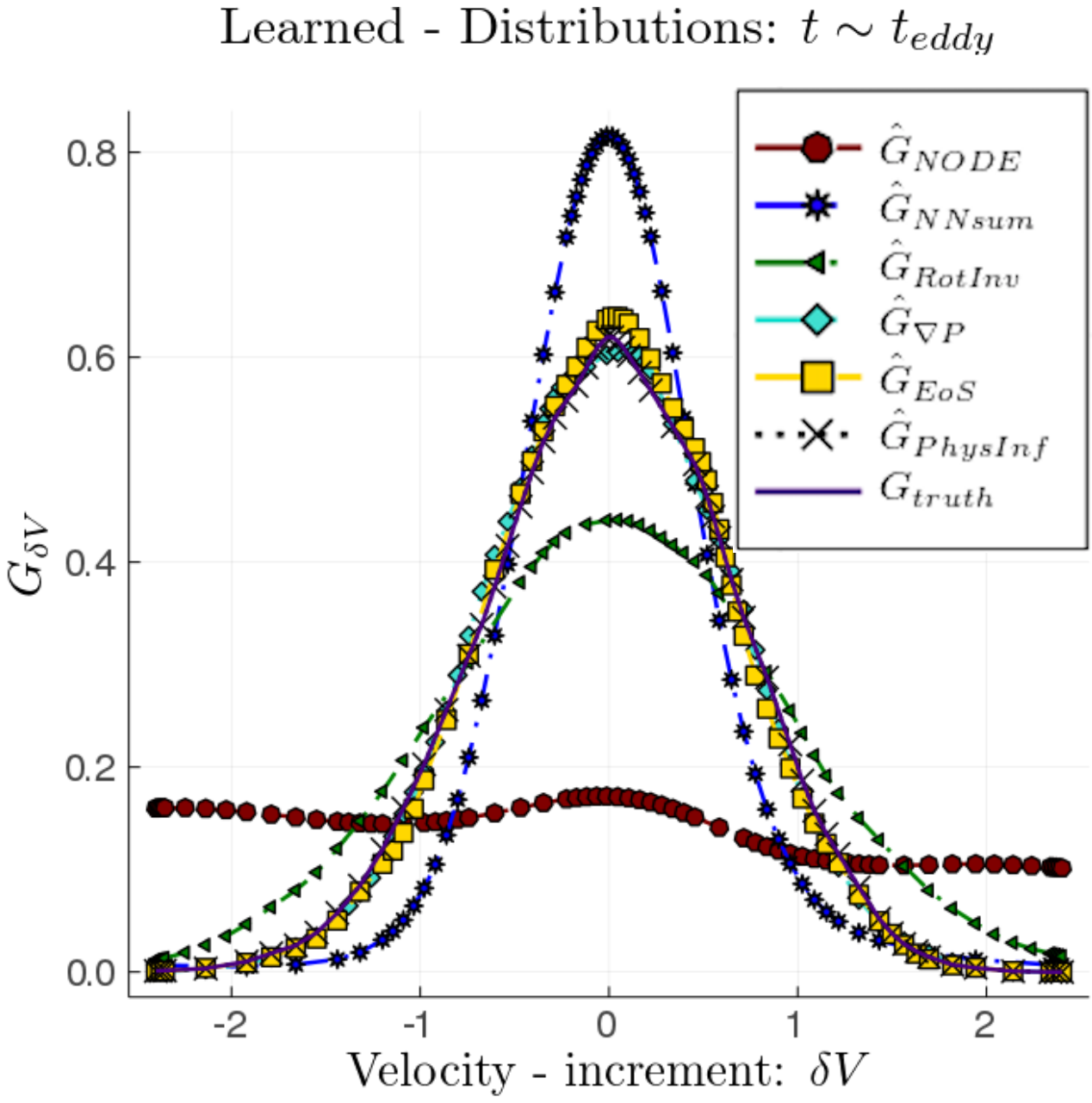}
\end{subfigure}
\caption{The velocity increment distribution learned over each model in the hierarchy Section \ref{sec:hiearchy} on the time scale $t_{\lambda}$  using the $L_{kl} + L_f$ loss function and the FSA method. We see that, on the learned time scale, all the models do well at capturing the small scale Lagrangian velocity increment statistics, however, only the more physics informed models do well at generalizing to larger time scales not seen in training.}
\label{fig:Gv_comp}
\end{figure}

\begin{figure}[ht]
\centering
\begin{subfigure}[b]{0.48\textwidth}
\centering
\includegraphics[width=1\textwidth]{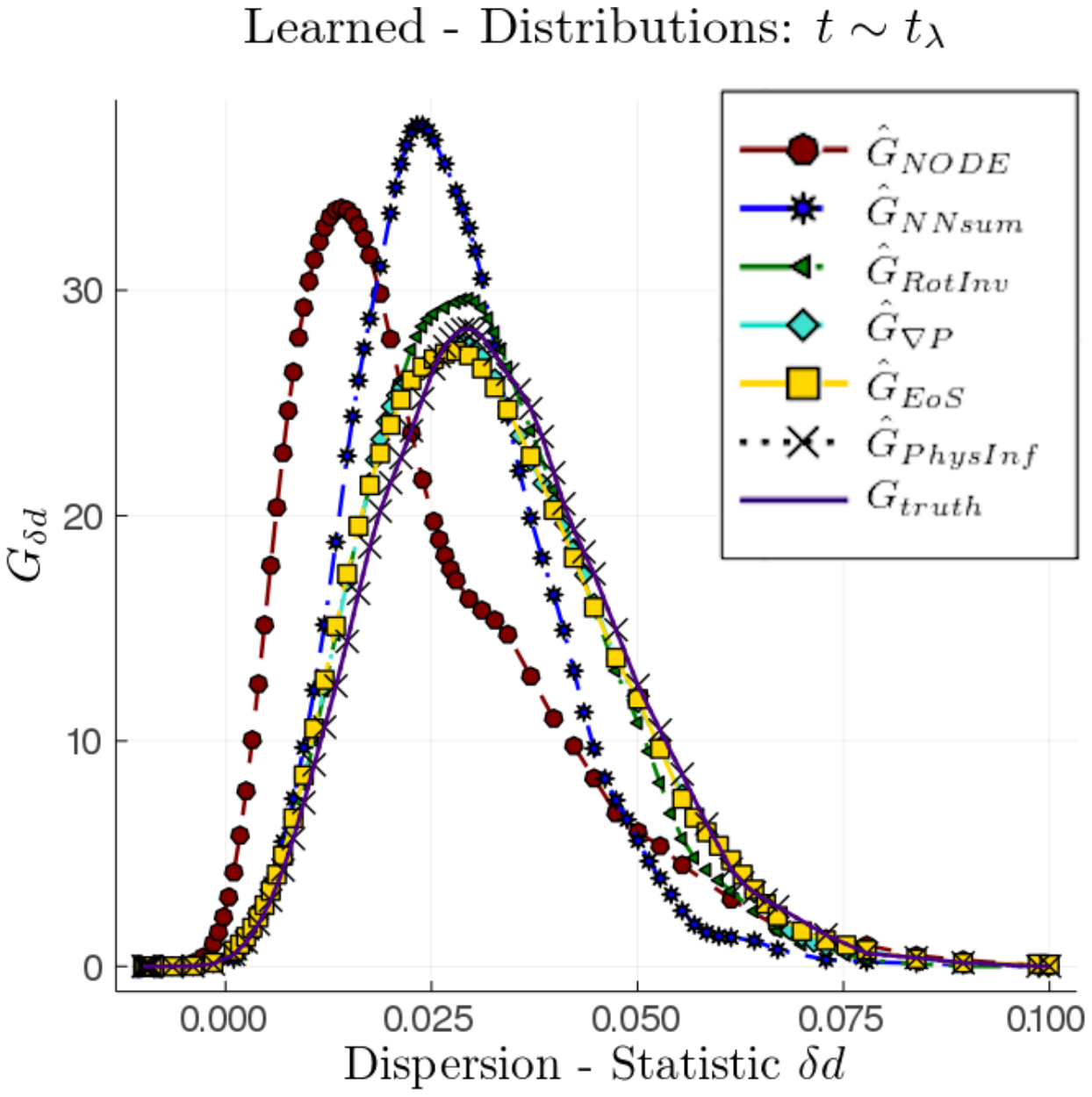}
\end{subfigure}
\begin{subfigure}[b]{0.48\textwidth}
\centering
\includegraphics[width=1\textwidth]{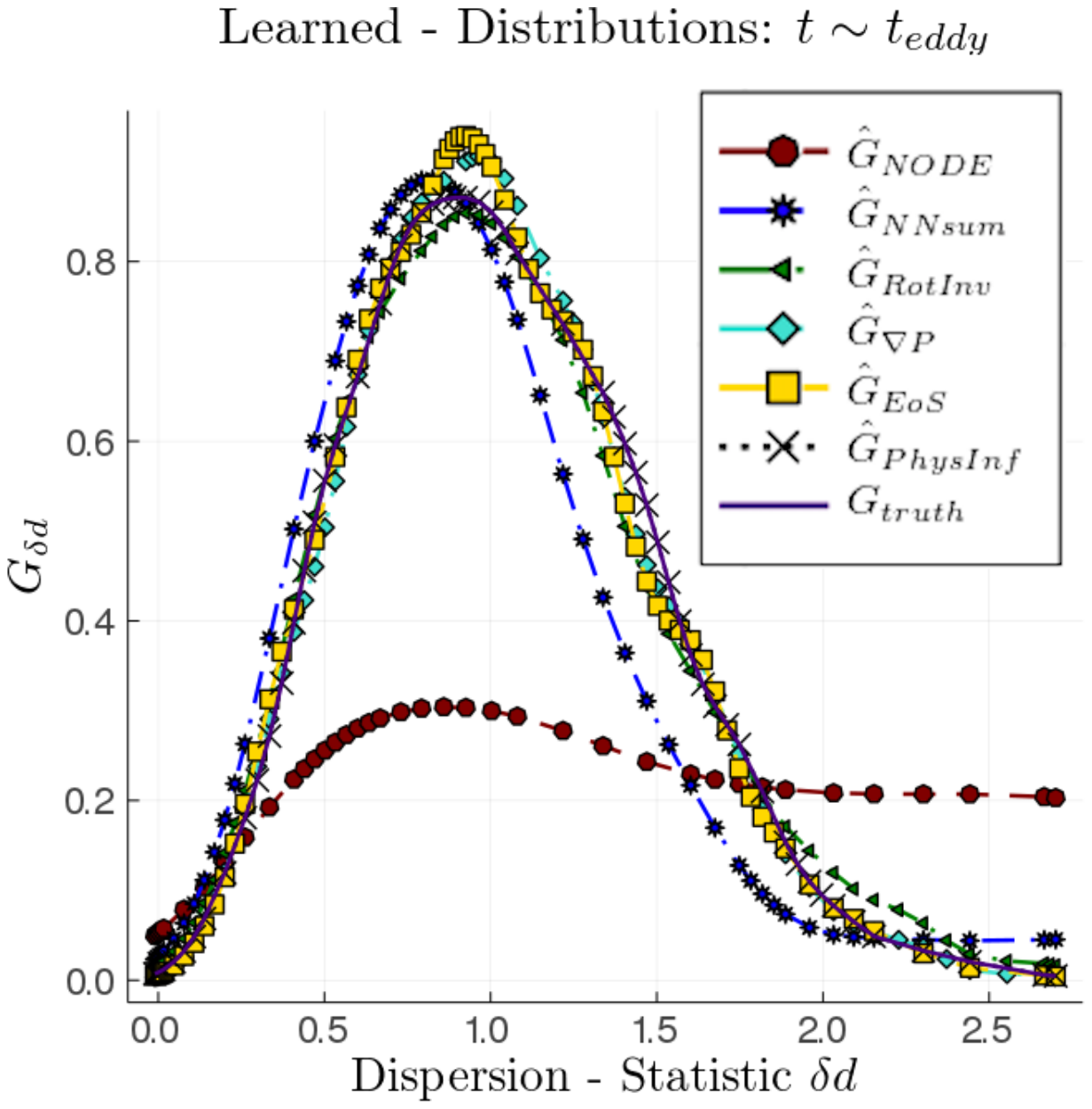}
\end{subfigure}
\caption{The single particle dispersion statistic distribution being used as a diagnostic check on each model in the hierarchy Section \ref{sec:hiearchy} on the learned time scale and on the eddy turn over time scale}
\label{fig:Gd_comp}
\end{figure}

In Fig. \ref{fig:Gv_comp}, we see that each model is capable of learning the parameters so that the underlying velocity increment distributions are approximated on the time scale on which learning takes place. Furthermore, In Fig. \ref{fig:Gd_comp} we test the trained models on their ability to reproduce the distribution associated with the single particle dispersion statistic \cite{yeung_borgas_2004} (showing how far the particle disperses from its initial condition on a set time scale). Here it is important to note that the dispersion statistic was not enforced in the $L_{kl}$ loss, and so Fig. \ref{fig:Gv_comp} illustrates an external diagnostic test. From this we find that without enforcing the dispersion statistic in the loss function, the less informed models (namely NODE, and NN summand) fail to reproduce the true dispersion statistics on both time scales ($t_{\lambda}$, and $t_{eddy}$).

These Figures show that the Lagrangian models introduced in Section \ref{sec:hiearchy} are learn-able (i.e. interpolated on the SPH training set). We arrive at these results applying the methods described in Section \ref{sec:fsa_asa} to a mixture of the loss functions $L_{kl} + L_f$ introduced in Section \ref{sec:Loss}.

\subsection{Generalizability (extrapolation capability) on SPH flow data}
\label{sec:extrapolate}
\label{sec:generl}

\begin{figure*}[hbt]
\centering
{\includegraphics[width=0.9\textwidth]{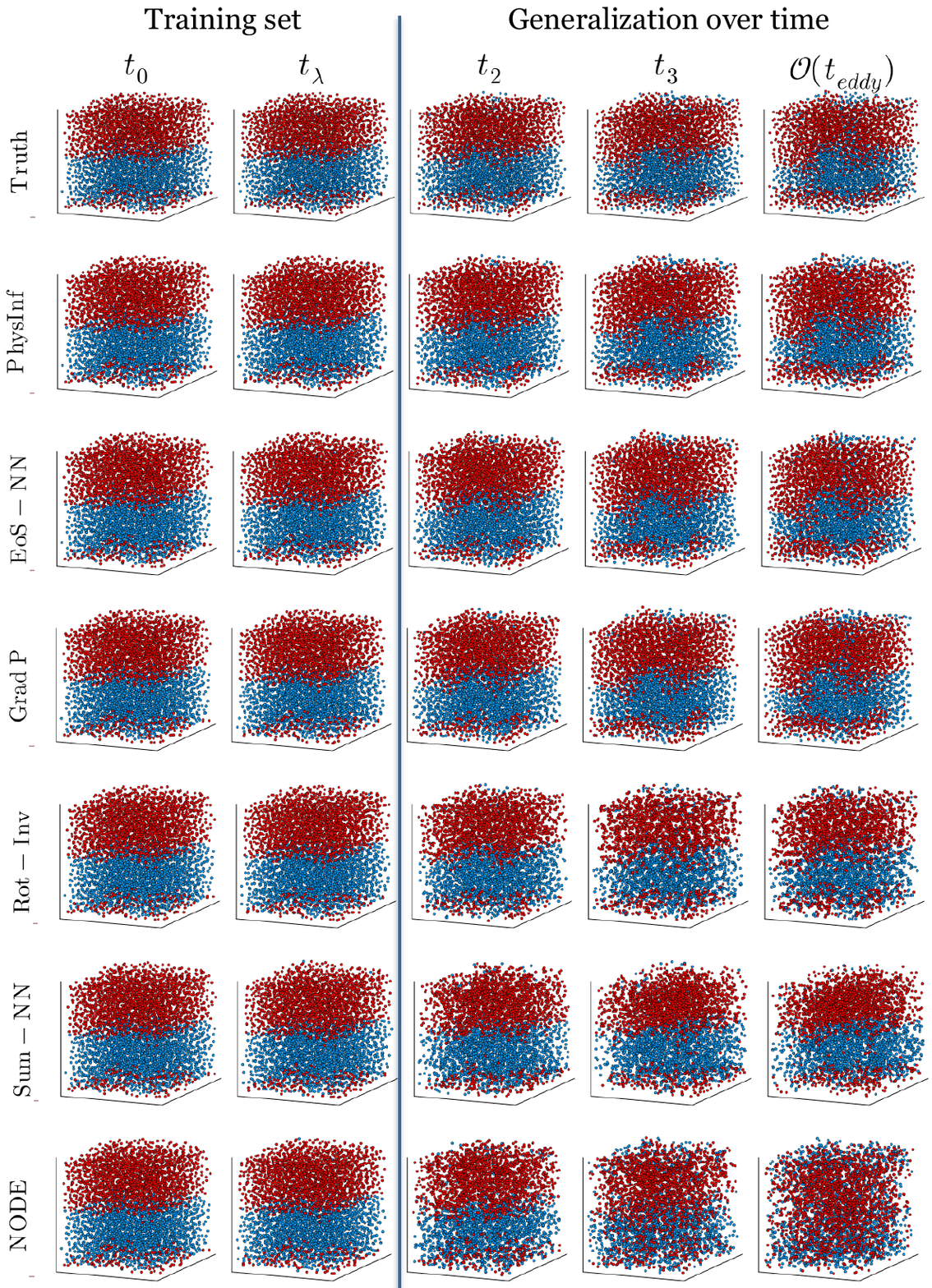}}

\caption{Snapshots of SPH particles evolving, comparing the ground truth SPH data to the learned models, where learning is done with the $L_{KL} + L_f$ loss function. Even though the training is occurring on the shortest (physically relevant) time scale $t_{\lambda}$ the more physics informed models are able to generalize to much longer time scales -- all the way to the longest (physically relevant) time scale, $t_{eddy}$ (which is the turnover time scale of the largest eddy of the flow).}
\label{fig:gen_t_qual_sph_pred}
\end{figure*}

 When the training is complete (i.e. when the loss function reaches its minimum, and conditions are set as described in Section \ref{sec:results}) we validate extrapolation capability of the models on a validation set, and test data set. The validation data set corresponds to data generated from the same initial conditions but are longer in duration (up to the eddy turn over time, see Fig. \ref{fig:gen_t_qual_sph_pred}).  The test set corresponds to flow derived from the setting corresponding to stronger turbulent Mach numbers (which we control by increasing intensity of the injection term, ${\bm f}_{ext}$, while keeping the integral, i.e. energy injection scale, constant, thus increasing the turbulent mach number, $M_t$). Furthermore, we measure the errors in the physical symmetries (\autoref{table:rot_tran_sph}), showing that including the SPH framework (even with NNs embedded within SPH) conserves linear and angular momentum.

\begin{table}[ht]
  \caption{Errors in Rotational and Translational Symmetries on trained models}
  \label{table:rot_tran_sph}
  \centering
  \begin{tabular}{llll}
    \toprule
    Trained Model  &   Rotational Error     &  Translational Error    \\
    \midrule
    NODE    &  4.59    &  6.57      \\
    NN Sum    &  5.42   &  1.51 \\
    Rot Inv    &   $1.21\times10^{-27}$  &    $4.22\times10^{-27}$  \\
    $\nabla P$    &    $1.20\times10^{-2}$      &    $1.48\times10^{-27}$   \\
    EoS NN    &     $3.49\times10^{-28}$      &  $1.18\times10^{-27}$    \\
    Phys Inf    &     $2.31\times10^{-27}$      &  $7.88\times10^{-27}$    \\
    \bottomrule
  \end{tabular}
\end{table}

\FloatBarrier

\end{document}